\documentclass[prb,twocolumn,notitlepage,superscriptaddress,nofootinbib]{revtex4-2}
\usepackage{amsmath,amssymb}
\usepackage[hidelinks,colorlinks,linkcolor=blue,
citecolor=blue,urlcolor=blue,pdfduplex=DuplexFlipLongEdge]{hyperref}
\usepackage{graphicx,siunitx}
\usepackage[dvipsnames]{xcolor}
\usepackage{footmisc}

\usepackage{pifont}

\usepackage[normalem]{ulem}

\usepackage{algorithm}
\usepackage{algpseudocode}
\usepackage{comment}
\usepackage{booktabs}
\usepackage{makecell}
\usepackage{ulem}
\usepackage{bm}

\usepackage{xcolor}
\DeclareUnicodeCharacter{3000}{\textcolor{red}{BAD!!}}

\newcommand{\be}{\begin{equation}}
\newcommand{\ee}{\end{equation}}
\newcommand{\bea}{\begin{eqnarray}}
\newcommand{\eea}{\end{eqnarray}}
\newcommand{\bi}{\begin{itemize}}
\newcommand{\ei}{\end{itemize}}

\renewcommand{\be}{\beta}

\newcommand{\bpm}{\begin{pmatrix}}
\newcommand{\epm}{\end{pmatrix}}

\usepackage{amsmath,amsfonts,amssymb,amsthm,epsfig,array}
\usepackage{dsfont}
\usepackage{slashed}
\usepackage{graphics}
\usepackage{float}
\usepackage{verbatim}
\usepackage{color}
\usepackage{tabularx}
\usepackage[mathscr]{euscript}
\usepackage{mathtools}

\renewcommand{\arraystretch}{1.5}

\newcommand{\bsl}[1]{\boldsymbol{#1}}

\newcommand{\bra}[1]{\langle #1|}
\newcommand{\ket}[1]{|#1 \rangle}
\newcommand{\braket}[2]{\left\langle #1 | #2  \right\rangle}

\newcommand{\nint}{\xi}

\newcommand{\dsZ}{\mathbb{Z}}

\renewcommand{\Re}{\mathop{\mathrm{Re}}}
\renewcommand{\Im}{\mathop{\mathrm{Im}}}

\newcommand{\refcite}[1]{Ref.\,\cite{#1}}

\newcommand{\eq}[1]{\begin{equation} #1 \end{equation}}

\newcommand{\eqa}[1]{\begin{align}\begin{split} #1 \end{split}\end{align}}

\usepackage{environ}
\NewEnviron{eqs}{%
\begin{equation}\begin{split}
    \BODY
\end{split}\end{equation}
}

\let\oldAA\AA
\renewcommand{\AA}{\text{\normalfont\oldAA}}

\newcommand{\ie}{{\emph{i.e.}}}
\newcommand{\eg}{{\emph{e.g.}}}

\algrenewcommand\algorithmicrequire{\textbf{Require:}}

\usepackage{cleveref}
\crefname{appendix}{App.}{Apps.}
\crefname{equation}{Eq.}{Eqs.}
\crefname{figure}{Fig.}{Figs.}
\crefname{table}{Tab.}{Tabs.}
\crefname{section}{Sec.}{Secs.}
\creflabelformat{appendix}{[#2#1#3]}

\begin{document}

\title{Reduced Density Matrices Through Machine Learning}

\author{Awwab A. Azam}
\thanks{These authors contributed equally to this work.}
\affiliation{Department of Electrical and Computer Engineering, University of Florida, Gainesville, USA}

\author{Lexu Zhao}
\thanks{These authors contributed equally to this work.}
\affiliation{Department of Physics, University of Florida, Gainesville, Florida 32611, USA}

\author{Jiabin Yu}
\email{yujiabin@ufl.edu}
\affiliation{Department of Physics, University of Florida, Gainesville, Florida 32611, USA}
\affiliation{Quantum Theory Project, University of Florida, Gainesville, FL, USA}

\begin{abstract}

$n$-particle reduced density matrices ($n$-RDMs) play a central role in understanding correlated phases of matter, but their calculation is often computationally inefficient for strongly-correlated states at large system sizes. In this work, we use neural network (NN) architectures to accelerate and even predict $n$-RDMs for large systems. Our underlying intuition is that, for gapped states, $n$-RDMs are often smooth functions over the Brillouin zone (BZ) and are therefore interpolable, allowing NNs trained on small-size systems to predict large-size ones. Building on this, we devise two NNs: (i) a self-attention NN that maps random RDMs to physical ones, and (ii) a Sinusoidal Representation Network (SIREN) that directly maps momentum-space coordinates to RDM values. 
We test the NNs on RDMs in three 2D models: the pair-pair correlation functions of the Richardson model of superconductivity, the translationally-invariant Hartree-Fock (HF) 1-RDM  in a four-band repulsive model, and the translation-breaking HF 1-RDM in the half-filled Hubbard model.
We find that a SIREN trained on a $6\times 6$ momentum mesh and a SIREN trained on $4$ tilted meshes (each of which has $12$ momentum points) can predict the $18\times 18$ pair-pair correlation function with a relative accuracy of $94.29\%$ and $93.77\%$, respectively.
NNs trained on $6\times 6$ and $8\times 8$ meshes provide high-quality initial guesses for $50\times 50$ translation-invariant HF and $30\times 30$ fully translation-breaking-allowed HF, reducing the required number of iterations by up to $91.63\%$ and $92.78\%$, respectively, compared to random initializations.
Our results illustrate the potential of NN-based methods for interpolable $n$-RDMs, which might open a new avenue for future research on strongly correlated phases.

\end{abstract}

\maketitle

\emph{Introduction}
Understanding strongly correlated phases of matter is a central topic in condensed matter physics. 
The rapid progress of neural-network (NN) methods has created new opportunities for studying strongly correlated quantum phases via NN-based wavefunction ansatzes
~\cite{Carleo_2017,Nomura_2017,Glasser_2018, Lou_2019_NN_Wavefunction, Choo_2019, Irikura_2020,Hibat_Allah_2020, Pfau_2020_NN_Wavefunction, Sharir_2020, Hermann_2020,PhysRevResearch.3.043126,li2021fermionicneuralnetworkeffective, Luo_2021, Lee_2021, Adams_2021, roth2021groupconvolutionalneuralnetworks, scherbela2021solvingelectronicschrodingerequation, Robledo_Moreno_2022, Pescia_2022, Li_2022, Li_2022_NNWavefunction, Guti_rrez_2022,vivas2022neuralnetworkquantumstatessystematic, luo2022gaugeequivariantneuralnetworks, chen2022simulating21dlatticequantum, Martyn_2023, Cassella_2023, Wilson_2023, Ren_2023, Roth_2023, li2023forwardlaplaciannewcomputational, luo2023pairingbasedgraphneuralnetwork, kim2023neuralnetworkquantumstatesultracold, Vonglehn_2023_selfattentionansatzabinitioquantum, Teng_2024_NN_ansatz_FQHE, Pescia_2024, Qian_2025, Goldshlager_2024, Lou_2024, chen2024antnbridgingautoregressiveneural, chen2025exactefficientrepresentationtotally, li2025deeplearningshedslight, Romero_2025, Geier_2025_ML_QMC, Luo_2025_ML_QMC_tMoTe2,
Gu_2025_Hubbard_NNansatz, chen2025neuralnetworkaugmentedpfaffianwavefunctions, tJ_NQS_model,
fu2025minimaluniversalrepresentationfermionic, roth2025superconductivitytwodimensionalhubbardmodel, lange2026simulatingsuperconductivitymixeddimensionaltparalleljparalleljperp},
which have been applied to various systems including Hubbard model ~\cite{Nomura_2017, PhysRevX.7.031038, Lou_2019_NN_Wavefunction, 
Ibarra_Garc_a_Padilla_2025,
PhysRevLett.121.167204, Irikura_2020, roth2025superconductivitytwodimensionalhubbardmodel}.
One notable type of such NN ansatze uses the self-attention mechanism~\cite{NIPS2017_3f5ee243}
to construct correlated orbitals from single-particle orbitals, expressing the many-body wavefunction as a linear combination of a small number of Slater determinants formed from these correlated orbitals ~\cite{Pfau_2020_NN_Wavefunction,Vonglehn_2023_selfattentionansatzabinitioquantum,Teng_2024_NN_ansatz_FQHE,Geier_2025_ML_QMC,ma2025transformerbasedneuralnetworksbackflow}. In this work, we propose a different way of using machine learning (or NNs) to understand strongly-correlated states---by focusing on their reduced density matrices. 

Given a many-body Hamiltonian, the zero-temperature $n$-particle reduced density matrix ($n$-RDM) is defined as
\eqa{
\label{eq:nRDM}
& O_{\bsl{k}_1\alpha_1,\bsl{k}_2\alpha_2,...,\bsl{k}_n\alpha_n, \bsl{k}_1'\alpha_1',\bsl{k}_2'\alpha_2',...,\bsl{k}_n'\alpha_n'} \\
& \quad = \left\langle c^\dagger_{\bsl{k}_1\alpha_1}c^\dagger_{\bsl{k}_2\alpha_2}...c^\dagger_{\bsl{k}_n\alpha_n}c_{\bsl{k}_1'\alpha_1'}c_{\bsl{k}_2'\alpha_2'}...c_{\bsl{k}_n'\alpha_n'}\right\rangle\ ,
}
where $\langle ... \rangle$ represents the average with respect to the ground state(s), and $c^\dagger_{\bsl {k}\alpha}$ creates a particle with Bloch momentum $\bsl{k}$ and intra-unit-cell degree of freedom $\alpha$ (which labels sub-lattice, orbital, spin, etc).

The $n$-RDM plays a vital role in understanding strongly correlated states \cite{RevModPhys.35.668,Kummer1967,ERDAHL1979147,ColemanYukalov2000,Zhao2004,PhysRevA.69.042511,Mazziotti_2006_2RDMReview,PhysRevA.74.012501,Mazziotti2007RDM,Schwerdtfeger2009Ising,PhysRevLett.105.213003,Mazziotti2012,Verstichel2012Hubbard,PhysRevLett.108.263002,PhysRevLett.122.013001,RubioGarcia2019XXZ,PhysRevLett.124.180603,active_space_2RDM_4_strong_corr,castillo2021effectivesolutionconvex1body,PhysRevLett.127.023001,Li2021ThreeParticle,Knight2022Ultracold,DePrince2024}. First, the Hamiltonians of realistic condensed matter systems almost always involve only one-body and two-body terms, implying that the 1-RDM and 2-RDM are sufficient to evaluate the many-body ground-state energy \cite{RevModPhys.35.668,GarrodPercus1964,PhysRevA.57.4219,PhysRevA.65.062511,Cances2006RDM,PhysRevLett.117.153001,deprince2023variationaldeterminationtwoelectronreduced}. This observation lies at the heart of semidefinite programming approaches~\cite{semidefinite_programming,Nakata2001,Zhao2004,PhysRevLett.93.213001,PhysRevA.72.032510,Mazziotti_2006_2RDMReview,Erdahl2007LowerBound,PhysRevLett.106.083001,PhysRevA.102.052819,PhysRevLett.130.153001,scheer2024hamiltonianbootstrap}, which iteratively compute the 2-RDM. Second, the order parameters that characterize spontaneous symmetry breaking---such as charge-density waves and magnetism---often depend only on the 1-RDM \cite{Giesbertz_2019}, which explains the effectiveness of the Hartree-Fock (HF) method~\cite{Szabo_1989_ModernQC}.

Given the importance of the $n$-RDMs, it would be transformative to design an NN that predicts $n$-RDMs for a given Hamiltonian, particularly for large systems where existing algorithms become computationally inefficient. Because semidefinite programming and the HF algorithm compute $n$-RDMs iteratively, the NN predictions (even if they are not exactly precise) can serve as an initial guess which is already very close to the converged state, thereby dramatically reducing the number of iterations required for convergence. Furthermore, 2-RDM methods (e.g. semidefinite programming) are inherently susceptible to converging to non-physical solutions, but comparing the converged solution to the NN predictions may help determine whether the converged solution is physical, since the NNs are trained on physical 2-RDMs and thus their predictions are likely to be physical.

In this work, we develop two such NNs: (i) a self-attention NN that maps random initial 1-RDMs to the predicted final 1-RDMs, and (ii) a Sinusoidal Representation Network (SIREN) \cite{sitzmann2020implicitneuralrepresentationsperiodic} that learns $n$-RDMs and interpolates them to denser $\bsl{k}$ meshes (\cref{sec:NN_arch_appendix}).
The underlying reason for the existence of such NNs lies in the interpolable nature of $n$-RDMs.
These NNs are trained on small system sizes, where the $n$-RDMs can be efficiently computed using existing methods, and are then used to predict the $n$-RDMs of much larger systems. 
We first benchmark SIREN on the pair-pair correlation functions of the superconducting ground states of the exactly-solvable Richardson model, and find that a SIREN trained only on the $6\times 6$ mesh can achieve a relative accuracy of $94.29\%$ for predictions on $18\times 18$. Additionally, a SIREN trained on four tilted meshes, each of which has $12$ $\bsl{k}$ points, achieves a relative accuracy of $93.77\%$ on $18\times 18$ (\cref{sec:RichardsonAppendix}).

We then benchmark the self-attention NN on the translationally-invariant HF calculations for a four-band model with short-range repulsion. Our results show that an NN trained only on $6 \times 6$ and $8 \times 8$ meshes can maintain a relatively low mean squared error (MSE, below $5\times10^{-3}$) for up to $50\times 50$. Furthermore, for momentum mesh sizes from $10\times 10, 12\times 12, \dots, 50\times 50$, using the NN predictions as the initial 1-RDM for HF calculations results in a $91.61\%$---$94.24\%$ reduction in the number of HF iterations required for convergence compared to the HF calculations with random initialization (\cref{sec:toymodelpart}).

We finally benchmark both NNs on the HF calculations that allow translation breaking in half-filled Hubbard model. We again compare the HF iteration numbers between using the NN-predicted final 1-RDM as initial 1-RDMs and using the random initial 1-RDMs.
We find a significant ($60.79\%$---$92.78\%$) reduction in the number of HF iterations for up to $30\times 30$ using SIRENs trained only on the $8 \times 8$ mesh, and a self-attention NN trained only on $8\times 8$ and $10\times 10$ meshes (\cref{app:hubbard}).
The remarkable reduction in iteration numbers for HF calculations---that forbid or allow translational breaking---demonstrates the potential of NNs for speeding up the calculation of or even the precise prediction of RDMs of large-size systems.

We note that our work is the first to construct NNs that predict large-size $n$-RDMs for strongly correlated phases, exploiting the interpolable structure of $n$-RDMs in momentum space.
In contrast to our work, previous developments of NNs that involve RDMs either focus on real-space 1-RDM in ab initio calculations (such as density functional theory) to establish maps between $1$-RDM and other quantities~\cite{shao2023machinelearningelectronicstructure,Ma2024MachineLH,Hazra_2024,Li_2022,Gong_2023} or 2-RDM for chemical molecules~\cite{Vogiatzis2023DataDriven,DelgadoGranados2025ML2RDM}.
Our general idea of using NNs for interpolable $n$-RDMs, exemplified by our implementations for pair–pair correlation functions and 1-RDMs, might open a new avenue for studying strongly correlated states.

\emph{Interpolability of $n$-RDM}
In this section, we discuss the interpolability of the $n$-RDM in the BZ. 
As shown in \cref{eq:nRDM}, the $n$-RDM is a function of $2n$ Bloch momenta; for a finite 2D system with $L\times L$ unit cells and periodic boundary conditions these momenta lie on an $L\times L$ mesh in the BZ, and increasing the system size to a commensurate $L'$ simply inserts additional momentum points between those of the original mesh (\cref{fig:interpolate_schematic}) while the Bloch momenta always occupy the same convex hull—the BZ.

If $n$-RDM is a smooth function of $2n$ Bloch momenta in the thermodynamic limit, then its values on extra momentum points can be interpolated from those on the original $L\times L$ grid. Namely, although the $n$-RDMs are extrapolated from small to large real-space system sizes, they are in fact interpolated in momentum space. 
This smoothness condition is naturally satisfied for gapped ground states, and even for gapless states certain combinations of $n$-RDMs (e.g., the pair–pair correlator in the Richardson model ~\cite{Richardson1963Restricted, RichardsonSherman1964, RevModPhys.76.643}) can remain smooth functions of the Bloch momenta.
Thus, smooth $n$-RDMs (or smooth combinations thereof) are ubiquitous in condensed-matter systems.

These properties are crucial for NNs to function effectively.
Our goal is to train a NN on small system sizes and use it to predict $n$-RDMs for large systems.
As large size RDM always lie within the fixed convex hull of the training distribution, predicting smooth $n$-RDMs for large systems from the training data on small systems is an interpolation (as opposed to extrapolation) task. 
Furthermore, smooth $n$-RDMs are continuous functions, so they can be well approximated by NNs (in principle with an arbitrary degree of precision \cite{cybenko1989approximation,hornik1989multilayer,pinkus1999approximation}, given no restrictions on the number of parameters).
Therefore, our task here is analogous to image-resolution enhancement, where extra pixel points are inserted to enhance the resolution---a problem where NNs already perform well \cite{Kim_2016_VDSR}, implying the feasibility of our approach.

\begin{figure}[t]
    \centering    \includegraphics[width=1.0\linewidth]{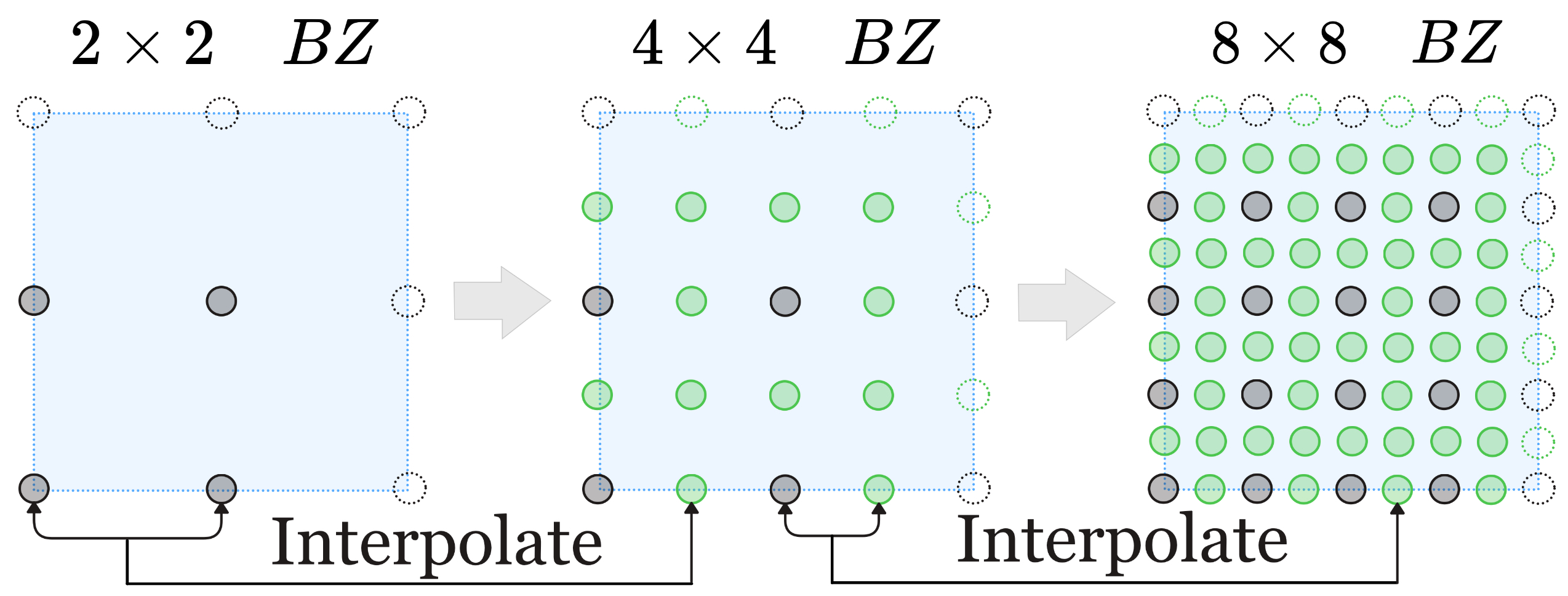}
    \caption{\textbf{Interpolability of $n$-RDMs in the BZ.}
    Momentum meshes for large systems can be generated by inserting new $\bsl{k}$ points between those of the small momentum meshes.
    Therefore, $n$-RDMs in momentum space are often interpolable.
    }\label{fig:interpolate_schematic}
\end{figure}

\emph{NN Architectures}\label{sec:NN_architectures}
We now discuss the NN architectures used in this work.
We will focus on the 1-RDM ($O_{\bsl k\alpha,\bsl k'\alpha'}=\langle c^\dagger_{\bsl k\alpha} c_{\bsl k'\alpha'}\rangle\ ,$ with $\alpha=1,2,...,\nint$) in momentum space and discuss the generalization at the end.
Particularly, we focus on fixed momentum transfer \(\bsl{k}-\bsl{k}'=\bsl{q}\), where $\bsl{q}=0$ means translation-invariance and we will include more than one $\bsl{q}$ for spontaneous translation-symmetry breaking cases.

The two NNs described in this section can be trained on small-size $O^{\bsl{q}}$ (generated via HF calculations) to predict large-size $O^{\bsl{q}}$.
We will only outline the key steps here, and leave details to \cref{sec:NN_arch_appendix}.
\begin{figure}[t]
    \centering \includegraphics[width=1.0\linewidth]{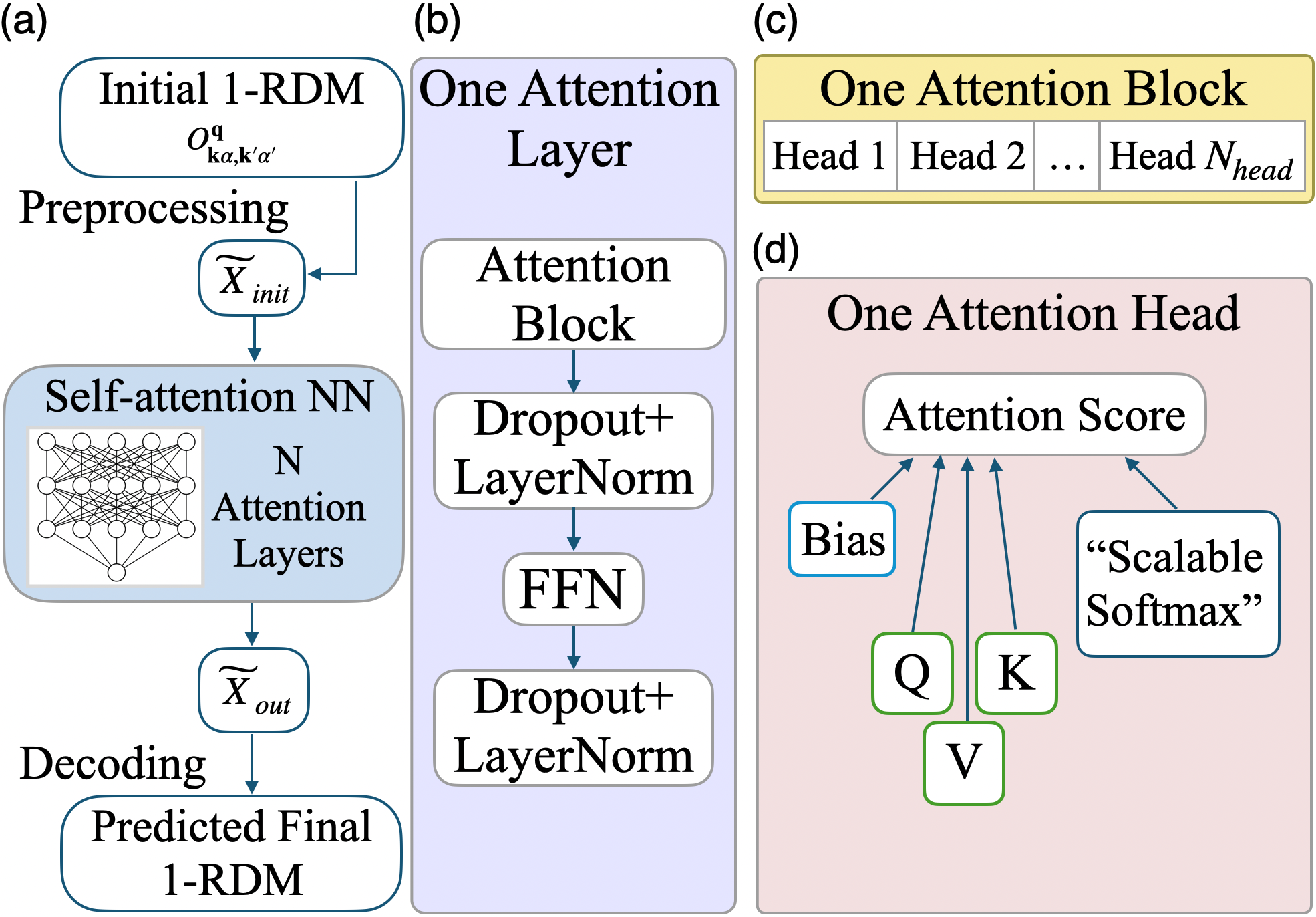}
    \caption{\textbf{Schematics of our self-attention NN.} (a) An initial 1–RDM on an \(L\times L\) BZ is preprocessed to \(\tilde{X}_{\mathrm{init}}\), transformed by the network to \(\tilde{X}_{\mathrm{out}}\), and decoded to the predicted final 1–RDM. (b) Each layer applies an attention block and a FFN with Dropout and LayerNorm. (c) The attention block includes $N_{\text{head}}$ parallel heads. (d) Each head uses \(Q,K,V\) with a learned periodic relative-position bias and a scalable softmax temperature (e.g., \(\propto \log L^{2}\)).}
\label{fig:transformer_drawing}
\end{figure}
The first NN architecture that we construct is a
self-attention NN (\cref{fig:transformer_drawing}) that maps a random initial 1-RDM $O^{\bsl{q}}_{\text{init}}$ to a predicted final 1-RDM $O^{\bsl{q}}_{\text{out}}$. It is trained on small \(L \times L\) using numerically generated pairs 
\(\{O^{\bsl{q}}_{\mathrm{init}}(\bsl{k})\}, \{O^{\bsl{q}}_{\mathrm{tgt}}(\bsl{k})\}\), where $L$ denotes the number of lattice sites along each side, and $O^{\bsl{q}}_{\text{tgt}}$ denotes the true final 1-RDM from the HF calculation.
The overall structure is shown in \cref{fig:transformer_drawing}(a).
We first preprocess the 1-RDM $O^{\bsl{q}}_{\bsl{k},\alpha\alpha'}$ into a set of standardized~\cite{Goodfellow-et-al-2016,LeCun1998EfficientBackprop} $2\xi^2$-dimensional real vectors $\tilde{\bsl x}_i$ at the $i^{\text{th}}$ Bloch momentum, which is then fed into the self-attention NN consisting of $N$ attention layers.
All the \(\tilde{\bsl{x}}_i\) are linearly transformed and stacked into the input matrix $H^{(0)}$ of size $L^2\times D$ for the first attention layer.
Subsequently, the $\ell^{\text{th}}$ attention layer maps $H^{(\ell-1)} \mapsto H^{(\ell)}$ with $l=1,2,..,N$, after which $H^{(N)}$ is decoded back to $O_{\mathrm{out}}^{\bsl q}(\bsl{k})$, which is the predicted final 1–RDM.
Within the ${\ell}^{\text{th}}$ attention layer (\cref{fig:transformer_drawing}(b)), we have an attention block, a feedforward network (FFN), Dropout, and LayerNorm, where the last two (Dropout and LayerNorm) are common NN layers to prevent overfitting and improve optimization stability \cite{JMLR:v15:srivastava14a,ba2016layernormalization}.
Each attention block uses $N_{\text{head}}$ parallel heads (\cref{fig:transformer_drawing}(c)); in head \(h\), the per-head width is \(d_k \equiv D/N_{\text{head}}\). We form $\alpha^{(h)} = H^{(\ell-1)}W_{\alpha}^{(h)}$, where $\alpha \in \{Q,K,V\}$ for the ``query", ``key", and ``value" matrices respectively, and each $W_{\alpha}^{(h)}\in \mathbb{R}^{D \times d_k}$ is learnable.

Our construction modifies the original self-attention mechanism \cite{NIPS2017_3f5ee243} in two major ways.
First, to encode the momentum information, we introduce a novel positional encoding scheme that uses a learnable bias $B_{ij}$ to increase or decrease the attention scores based on the momentum-space separation between $\bsl{k}_i$ and $\bsl{k}_j$ (\cref{fig:transformer_drawing}(d)).
Second, we also use a learnable scalar $s$ \cite{nakanishi2025scalablesoftmaxsuperiorattention} to avoid ``attention fading" and better enable the NN to handle variable-size inputs.

\begin{figure}[t]
    \centering
    \includegraphics[width=1.0\linewidth]{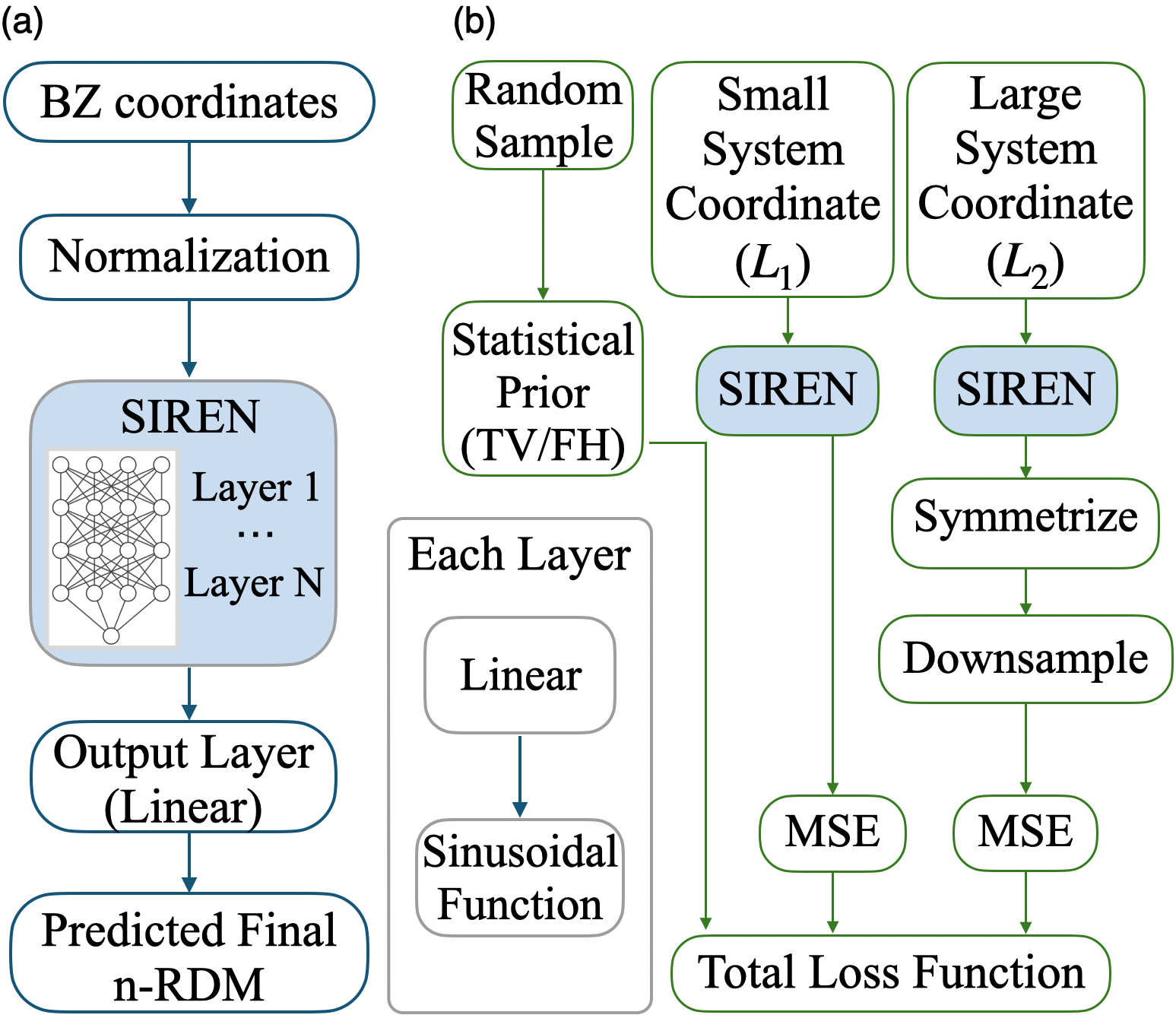}
    \caption{\textbf{Schematic of our SIREN.} (a) Normalize BZ coordinates to $\bsl v(\bsl{k}) \in [-1,1]^2$, then pass $\bsl v$ through layers of linear and sine functions to predict one component of the final $n$-RDM (\eg, a specific $\alpha,\alpha'$ component for 1-RDM); aggregating components reconstructs the predicted final $n$-RDM. 
    (b) We train the NN using a total loss consisting of: (i) the MSE on a small $L_1 \times L_1$ mesh, (ii) a loss term computed by evaluating the SIREN on a larger $L_2 \times L_2$ mesh, symmetrizing,  downsampling back to $L_1 \times L_1$ and taking the MSE with the true final RDM for the $L_1\times L_1$ mesh, and (iii) an optional statistical prior ($\mathrm{TV}$ or $\mathrm{FH}$) on randomly chosen coordinates $\bsl u_i\in[-1,1]^2$ for regularization.
    Therefore, only true final RDMs for the small $L_1\times L_1$ mesh are used as training data.}
    \label{fig:SIREN_drawing}
\end{figure}

The second NN architecture we use is SIREN \cite{sitzmann2020implicitneuralrepresentationsperiodic}, which represents the 1–RDM as a set of smooth, periodic functions of $\bsl{k}$ coordinates, thereby enabling interpolation to denser $\bsl{k}$ meshes.
The overall structure is shown in \cref{fig:SIREN_drawing}(a).
We first map BZ $\bsl{k}$ points to normalized coordinates $\bsl v(\bsl{k}) \in [-1,1]^2$.
Then, for each fixed \((\alpha, \alpha')\), the real and imaginary parts of $O^{\bsl{q}}_{\bsl{k},\alpha\alpha'}$ are a function of $\bsl{v}$, and we train a scalar function $\Phi_{\theta}$ such that $\Phi_\theta(\bsl{v}) \approx \Re[O^{\bsl{q}}_{\alpha\alpha'}(\bsl{k})]$ or $\Im[O^{\bsl{q}}_{\alpha\alpha'}(\bsl{k})]\ $.
Specifically, we construct each $\Phi_{\theta}$ using 4 hidden layers of sinusoidal functions followed by a linear function, and finally assemble $\Phi_{\theta}$ for different \((\alpha,\alpha')\) to give the final predicted \(O^{\bsl{q}}(\bsl{k_i})\).
To train the SIREN on a $L_1\times L_1$ mesh, we introduce a custom loss function with three parts (\cref{fig:SIREN_drawing}(b)): (i) the mean-squared error (MSE) on the smaller system of size $L_1$, (ii) a novel consistency term that evaluates $\Phi_\theta$ on a denser mesh of size $L_2$, averages over a chosen symmetry subgroup $G$, downsamples back to size $L_1$, and then computes the MSE with the true final $L_1\times L_1$ 1-RDM, and (iii) a statistical prior consisting of TV/Hessian regularization at a randomly sampled set of points \cite{sitzmann2020implicitneuralrepresentationsperiodic}.
It is straightforward to generalize our NN architectures to any special combination of $n$-RDM (with $n>1$) that only depends on one Bloch momentum (with other momenta fixed just like $O^{\bsl{q}}_{\bsl{k},\alpha\alpha'}$).
Beyond one-momentum quantities, the SIREN can be applied to the pair-pair correlation function (a special $2$-RDM) for superconductors, which depends on two momenta, as we show in the subsequent section.
In general, we expect the SIREN can naturally be extended to quantities that are a smooth function of more than two momenta.

\begin{figure}[t]
    \centering
    \includegraphics[width=1.0\linewidth]{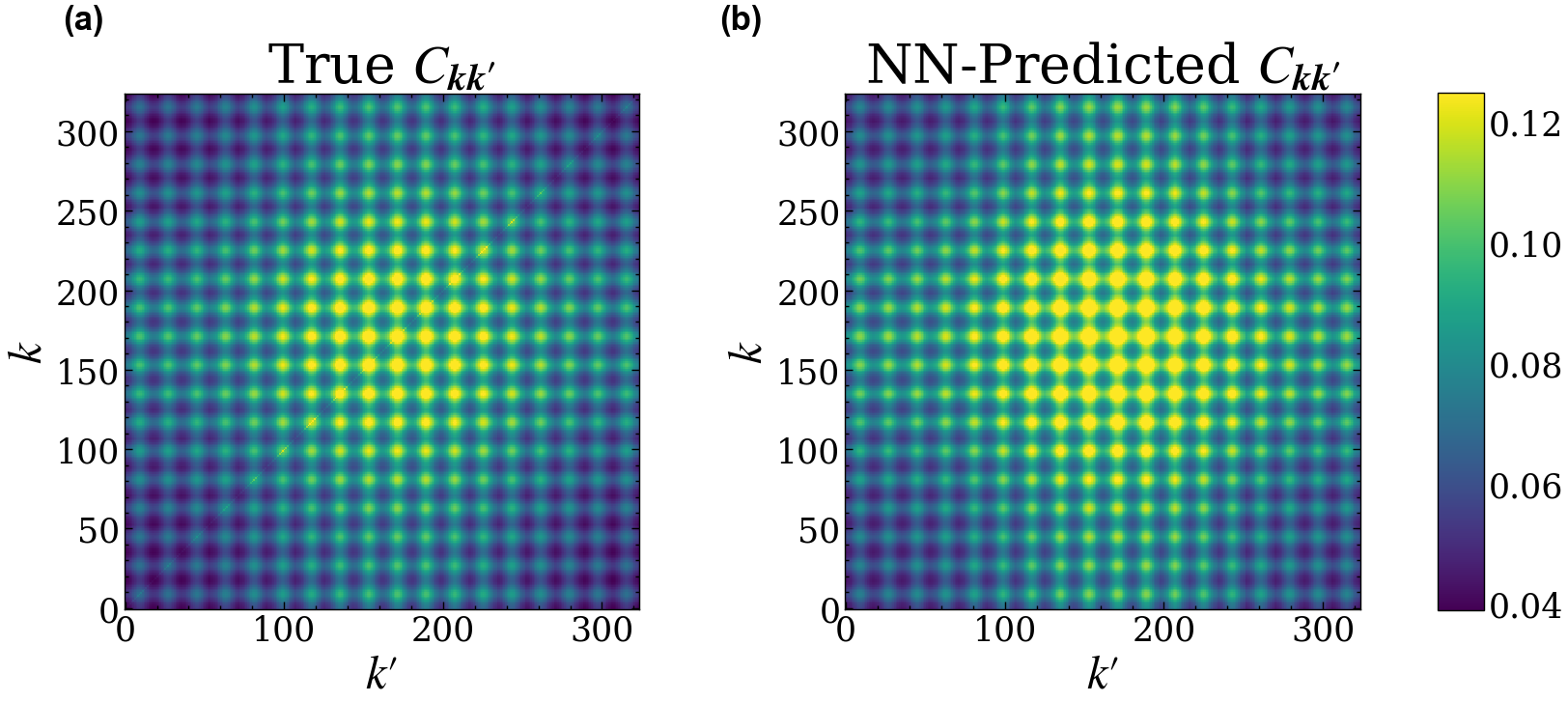}
    \caption{Comparison between (a) the true final $C_{\bsl{k}\bsl{k}'}$ for $L=18$ and (b) the predicted final $L=18$ $C_{\bsl{k}\bsl{k}'}$ given by a SIREN trained on $L=6$. Here the indices \(k,k'\in\{0,\dots,L^{2}-1\}\) enumerate the \(L\times L\) BZ points by flattening \((\ell_{1},\ell_{2})\) with \(\ell_{1,2}\in\{0,\dots,L-1\}\) via \(k=\ell_{1}+L\,\ell_{2}\); the corresponding momentum vector is $\bsl{k}$, and similarly for \(k'\).}
    \label{fig:richardson_comparison_maintext}
\end{figure}

\emph{Pair-Pair Correlation Functions}
In this section, we benchmark the SIREN on the pair-pairing correlation function for superconductivity.
As a benchmark, we consider the exactly-solvable Richardson model ~\cite{Richardson1963Restricted, RichardsonSherman1964, RevModPhys.76.643} in 2 dimensions on an $L \times L$ $\bsl{k}$ mesh
\begin{equation}
H = \sum_{\bsl k,s}\varepsilon_{\bsl k}\, c^\dagger_{\bsl k s} c_{\bsl k s}
+ \frac{u}{L^2}\sum_{\bsl k} A_{\bsl k}^\dagger\sum_{\bsl k'} A_{\bsl k'}\ ,
\label{eq:rich_h}
\end{equation} 
where $L^2$ is the number of lattice sites, $ \varepsilon_{\bsl{k}}= t(\cos k_x+\cos k_y)$ with $t= 0.1$, $u=-1$, and $c^\dagger_{\bsl k s}$ creates an electron with Bloch momentum $\bsl{k}$ and spin $s$. We work in the number-conserving case, where the pair–pair correlation function is just the Cooper–channel matrix channel of 2-RDM $C_{\bsl{k}\bsl{k}'} = O_{\bsl{k}\uparrow,-\bsl{k}\downarrow,\bsl{k}'\uparrow,-\bsl{k}'\downarrow}= \big\langle A^\dagger_{\bsl{k}}\,A_{\bsl{k}'} \big\rangle$, as $\langle A_{\bsl{k}}\rangle= \langle c^\dagger_{\bsl{k}\uparrow}\,c^\dagger_{-\bsl{k}\downarrow}\rangle=0$.
The model is exactly solved for electron filling $1/6$ for $L=6$, $L=12$, and $L=18$ (see details in \cref{sec:RichardsonAppendix}).
We then trained a SIREN on the obtained $C_{\bsl k\bsl k'}$ for $L=6$, and tested the precision of the prediction on larger sizes up to $L=18$ by evaluating the relative accuracy $r_n = \left(1-\frac{\sqrt{\text{MSE}}}{\text{max}(C) - \text{min}(C)}\right)$. (Here ``MSE'' stands for the mean squared error between the NN-predicted final and the true final correlation function, as discussed in \cref{sec:RichardsonAppendix}.)
\begin{figure}[t]
    \centering
    \includegraphics[width=1.0\linewidth]{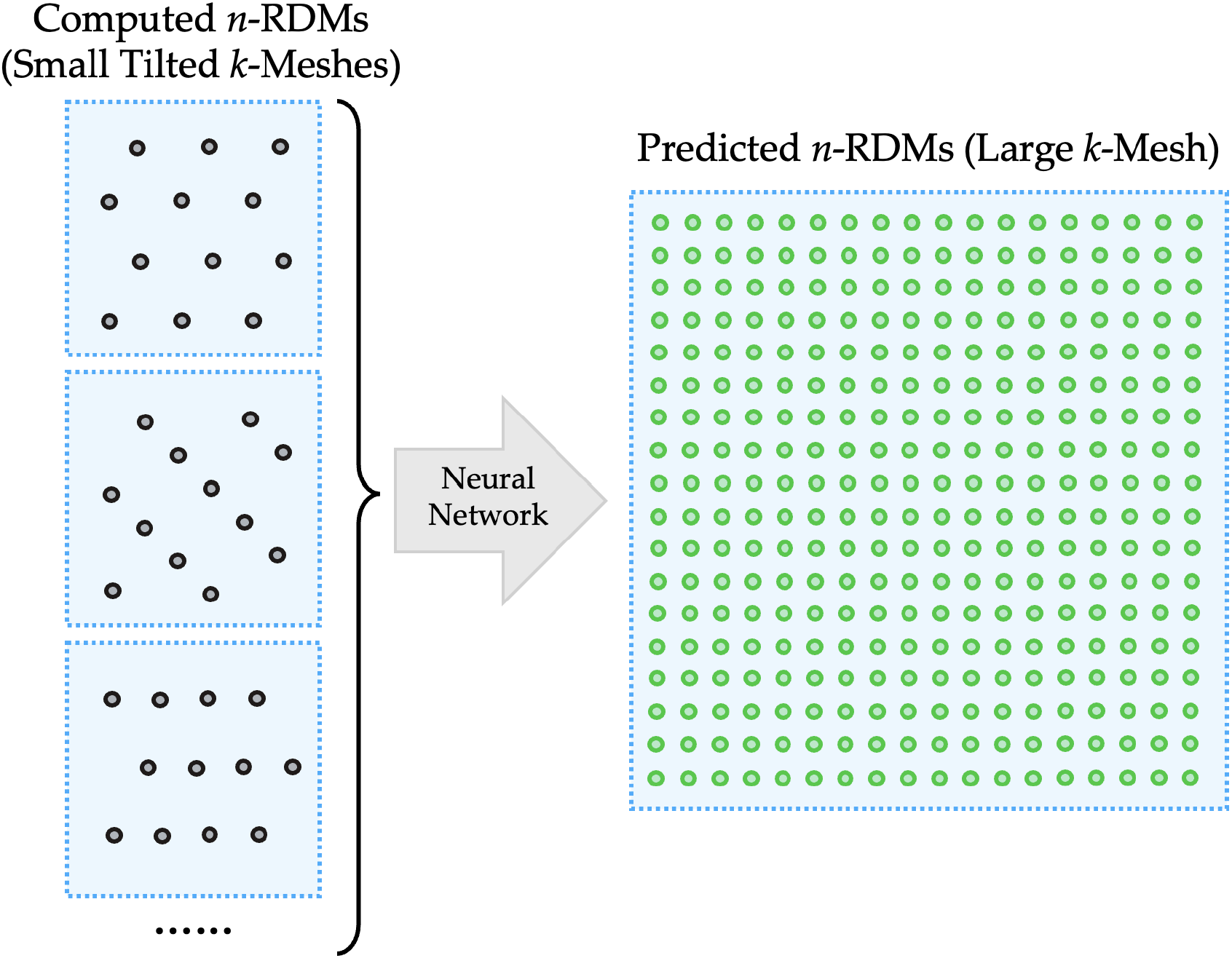}
    \caption{Instead of training on the pair-pair correlation function for a single square lattice ($\eg$ $L\times L$ for $L=6$), for efficiency we can train a NN (implemented by a SIREN in our case) on $C_{\bsl{k}\bsl{k}'}$ for multiple tilted meshes, each of which has a smaller number ($\eg$ $12$) of $\bsl{k}$ points, and use it to predict $C_{\bsl{k}\bsl{k}'}$ for a much larger size ($\eg$ $18\times18$) without a significant drop in accuracy.}
    \label{fig:tilted_mesh_figure}
\end{figure}
Remarkably, the SIREN trained only on $L=6$ can predict $C_{\bsl{k}\bsl{k}'}$ for a larger size ($L=18$) with a rather high accuracy of $r_n=94.29\%$ (see \cref{fig:richardson_comparison_maintext} for comparison), verifying the validity of our method.

To further demonstrate the efficiency of our method, we consider smaller training meshes. Because the dimension of the Hilbert space scales exponentially with system size, it is often computationally much cheaper to calculate RDMs for several smaller momentum meshes rather than for a single larger mesh.
Building on this insight, we trained an additional SIREN on $C_{\bsl{k}\bsl{k}'}$ for 4 tilted meshes (see \cref{sec:RichardsonAppendix}), each of which has $12$ $\bsl{k}$ points, as depicted in \cref{fig:tilted_mesh_figure}. The trained SIREN is then able to predict $C_{\bsl{k}\bsl{k}'}$ for the $18\times18$ mesh with a relative accuracy of $93.77\%$, which further underscores the efficiency of our method.

\emph{ Translationally-Invariant HF}
We further benchmark our self-attention NN on translationally-invariant HF for the 1-RDM.
The final true 1-RDM is calculated via the HF algorithm starting from a random initial 1-RDM. The model we consider is a $L \times L$ square lattice with a four-component spinor per unit cell, short-range hopping terms, and repulsive interaction.

We choose the hopping such that the lower two single-particle bands have a nontrivial time-reversal–invariant ${Z}_2$ topology~\cite{Kane2005Z2}.
We perform the translationally-invariant HF calculations at filling $\nu=1$ for $L=6, 8,  10, 12, \dots 50$. Using this data, we train a self-attention NN on $10,000$ (initial 1-RDM, true final 1-RDM) pairs for each of $L=6, 8$ (see \cref{sec:toymodelpart}). In order to evaluate its effectiveness, we feed $1,000$ random initial 1-RDMs for each of $L=10, 12, \dots 50$ into the trained NN to obtain the corresponding predicted final 1-RDMs. We then use those predicted final 1-RDMs as initial conditions for the HF calculation and record the number of iterations to converge, and compare it with HF started from a random initialization.
Particularly, we find that it takes an average of $8.057$ iteration steps to converge when using the predicted final 1-RDMs for $L=50$ as the initial 1-RDMs for HF, as compared to an average of $96.301$ iterations with random initial 1-RDMs. 
Thus, our NN, when trained on $L=6,8$, results in a $91.63$\% reduction in the number of iterations required for HF convergence for $L=50$.
The percent reduction for other system sizes is shown in \cref{fig:toyModel_reduction}. Notably, they are all above $90\%$.
(see \cref{sec:toymodelpart} for details).
\begin{figure}[t]
    \centering
    \includegraphics[width=1\linewidth]{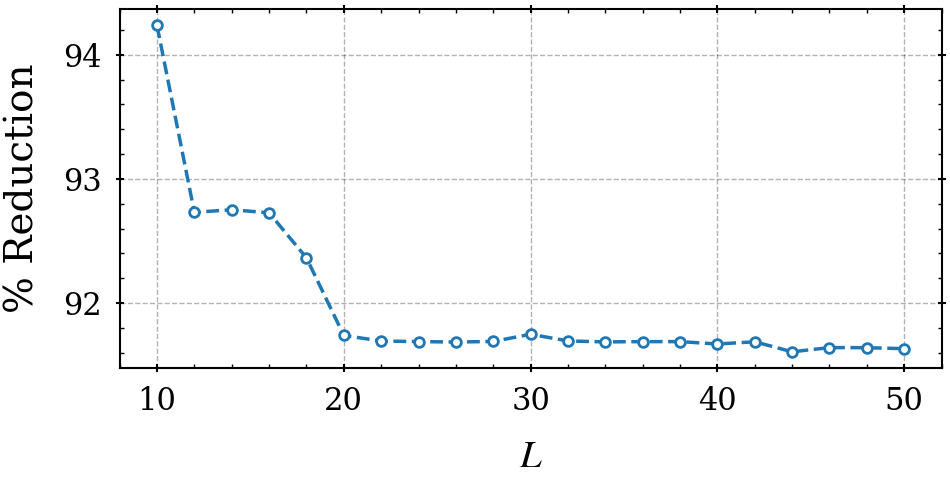}
    \caption{\textbf{NN for translation-invariant HF.} Here, ``\% Reduction" is the percent difference between the number of HF iterations using a predicted final 1-RDM from the self-attention NN as the initial 1-RDM as compared to using a random initial 1-RDM, and is plotted as a function of the system size $L$ for the four-band  model. See \cref{sec:toymodelpart} for the exact values.
    The HF calculation is forced to preserve the lattice translation symmetry.}
    \label{fig:toyModel_reduction}
\end{figure}

\begin{figure}[t]
    \centering \includegraphics[width=1.0\linewidth]{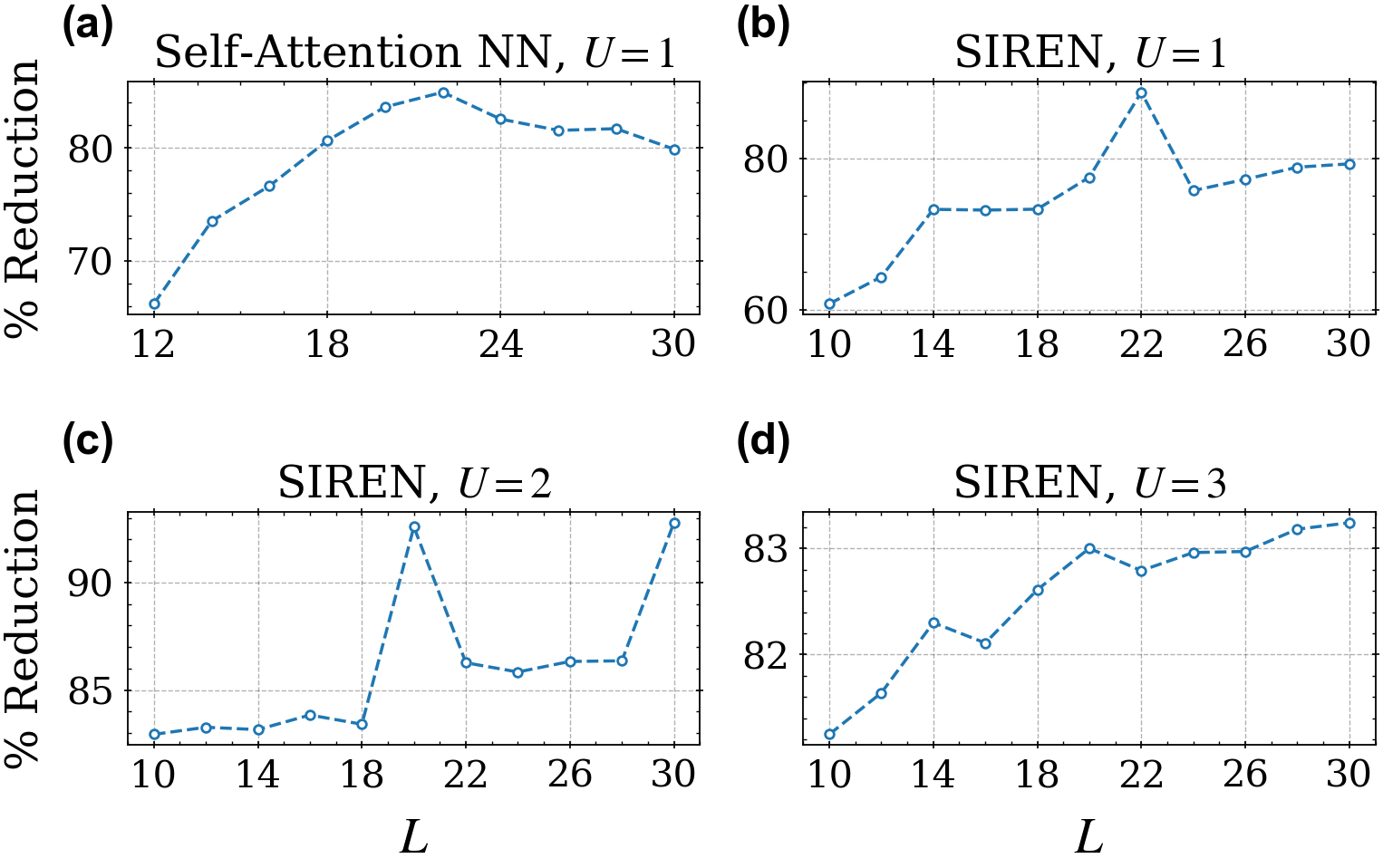}
    \caption{\textbf{NN for the Hubbard model.} Summary of results for the Hubbard model using a self-attention NN ($U=1$) and SIRENs ($U=1,2,3$) showing the percent reduction in the number of  HF iterations  as a function of the system size $L$. See \cref{app:hubbard} for the exact values.
    The HF calculations here allow any kind of breaking of lattice translation symmetries that are compatible with the momentum mesh.}    \label{fig:hubbard_reduction}
\end{figure}

\emph{HF Allowing Translational Breaking}
Finally, we benchmark both NNs for HF calculations that allow any kind of translation breaking that is compatible with the momentum meshes.
We focus on the Hubbard model on a 2D square lattice \cite{1963RSPSA.276..238H, 1964RSPSA.277..237H} with onsite $U$ and hopping $t=1$.
We consider the half filling and assume that a spin $U(1)$ is conserved.
Then, the 1-RDM that serves as the order parameters is spin-resolved: $[O_\sigma]_{\bsl k\bsl k'} = \langle c^{\dagger}_{\bsl k'\sigma} c_{\bsl k\sigma}\rangle,
  \quad \sigma\in\{\uparrow,\downarrow\}$.
We perform zero-total-spin HF calculations for $L=8,10,\dots,20$ and $U=1,2,3$, allowing all translation breaking that is compatible with the momentum mesh.
The lowest-energy converged $[O_\sigma]_{\bsl k\bsl k'} $ are not zero only when $\bsl{k}'-\bsl{k}$ is $(\pi,\pi)$ and $(0,0)$, which is consistent with the $(\pi,\pi)$-antiferromagnetism at half filling \cite{PhysRevLett.62.591}.
We train both a self-attention NN and a SIREN. The self-attention NN was trained on $L=8$ and $L=10$ for $U=1$ only (due to limited training data), and tested on $L=12, \dots 30$. The SIRENs were trained on $L=8$ for $U=1,2,3$, and tested on $L=10, \dots 30$. For evaluation, we again compute the percent reduction in the number of HF iterations using the predicted final 1-RDM as the initial guess versus using a random initial 1-RDM, and observe a dramatic reduction $(73.27\%$---$92.78\%)$ at larger sizes ($\ie,\, L\geq18$).
\cref{fig:hubbard_reduction} shows the percent reduction plotted as a function of $L$ for different values of $U$.

\emph{Conclusion and Discussion:} In summary, we showed that large-size $n$-RDMs can be predicted by NNs trained on small-size $n$-RDMs, with benchmarks on the pair-pair correlation function and translationally-invariant/breaking HF.
One immediate future step is to use these NNs to help predict new correlated states in other systems.
Certainly, the development of NNs for general $2$-RDMs (or $n$-RDM with $n>2$) is an important future direction.
The application of our NNs to strongly correlated systems, such as moiré materials, is also expected to yield useful results.

\emph{Acknowledgments:} J.Y.'s work is supported by startup funds at University of Florida. A.A.A.'s work is supported by the AI Scholars program at the University of Florida and by the UF CCMS Undergraduate Fellowship. The authors acknowledge UFIT Research Computing for providing computational resources and support.

\nocite{TPE_paper,loshchilov2017sgdr,pmlr-v9-glorot10a,kaiming}

\bibliographystyle{apsrev4-2}
\bibliography{references,bibfile_references.bib}

@misc{Gu_2025_Hubbard_NNansatz,
      title={Solving the Hubbard model with Neural Quantum States}, 
      author={Yuntian Gu and Wenrui Li and Heng Lin and Bo Zhan and Ruichen Li and Yifei Huang and Di He and Yantao Wu and Tao Xiang and Mingpu Qin and Liwei Wang and Dingshun Lv},
      year={2025},
      eprint={2507.02644},
      archivePrefix={arXiv},
      primaryClass={cond-mat.str-el},
      url={https://arxiv.org/abs/2507.02644}, 
}

@Article{Li_2022_NNWavefunction,
author={Li, Xiang
and Li, Zhe
and Chen, Ji},
title={Ab initio calculation of real solids via neural network ansatz},
journal={Nature Communications},
year={2022},
month={Dec},
day={22},
volume={13},
number={1},
pages={7895},
issn={2041-1723},
doi={10.1038/s41467-022-35627-1},
url={https://doi.org/10.1038/s41467-022-35627-1}
}

@article{Lou_2019_NN_Wavefunction,
  title = {Backflow Transformations via Neural Networks for Quantum Many-Body Wave Functions},
  author = {Luo, Di and Clark, Bryan K.},
  journal = {Phys. Rev. Lett.},
  volume = {122},
  issue = {22},
  pages = {226401},
  numpages = {6},
  year = {2019},
  month = {Jun},
  publisher = {American Physical Society},
  doi = {10.1103/PhysRevLett.122.226401},
  url = {https://link.aps.org/doi/10.1103/PhysRevLett.122.226401}
}

@misc{Vonglehn_2023_selfattentionansatzabinitioquantum,
      title={A Self-Attention Ansatz for Ab-initio Quantum Chemistry}, 
      author={Ingrid von Glehn and James S. Spencer and David Pfau},
      year={2023},
      eprint={2211.13672},
      archivePrefix={arXiv},
      primaryClass={physics.chem-ph},
      url={https://arxiv.org/abs/2211.13672}, 
}

@article{Pfau_2020_NN_Wavefunction,
  title = {Ab initio solution of the many-electron Schr\"odinger equation with deep neural networks},
  author = {Pfau, David and Spencer, James S. and Matthews, Alexander G. D. G. and Foulkes, W. M. C.},
  journal = {Phys. Rev. Res.},
  volume = {2},
  issue = {3},
  pages = {033429},
  numpages = {20},
  year = {2020},
  month = {Sep},
  publisher = {American Physical Society},
  doi = {10.1103/PhysRevResearch.2.033429},
  url = {https://link.aps.org/doi/10.1103/PhysRevResearch.2.033429}
}

@misc{Teng_2024_NN_ansatz_FQHE,
      title={Solving and visualizing fractional quantum Hall wavefunctions with neural network}, 
      author={Yi Teng and David D. Dai and Liang Fu},
      year={2024},
      eprint={2412.00618},
      archivePrefix={arXiv},
      primaryClass={cond-mat.str-el},
      url={https://arxiv.org/abs/2412.00618}, 
}

@misc{nakanishi2025scalablesoftmaxsuperiorattention,
      title={Scalable-Softmax Is Superior for Attention}, 
      author={Ken M. Nakanishi},
      year={2025},
      eprint={2501.19399},
      archivePrefix={arXiv},
      primaryClass={cs.CL},
      url={https://arxiv.org/abs/2501.19399}, 
}

@misc{zhou2024transformersachievelengthgeneralization,
      title={Transformers Can Achieve Length Generalization But Not Robustly}, 
      author={Yongchao Zhou and Uri Alon and Xinyun Chen and Xuezhi Wang and Rishabh Agarwal and Denny Zhou},
      year={2024},
      eprint={2402.09371},
      archivePrefix={arXiv},
      primaryClass={cs.LG},
      url={https://arxiv.org/abs/2402.09371}, 
}

@article{Li_2024_FIRE,
  title={Functional Interpolation for Relative Positions Improves Long Context Transformers},
  author={Li, Shanda and You, Chong and Guruganesh, Guru and Ainslie, Joshua and Ontanon, Santiago and Zaheer, Manzil and Sanghai, Sumit and Yang, Yiming and Kumar, Sanjiv and Bhojanapalli, Srinadh},
  journal={arXiv preprint arXiv:2310.04418},
  year={2023},
  url={https://arxiv.org/abs/2310.04418}
}

@article{Mazziotti_2006_2RDMReview,
  author  = {David A. Mazziotti},
  title   = {Quantum Chemistry without Wave Functions: Two-Electron Reduced Density Matrices},
  journal = {Accounts of Chemical Research},
  volume  = {39},
  number  = {3},
  pages   = {207--215},
  year    = {2006},
  doi     = {10.1021/ar050029d},
  url     = {https://pubs.acs.org/doi/10.1021/ar050029d}
}

@book{Szabo_1989_ModernQC,
  author = {Szabo, Attila and Ostlund, Neil S.},
  title = {Modern Quantum Chemistry: Introduction to Advanced Electronic Structure Theory},
  publisher = {Dover Publications},
  year = {1989},
  isbn = {9780486691862}
}

@misc{Luo_2025_ML_QMC_tMoTe2,
      title={Solving fractional electron states in twisted MoTe$_2$ with deep neural network}, 
      author={Di Luo and Timothy Zaklama and Liang Fu},
      year={2025},
      eprint={2503.13585},
      archivePrefix={arXiv},
      primaryClass={cond-mat.str-el},
      url={https://arxiv.org/abs/2503.13585}, 
}

@misc{Geier_2025_ML_QMC,
      title={Is attention all you need to solve the correlated electron problem?}, 
      author={Max Geier and Khachatur Nazaryan and Timothy Zaklama and Liang Fu},
      year={2025},
      eprint={2502.05383},
      archivePrefix={arXiv},
      primaryClass={cond-mat.str-el},
      url={https://arxiv.org/abs/2502.05383}, 
}

@inproceedings{Kim_2016_VDSR,
  author    = {Jiwon Kim and Jung Kwon Lee and Kyoung Mu Lee},
  title     = {Accurate Image Super-Resolution Using Very Deep Convolutional Networks},
  booktitle = {Proceedings of the IEEE Conference on Computer Vision and Pattern Recognition (CVPR)},
  year      = {2016},
  pages     = {1646--1654},
  doi       = {10.1109/CVPR.2016.182},
  url       = {https://ieeexplore.ieee.org/document/7780551}
}

@article{Kane2005Z2,
  title = {${Z}_{2}$ Topological Order and the Quantum Spin Hall Effect},
  author = {Kane, C. L. and Mele, E. J.},
  journal = {Phys. Rev. Lett.},
  volume = {95},
  issue = {14},
  pages = {146802},
  numpages = {4},
  year = {2005},
  month = {Sep},
  publisher = {American Physical Society},
  doi = {10.1103/PhysRevLett.95.146802},
  url = {https://link.aps.org/doi/10.1103/PhysRevLett.95.146802}
}

@inproceedings{NIPS2017_3f5ee243,
 author = {Vaswani, Ashish and Shazeer, Noam and Parmar, Niki and Uszkoreit, Jakob and Jones, Llion and Gomez, Aidan N and Kaiser, \L ukasz and Polosukhin, Illia},
 booktitle = {Advances in Neural Information Processing Systems},
 editor = {I. Guyon and U. Von Luxburg and S. Bengio and H. Wallach and R. Fergus and S. Vishwanathan and R. Garnett},
 pages = {},
 publisher = {Curran Associates, Inc.},
 title = {Attention is All you Need},
 url = {https://proceedings.neurips.cc/paper_files/paper/2017/file/3f5ee243547dee91fbd053c1c4a845aa-Paper.pdf},
 volume = {30},
 year = {2017}
}

@misc{shaw2018selfattentionrelativepositionrepresentations,
      title={Self-Attention with Relative Position Representations}, 
      author={Peter Shaw and Jakob Uszkoreit and Ashish Vaswani},
      year={2018},
      eprint={1803.02155},
      archivePrefix={arXiv},
      primaryClass={cs.CL},
      url={https://arxiv.org/abs/1803.02155}, 
}

@misc{chen2025exactefficientrepresentationtotally,
      title={Exact and Efficient Representation of Totally Anti-Symmetric Functions}, 
      author={Ziang Chen and Jianfeng Lu},
      year={2025},
      eprint={2311.05064},
      archivePrefix={arXiv},
      primaryClass={math.CA},
      url={https://arxiv.org/abs/2311.05064}, 
}

@misc{fu2025minimaluniversalrepresentationfermionic,
      title={A minimal and universal representation of fermionic wavefunctions (fermions = bosons + one)}, 
      author={Liang Fu},
      year={2025},
      eprint={2510.11431},
      archivePrefix={arXiv},
      primaryClass={cond-mat.str-el},
      url={https://arxiv.org/abs/2510.11431}, 
}

@misc{sitzmann2020implicitneuralrepresentationsperiodic,
      title={Implicit Neural Representations with Periodic Activation Functions}, 
      author={Vincent Sitzmann and Julien N. P. Martel and Alexander W. Bergman and David B. Lindell and Gordon Wetzstein},
      year={2020},
      eprint={2006.09661},
      archivePrefix={arXiv},
      primaryClass={cs.CV},
      url={https://arxiv.org/abs/2006.09661}, 
}

@inproceedings{NeRF,
author = {Mildenhall, Ben and Srinivasan, Pratul P. and Tancik, Matthew and Barron, Jonathan T. and Ramamoorthi, Ravi and Ng, Ren},
title = {NeRF: Representing Scenes as Neural Radiance Fields for View Synthesis},
year = {2020},
isbn = {978-3-030-58451-1},
publisher = {Springer-Verlag},
address = {Berlin, Heidelberg},
url = {https://doi.org/10.1007/978-3-030-58452-8_24},
doi = {10.1007/978-3-030-58452-8_24},
booktitle = {Computer Vision – ECCV 2020: 16th European Conference, Glasgow, UK, August 23–28, 2020, Proceedings, Part I},
pages = {405–421},
numpages = {17},
location = {Glasgow, United Kingdom}
}

@article{JMLR:v15:srivastava14a,
  author  = {Nitish Srivastava and Geoffrey Hinton and Alex Krizhevsky and Ilya Sutskever and Ruslan Salakhutdinov},
  title   = {Dropout: A Simple Way to Prevent Neural Networks from Overfitting},
  journal = {Journal of Machine Learning Research},
  year    = {2014},
  volume  = {15},
  number  = {56},
  pages   = {1929--1958},
  url     = {http://jmlr.org/papers/v15/srivastava14a.html}
}

@misc{hendrycks2023gaussianerrorlinearunits,
      title={Gaussian Error Linear Units (GELUs)}, 
      author={Dan Hendrycks and Kevin Gimpel},
      year={2023},
      eprint={1606.08415},
      archivePrefix={arXiv},
      primaryClass={cs.LG},
      url={https://arxiv.org/abs/1606.08415}, 
}

@misc{ba2016layernormalization,
      title={Layer Normalization}, 
      author={Jimmy Lei Ba and Jamie Ryan Kiros and Geoffrey E. Hinton},
      year={2016},
      eprint={1607.06450},
      archivePrefix={arXiv},
      primaryClass={stat.ML},
      url={https://arxiv.org/abs/1607.06450}, 
}

@misc{loshchilov2019decoupledweightdecayregularization,
      title={Decoupled Weight Decay Regularization}, 
      author={Ilya Loshchilov and Frank Hutter},
      year={2019},
      eprint={1711.05101},
      archivePrefix={arXiv},
      primaryClass={cs.LG},
      url={https://arxiv.org/abs/1711.05101}, 
}

@inproceedings{
loshchilov2017sgdr,
title={{SGDR}: Stochastic Gradient Descent with Warm Restarts},
author={Ilya Loshchilov and Frank Hutter},
booktitle={International Conference on Learning Representations},
year={2017},
url={https://openreview.net/forum?id=Skq89Scxx}
}

@INPROCEEDINGS{kaiming,
  author={He, Kaiming and Zhang, Xiangyu and Ren, Shaoqing and Sun, Jian},
  booktitle={2015 IEEE International Conference on Computer Vision (ICCV)}, 
  title={Delving Deep into Rectifiers: Surpassing Human-Level Performance on ImageNet Classification}, 
  year={2015},
  volume={},
  number={},
  pages={1026-1034},
  doi={10.1109/ICCV.2015.123}}

@InProceedings{pmlr-v9-glorot10a,
  title = 	 {Understanding the difficulty of training deep feedforward neural networks},
  author = 	 {Glorot, Xavier and Bengio, Yoshua},
  booktitle = 	 {Proceedings of the Thirteenth International Conference on Artificial Intelligence and Statistics},
  pages = 	 {249--256},
  year = 	 {2010},
  editor = 	 {Teh, Yee Whye and Titterington, Mike},
  volume = 	 {9},
  series = 	 {Proceedings of Machine Learning Research},
  address = 	 {Chia Laguna Resort, Sardinia, Italy},
  month = 	 {13--15 May},
  publisher =    {PMLR},
  url = 	 {https://proceedings.mlr.press/v9/glorot10a.html}
}

@article{Richardson1963Restricted,
  author  = {R. W. Richardson},
  title   = {A restricted class of exact eigenstates of the pairing-force Hamiltonian},
  journal = {Physics Letters},
  volume  = {3},
  number  = {6},
  pages   = {277--279},
  year    = {1963},
  issn    = {0031-9163},
  doi     = {10.1016/0031-9163(63)90259-2},
  url     = {https://www.sciencedirect.com/science/article/pii/0031916363902592}
}

@article{RichardsonSherman1964,
  author  = {R. W. Richardson and N. Sherman},
  title   = {Exact eigenstates of the pairing-force Hamiltonian},
  journal = {Nuclear Physics},
  volume  = {52},
  pages   = {221--238},
  year    = {1964},
  issn    = {0029-5582},
  doi     = {10.1016/0029-5582(64)90687-X},
  url     = {https://www.sciencedirect.com/science/article/pii/002955826490687X}
}

@article{PhysRev.108.1175,
  title = {Theory of Superconductivity},
  author = {Bardeen, J. and Cooper, L. N. and Schrieffer, J. R.},
  journal = {Phys. Rev.},
  volume = {108},
  issue = {5},
  pages = {1175--1204},
  numpages = {0},
  year = {1957},
  month = {Dec},
  publisher = {American Physical Society},
  doi = {10.1103/PhysRev.108.1175},
  url = {https://link.aps.org/doi/10.1103/PhysRev.108.1175}
}

@article{RevModPhys.76.643,
  title = {Colloquium: Exactly solvable Richardson-Gaudin models for many-body quantum systems},
  author = {Dukelsky, J. and Pittel, S. and Sierra, G.},
  journal = {Rev. Mod. Phys.},
  volume = {76},
  issue = {3},
  pages = {643--662},
  numpages = {0},
  year = {2004},
  month = {Aug},
  publisher = {American Physical Society},
  doi = {10.1103/RevModPhys.76.643},
  url = {https://link.aps.org/doi/10.1103/RevModPhys.76.643}
}

@ARTICLE{1963RSPSA.276..238H,
       author = {{Hubbard}, J.},
        title = "{Electron Correlations in Narrow Energy Bands}",
      journal = {Proceedings of the Royal Society of London Series A},
         year = 1963,
        month = nov,
       volume = {276},
       number = {1365},
        pages = {238-257},
          doi = {10.1098/rspa.1963.0204},
       adsurl = {https://ui.adsabs.harvard.edu/abs/1963RSPSA.276..238H}
}

@ARTICLE{1964RSPSA.277..237H,
       author = {{Hubbard}, J.},
        title = "{Electron Correlations in Narrow Energy Bands. II. The Degenerate Band Case}",
      journal = {Proceedings of the Royal Society of London Series A},
         year = 1964,
        month = jan,
       volume = {277},
       number = {1369},
        pages = {237-259},
          doi = {10.1098/rspa.1964.0019},
       adsurl = {https://ui.adsabs.harvard.edu/abs/1964RSPSA.277..237H}
}

@article{PhysRevX.7.031038,
  title = {Machine Learning Phases of Strongly Correlated Fermions},
  author = {Ch'ng, Kelvin and Carrasquilla, Juan and Melko, Roger G. and Khatami, Ehsan},
  journal = {Phys. Rev. X},
  volume = {7},
  issue = {3},
  pages = {031038},
  numpages = {9},
  year = {2017},
  month = {Aug},
  publisher = {American Physical Society},
  doi = {10.1103/PhysRevX.7.031038},
  url = {https://link.aps.org/doi/10.1103/PhysRevX.7.031038}
}

@article{Ibarra_Garc_a_Padilla_2025,
   title={Autoregressive neural quantum states of Fermi Hubbard models},
   volume={7},
   ISSN={2643-1564},
   url={http://dx.doi.org/10.1103/PhysRevResearch.7.013122},
   DOI={10.1103/physrevresearch.7.013122},
   number={1},
   journal={Physical Review Research},
   publisher={American Physical Society (APS)},
   author={Ibarra-García-Padilla, Eduardo and Lange, Hannah and Melko, Roger G. and Scalettar, Richard T. and Carrasquilla, Juan and Bohrdt, Annabelle and Khatami, Ehsan},
   year={2025},
   month=feb }

@article{PhysRevLett.121.167204,
  title = {Symmetries and Many-Body Excitations with Neural-Network Quantum States},
  author = {Choo, Kenny and Carleo, Giuseppe and Regnault, Nicolas and Neupert, Titus},
  journal = {Phys. Rev. Lett.},
  volume = {121},
  issue = {16},
  pages = {167204},
  numpages = {6},
  year = {2018},
  month = {Oct},
  publisher = {American Physical Society},
  doi = {10.1103/PhysRevLett.121.167204},
  url = {https://link.aps.org/doi/10.1103/PhysRevLett.121.167204}
}

@article{Carleo_2017,
   title={Solving the quantum many-body problem with artificial neural networks},
   volume={355},
   ISSN={1095-9203},
   url={http://dx.doi.org/10.1126/science.aag2302},
   DOI={10.1126/science.aag2302},
   number={6325},
   journal={Science},
   publisher={American Association for the Advancement of Science (AAAS)},
   author={Carleo, Giuseppe and Troyer, Matthias},
   year={2017},
   month=feb, pages={602–606} }

@misc{vivas2022neuralnetworkquantumstatessystematic,
      title={Neural-Network Quantum States: A Systematic Review}, 
      author={David R. Vivas and Javier Madroñero and Victor Bucheli and Luis O. Gómez and John H. Reina},
      year={2022},
      eprint={2204.12966},
      archivePrefix={arXiv},
      primaryClass={quant-ph},
      url={https://arxiv.org/abs/2204.12966}, 
}

@article{Sharir_2020,
   title={Deep Autoregressive Models for the Efficient Variational Simulation of Many-Body Quantum Systems},
   volume={124},
   ISSN={1079-7114},
   url={http://dx.doi.org/10.1103/PhysRevLett.124.020503},
   DOI={10.1103/physrevlett.124.020503},
   number={2},
   journal={Physical Review Letters},
   publisher={American Physical Society (APS)},
   author={Sharir, Or and Levine, Yoav and Wies, Noam and Carleo, Giuseppe and Shashua, Amnon},
   year={2020},
   month=jan }

@article{Guti_rrez_2022,
   title={Real time evolution with neural-network quantum states},
   volume={6},
   ISSN={2521-327X},
   url={http://dx.doi.org/10.22331/q-2022-01-20-627},
   DOI={10.22331/q-2022-01-20-627},
   journal={Quantum},
   publisher={Verein zur Forderung des Open Access Publizierens in den Quantenwissenschaften},
   author={Gutiérrez, Irene López and Mendl, Christian B.},
   year={2022},
   month=jan, pages={627} }

@article{Choo_2019,
   title={Two-dimensional frustrated 
<mml:math xmlns:mml="http://www.w3.org/1998/Math/MathML"><mml:mrow><mml:msub><mml:mi>J</mml:mi><mml:mn>1</mml:mn></mml:msub><mml:mtext>−</mml:mtext><mml:msub><mml:mi>J</mml:mi><mml:mn>2</mml:mn></mml:msub></mml:mrow></mml:math>
 model studied with neural network quantum states},
   volume={100},
   ISSN={2469-9969},
   url={http://dx.doi.org/10.1103/PhysRevB.100.125124},
   DOI={10.1103/physrevb.100.125124},
   number={12},
   journal={Physical Review B},
   publisher={American Physical Society (APS)},
   author={Choo, Kenny and Neupert, Titus and Carleo, Giuseppe},
   year={2019},
   month=sep }

@misc{deprince2023variationaldeterminationtwoelectronreduced,
      title={Variational Determination of the Two-Electron Reduced Density Matrix: A Tutorial Review}, 
      author={A. Eugene DePrince III},
      year={2023},
      eprint={2310.10746},
      archivePrefix={arXiv},
      primaryClass={physics.chem-ph},
      url={https://arxiv.org/abs/2310.10746}, 
}

@article{PhysRevLett.93.213001,
  title = {Realization of Quantum Chemistry without Wave Functions through First-Order Semidefinite Programming},
  author = {Mazziotti, David A.},
  journal = {Phys. Rev. Lett.},
  volume = {93},
  issue = {21},
  pages = {213001},
  numpages = {4},
  year = {2004},
  month = {Nov},
  publisher = {American Physical Society},
  doi = {10.1103/PhysRevLett.93.213001},
  url = {https://link.aps.org/doi/10.1103/PhysRevLett.93.213001}
}

@article{PhysRevLett.106.083001,
  title = {Large-Scale Semidefinite Programming for Many-Electron Quantum Mechanics},
  author = {Mazziotti, David A.},
  journal = {Phys. Rev. Lett.},
  volume = {106},
  issue = {8},
  pages = {083001},
  numpages = {4},
  year = {2011},
  month = {Feb},
  publisher = {American Physical Society},
  doi = {10.1103/PhysRevLett.106.083001},
  url = {https://link.aps.org/doi/10.1103/PhysRevLett.106.083001}
}

@misc{scheer2024hamiltonianbootstrap,
      title={Hamiltonian Bootstrap}, 
      author={Michael G. Scheer},
      year={2024},
      eprint={2410.00810},
      archivePrefix={arXiv},
      primaryClass={cond-mat.str-el},
      url={https://arxiv.org/abs/2410.00810}, 
}

@article{Giesbertz_2019,
   title={One-body reduced density-matrix functional theory in finite basis sets at elevated temperatures},
   volume={806},
   ISSN={0370-1573},
   url={http://dx.doi.org/10.1016/j.physrep.2019.01.010},
   DOI={10.1016/j.physrep.2019.01.010},
   journal={Physics Reports},
   publisher={Elsevier BV},
   author={Giesbertz, Klaas J.H. and Ruggenthaler, Michael},
   year={2019},
   month=may, pages={1–47} }

@book{Goodfellow-et-al-2016,
    title={Deep Learning},
    author={Ian Goodfellow and Yoshua Bengio and Aaron Courville},
    publisher={MIT Press},
    note={\url{http://www.deeplearningbook.org}},
    year={2016}
}

@incollection{LeCun1998EfficientBackprop,
  author    = {Yann LeCun and L{\'e}on Bottou and Genevieve B. Orr and Klaus-Robert M{\"u}ller},
  title     = {Efficient BackProp},
  booktitle = {Neural Networks: Tricks of the Trade},
  editor    = {Genevieve B. Orr and Klaus-Robert M{\"u}ller},
  series    = {Lecture Notes in Computer Science},
  volume    = {1524},
  pages     = {9--50},
  publisher = {Springer},
  address   = {Berlin, Heidelberg},
  year      = {1998},
  doi       = {10.1007/3-540-49430-8_2}
}

@article{RevModPhys.35.668,
  title = {Structure of Fermion Density Matrices},
  author = {Coleman, A. J.},
  journal = {Rev. Mod. Phys.},
  volume = {35},
  issue = {3},
  pages = {668--686},
  numpages = {0},
  year = {1963},
  month = {Jul},
  publisher = {American Physical Society},
  doi = {10.1103/RevModPhys.35.668},
  url = {https://link.aps.org/doi/10.1103/RevModPhys.35.668}
}

@article{PhysRevA.72.032510,
  title = {Variational two-electron reduced density matrix theory for many-electron atoms and molecules: Implementation of the spin- and symmetry-adapted ${T}_{2}$ condition through first-order semidefinite programming},
  author = {Mazziotti, David A.},
  journal = {Phys. Rev. A},
  volume = {72},
  issue = {3},
  pages = {032510},
  numpages = {11},
  year = {2005},
  month = {Sep},
  publisher = {American Physical Society},
  doi = {10.1103/PhysRevA.72.032510},
  url = {https://link.aps.org/doi/10.1103/PhysRevA.72.032510}
}

@article{Nakata2001,
  author  = {Nakata, Maho and Nakatsuji, Hiroshi and Ehara, Masahiro and Fukuda, Mitsuhiro and Nakata, Kazuhide and Fujisawa, Katsuki},
  title   = {Variational calculations of fermion second-order reduced density matrices by semidefinite programming algorithm},
  journal = {The Journal of Chemical Physics},
  year    = {2001},
  volume  = {114},
  number  = {19},
  pages   = {8282--8292},
  doi     = {10.1063/1.1360199}
}

@article{Zhao2004,
  author  = {Zhao, Zhengji and Braams, Bastiaan J. and Fukuda, Mituhiro and Overton, Michael L. and Percus, Jerome K.},
  title   = {The reduced density matrix method for electronic structure calculations and the role of three-index representability conditions},
  journal = {The Journal of Chemical Physics},
  year    = {2004},
  volume  = {120},
  number  = {5},
  pages   = {2095--2104},
  doi     = {10.1063/1.1636721}
}

@article{PhysRevLett.108.263002,
  title = {Structure of Fermionic Density Matrices: Complete $N$-Representability Conditions},
  author = {Mazziotti, David A.},
  journal = {Phys. Rev. Lett.},
  volume = {108},
  issue = {26},
  pages = {263002},
  numpages = {5},
  year = {2012},
  month = {Jun},
  publisher = {American Physical Society},
  doi = {10.1103/PhysRevLett.108.263002},
  url = {https://link.aps.org/doi/10.1103/PhysRevLett.108.263002}
}

@misc{shao2023machinelearningelectronicstructure,
      title={Machine Learning Electronic Structure Methods Based On The One-Electron Reduced Density Matrix}, 
      author={Xuecheng Shao and Lukas Paetow and Mark E. Tuckerman and Michele Pavanello},
      year={2023},
      eprint={2302.10741},
      archivePrefix={arXiv},
      primaryClass={physics.chem-ph},
      url={https://arxiv.org/abs/2302.10741}, 
}

@article{Ma2024MachineLH,
  title={Machine Learning-Assisted Hartree–Fock Approach for Energy Level Calculations in the Neutral Ytterbium Atom},
  author={Kaichen Ma and Chen Yang and Junyao Zhang and Yunfei Li and Gang Jiang and Junjie Chai},
  journal={Entropy},
  year={2024},
  volume={26},
  url={https://api.semanticscholar.org/CorpusID:273947564}
}

@article{DelgadoGranados2025ML2RDM,
  title   = {Machine Learning of Two-Electron Reduced Density Matrices for Many-Body Problems},
  author  = {Delgado-Granados, Luis H. and Sager-Smith, LeeAnn M. and Trifonova, Kristina and Mazziotti, David A.},
  journal = {The Journal of Physical Chemistry Letters},
  year    = {2025},
  volume  = {16},
  number  = {9},
  pages   = {2231--2237},
  doi     = {10.1021/acs.jpclett.4c03366}
}

@article{Hazra_2024,
   title={Predicting the One-Particle Density Matrix with Machine Learning},
   volume={20},
   ISSN={1549-9626},
   url={http://dx.doi.org/10.1021/acs.jctc.4c00042},
   DOI={10.1021/acs.jctc.4c00042},
   number={11},
   journal={Journal of Chemical Theory and Computation},
   publisher={American Chemical Society (ACS)},
   author={Hazra, S. and Patil, U. and Sanvito, S.},
   year={2024},
   month=may, pages={4569–4578} }

@article{Mazziotti2012,
  author  = {Mazziotti, David A.},
  title   = {Two-electron reduced density matrix as the basic variable in many-electron quantum chemistry and physics},
  journal = {Chemical Reviews},
  year    = {2012},
  volume  = {112},
  number  = {1},
  pages   = {244--262},
  doi     = {10.1021/cr2000493}
}

@article{DePrince2024,
  author  = {DePrince, A. Eugene III},
  title   = {Variational determination of the two-electron reduced density matrix: A tutorial review},
  journal = {WIREs: Computational Molecular Science},
  year    = {2024},
  volume  = {14},
  number  = {1},
  pages   = {e1702},
  doi     = {10.1002/wcms.1702}
}

@article{GarrodPercus1964,
  author  = {Garrod, Claude and Percus, Jerome K.},
  title   = {Reduction of the N‐Particle Variational Problem},
  journal = {Journal of Mathematical Physics},
  year    = {1964},
  volume  = {5},
  number  = {12},
  pages   = {1756--1776},
  doi     = {10.1063/1.1704098}
}

@book{Rasmussen2006GPML,
  author    = {Carl Edward Rasmussen and Christopher K. I. Williams},
  title     = {Gaussian Processes for Machine Learning},
  year      = {2006},
  publisher = {MIT Press},
  address   = {Cambridge, MA},
  url       = {http://www.gaussianprocess.org/gpml/chapters/RW.pdf}
}

@article{Gneiting2002Compact,
  author  = {Tilmann Gneiting},
  title   = {Compactly Supported Correlation Functions},
  journal = {Journal of Multivariate Analysis},
  year    = {2002},
  volume  = {83},
  number  = {2},
  pages   = {493--508},
  doi     = {10.1006/jmva.2001.2056}
}

@book{Buhmann2003RBF,
  author    = {Martin D. Buhmann},
  title     = {Radial Basis Functions: Theory and Implementations},
  series    = {Cambridge Monographs on Applied and Computational Mathematics},
  volume    = {12},
  publisher = {Cambridge University Press},
  address   = {Cambridge},
  year      = {2003},
  doi       = {10.1017/CBO9780511550497}
}

@misc{press2022trainshorttestlong,
      title={Train Short, Test Long: Attention with Linear Biases Enables Input Length Extrapolation}, 
      author={Ofir Press and Noah A. Smith and Mike Lewis},
      year={2022},
      eprint={2108.12409},
      archivePrefix={arXiv},
      primaryClass={cs.CL},
      url={https://arxiv.org/abs/2108.12409}, 
}

@misc{bahdanau2016neuralmachinetranslationjointly,
      title={Neural Machine Translation by Jointly Learning to Align and Translate}, 
      author={Dzmitry Bahdanau and Kyunghyun Cho and Yoshua Bengio},
      year={2016},
      eprint={1409.0473},
      archivePrefix={arXiv},
      primaryClass={cs.CL},
      url={https://arxiv.org/abs/1409.0473}, 
}

@inproceedings{Bridle1989ProbabilisticIO,
  title={Probabilistic Interpretation of Feedforward Classification Network Outputs, with Relationships to Statistical Pattern Recognition},
  author={John Scott Bridle},
  booktitle={NATO Neurocomputing},
  year={1989},
  url={https://api.semanticscholar.org/CorpusID:59636530}
}

@misc{kingma2017adammethodstochasticoptimization,
      title={Adam: A Method for Stochastic Optimization}, 
      author={Diederik P. Kingma and Jimmy Ba},
      year={2017},
      eprint={1412.6980},
      archivePrefix={arXiv},
      primaryClass={cs.LG},
      url={https://arxiv.org/abs/1412.6980}, 
}

@article{Li_2022,
   title={Deep-learning density functional theory Hamiltonian for efficient ab initio electronic-structure calculation},
   volume={2},
   ISSN={2662-8457},
   url={http://dx.doi.org/10.1038/s43588-022-00265-6},
   DOI={10.1038/s43588-022-00265-6},
   number={6},
   journal={Nature Computational Science},
   publisher={Springer Science and Business Media LLC},
   author={Li, He and Wang, Zun and Zou, Nianlong and Ye, Meng and Xu, Runzhang and Gong, Xiaoxun and Duan, Wenhui and Xu, Yong},
   year={2022},
   month=jun, pages={367–377} }

@article{Gong_2023,
   title={General framework for E(3)-equivariant neural network representation of density functional theory Hamiltonian},
   volume={14},
   ISSN={2041-1723},
   url={http://dx.doi.org/10.1038/s41467-023-38468-8},
   DOI={10.1038/s41467-023-38468-8},
   number={1},
   journal={Nature Communications},
   publisher={Springer Science and Business Media LLC},
   author={Gong, Xiaoxun and Li, He and Zou, Nianlong and Xu, Runzhang and Duan, Wenhui and Xu, Yong},
   year={2023},
   month=may }

@InProceedings{TPE_paper,
  title = 	 {Making a Science of Model Search: Hyperparameter Optimization in Hundreds of Dimensions for Vision Architectures},
  author = 	 {Bergstra, James and Yamins, Daniel and Cox, David},
  booktitle = 	 {Proceedings of the 30th International Conference on Machine Learning},
  pages = 	 {115--123},
  year = 	 {2013},
  editor = 	 {Dasgupta, Sanjoy and McAllester, David},
  volume = 	 {28},
  number =       {1},
  series = 	 {Proceedings of Machine Learning Research},
  address = 	 {Atlanta, Georgia, USA},
  month = 	 {17--19 Jun},
  publisher =    {PMLR},
  url = 	 {https://proceedings.mlr.press/v28/bergstra13.html}
}

@article{keys1981cubic,
  title={Cubic convolution interpolation for digital image processing},
  author={Keys, Robert G.},
  journal={IEEE Transactions on Acoustics, Speech, and Signal Processing},
  volume={29},
  number={6},
  pages={1153--1160},
  year={1981},
  publisher={IEEE},
  doi={10.1109/TASSP.1981.1163711}
}

@inproceedings{mitchell1988reconstruction,
  title={Reconstruction filters in computer graphics},
  author={Mitchell, Don P. and Netravali, Arun N.},
  booktitle={Proceedings of the 15th Annual Conference on Computer Graphics and Interactive Techniques (SIGGRAPH '88)},
  pages={221--228},
  year={1988},
  publisher={ACM},
  address={Atlanta, Georgia, USA},
  doi={10.1145/54852.378514}
}

@misc{li2021fermionicneuralnetworkeffective,
      title={Fermionic Neural Network with Effective Core Potential}, 
      author={Xiang Li and Cunwei Fan and Weiluo Ren and Ji Chen},
      year={2021},
      eprint={2108.11661},
      archivePrefix={arXiv},
      primaryClass={physics.chem-ph},
      url={https://arxiv.org/abs/2108.11661}, 
}

@article{PhysRevLett.62.591,
  title = {Antiferromagnetism in the Two-Dimensional Hubbard Model},
  author = {Hirsch, J. E. and Tang, S.},
  journal = {Phys. Rev. Lett.},
  volume = {62},
  issue = {5},
  pages = {591--594},
  numpages = {0},
  year = {1989},
  month = {Jan},
  publisher = {American Physical Society},
  doi = {10.1103/PhysRevLett.62.591},
  url = {https://link.aps.org/doi/10.1103/PhysRevLett.62.591}
}

@article{Lee_2021,
   title={Neural-network variational quantum algorithm for simulating many-body dynamics},
   volume={3},
   ISSN={2643-1564},
   url={http://dx.doi.org/10.1103/PhysRevResearch.3.023095},
   DOI={10.1103/physrevresearch.3.023095},
   number={2},
   journal={Physical Review Research},
   publisher={American Physical Society (APS)},
   author={Lee, Chee Kong and Patil, Pranay and Zhang, Shengyu and Hsieh, Chang Yu},
   year={2021},
   month=may }

@article{Irikura_2020,
   title={Neural-network quantum states at finite temperature},
   volume={2},
   ISSN={2643-1564},
   url={http://dx.doi.org/10.1103/PhysRevResearch.2.013284},
   DOI={10.1103/physrevresearch.2.013284},
   number={1},
   journal={Physical Review Research},
   publisher={American Physical Society (APS)},
   author={Irikura, Naoki and Saito, Hiroki},
   year={2020},
   month=mar }

@article{Hibat_Allah_2020,
   title={Recurrent neural network wave functions},
   volume={2},
   ISSN={2643-1564},
   url={http://dx.doi.org/10.1103/PhysRevResearch.2.023358},
   DOI={10.1103/physrevresearch.2.023358},
   number={2},
   journal={Physical Review Research},
   publisher={American Physical Society (APS)},
   author={Hibat-Allah, Mohamed and Ganahl, Martin and Hayward, Lauren E. and Melko, Roger G. and Carrasquilla, Juan},
   year={2020},
   month=jun }

@misc{luo2022gaugeequivariantneuralnetworks,
      title={Gauge Equivariant Neural Networks for 2+1D U(1) Gauge Theory Simulations in Hamiltonian Formulation}, 
      author={Di Luo and Shunyue Yuan and James Stokes and Bryan K. Clark},
      year={2022},
      eprint={2211.03198},
      archivePrefix={arXiv},
      primaryClass={hep-lat},
      url={https://arxiv.org/abs/2211.03198}, 
}

@misc{luo2023pairingbasedgraphneuralnetwork,
      title={Pairing-based graph neural network for simulating quantum materials}, 
      author={Di Luo and David D. Dai and Liang Fu},
      year={2023},
      eprint={2311.02143},
      archivePrefix={arXiv},
      primaryClass={cond-mat.str-el},
      url={https://arxiv.org/abs/2311.02143}, 
}

@article{Qian_2025,
   title={Describing Landau Level Mixing in Fractional Quantum Hall States with Deep Learning},
   volume={134},
   ISSN={1079-7114},
   url={http://dx.doi.org/10.1103/PhysRevLett.134.176503},
   DOI={10.1103/physrevlett.134.176503},
   number={17},
   journal={Physical Review Letters},
   publisher={American Physical Society (APS)},
   author={Qian, Yubing and Zhao, Tongzhou and Zhang, Jianxiao and Xiang, Tao and Li, Xiang and Chen, Ji},
   year={2025},
   month=apr }

@misc{li2025deeplearningshedslight,
      title={Deep Learning Sheds Light on Integer and Fractional Topological Insulators}, 
      author={Xiang Li and Yixiao Chen and Bohao Li and Haoxiang Chen and Fengcheng Wu and Ji Chen and Weiluo Ren},
      year={2025},
      eprint={2503.11756},
      archivePrefix={arXiv},
      primaryClass={cond-mat.str-el},
      url={https://arxiv.org/abs/2503.11756}, 
}

@article{Roth_2023,
   title={High-accuracy variational Monte Carlo for frustrated magnets with deep neural networks},
   volume={108},
   ISSN={2469-9969},
   url={http://dx.doi.org/10.1103/PhysRevB.108.054410},
   DOI={10.1103/physrevb.108.054410},
   number={5},
   journal={Physical Review B},
   publisher={American Physical Society (APS)},
   author={Roth, Christopher and Szabó, Attila and MacDonald, Allan H.},
   year={2023},
   month=aug }

@misc{li2023forwardlaplaciannewcomputational,
      title={Forward Laplacian: A New Computational Framework for Neural Network-based Variational Monte Carlo}, 
      author={Ruichen Li and Haotian Ye and Du Jiang and Xuelan Wen and Chuwei Wang and Zhe Li and Xiang Li and Di He and Ji Chen and Weiluo Ren and Liwei Wang},
      year={2023},
      eprint={2307.08214},
      archivePrefix={arXiv},
      primaryClass={physics.comp-ph},
      url={https://arxiv.org/abs/2307.08214}, 
}

@article{Ren_2023,
   title={Towards the ground state of molecules via diffusion Monte Carlo on neural networks},
   volume={14},
   ISSN={2041-1723},
   url={http://dx.doi.org/10.1038/s41467-023-37609-3},
   DOI={10.1038/s41467-023-37609-3},
   number={1},
   journal={Nature Communications},
   publisher={Springer Science and Business Media LLC},
   author={Ren, Weiluo and Fu, Weizhong and Wu, Xiaojie and Chen, Ji},
   year={2023},
   month=apr }

@article{Goldshlager_2024,
   title={A Kaczmarz-inspired approach to accelerate the optimization of neural network wavefunctions},
   volume={516},
   ISSN={0021-9991},
   url={http://dx.doi.org/10.1016/j.jcp.2024.113351},
   DOI={10.1016/j.jcp.2024.113351},
   journal={Journal of Computational Physics},
   publisher={Elsevier BV},
   author={Goldshlager, Gil and Abrahamsen, Nilin and Lin, Lin},
   year={2024},
   month=nov, pages={113351} }

@article{PhysRevResearch.3.043126,
  title = {Determinant-free fermionic wave function using feed-forward neural networks},
  author = {Inui, Koji and Kato, Yasuyuki and Motome, Yukitoshi},
  journal = {Phys. Rev. Res.},
  volume = {3},
  issue = {4},
  pages = {043126},
  numpages = {9},
  year = {2021},
  month = {Nov},
  publisher = {American Physical Society},
  doi = {10.1103/PhysRevResearch.3.043126},
  url = {https://link.aps.org/doi/10.1103/PhysRevResearch.3.043126}
}

@article{Adams_2021,
   title={Variational Monte Carlo Calculations of 
<mml:math xmlns:mml="http://www.w3.org/1998/Math/MathML" display="inline"><mml:mrow><mml:mi>A</mml:mi><mml:mo>≤</mml:mo><mml:mn>4</mml:mn></mml:mrow></mml:math>
 Nuclei with an Artificial Neural-Network Correlator Ansatz},
   volume={127},
   ISSN={1079-7114},
   url={http://dx.doi.org/10.1103/PhysRevLett.127.022502},
   DOI={10.1103/physrevlett.127.022502},
   number={2},
   journal={Physical Review Letters},
   publisher={American Physical Society (APS)},
   author={Adams, Corey and Carleo, Giuseppe and Lovato, Alessandro and Rocco, Noemi},
   year={2021},
   month=jul }

@misc{scherbela2021solvingelectronicschrodingerequation,
      title={Solving the electronic Schr\"odinger equation for multiple nuclear geometries with weight-sharing deep neural networks}, 
      author={Michael Scherbela and Rafael Reisenhofer and Leon Gerard and Philipp Marquetand and Philipp Grohs},
      year={2021},
      eprint={2105.08351},
      archivePrefix={arXiv},
      primaryClass={physics.comp-ph},
      url={https://arxiv.org/abs/2105.08351}, 
}

@article{Wilson_2023,
   title={Neural network ansatz for periodic wave functions and the homogeneous electron gas},
   volume={107},
   ISSN={2469-9969},
   url={http://dx.doi.org/10.1103/PhysRevB.107.235139},
   DOI={10.1103/physrevb.107.235139},
   number={23},
   journal={Physical Review B},
   publisher={American Physical Society (APS)},
   author={Wilson, Max and Moroni, Saverio and Holzmann, Markus and Gao, Nicholas and Wudarski, Filip and Vegge, Tejs and Bhowmik, Arghya},
   year={2023},
   month=jun }

@article{Lou_2024,
   title={Neural Wave Functions for Superfluids},
   volume={14},
   ISSN={2160-3308},
   url={http://dx.doi.org/10.1103/PhysRevX.14.021030},
   DOI={10.1103/physrevx.14.021030},
   number={2},
   journal={Physical Review X},
   publisher={American Physical Society (APS)},
   author={Lou, Wan Tong and Sutterud, Halvard and Cassella, Gino and Foulkes, W. M. C. and Knolle, Johannes and Pfau, David and Spencer, James S.},
   year={2024},
   month=may }

@misc{kim2023neuralnetworkquantumstatesultracold,
      title={Neural-network quantum states for ultra-cold Fermi gases}, 
      author={Jane Kim and Gabriel Pescia and Bryce Fore and Jannes Nys and Giuseppe Carleo and Stefano Gandolfi and Morten Hjorth-Jensen and Alessandro Lovato},
      year={2023},
      eprint={2305.08831},
      archivePrefix={arXiv},
      primaryClass={cond-mat.quant-gas},
      url={https://arxiv.org/abs/2305.08831}, 
}

@article{Pescia_2024,
   title={Message-passing neural quantum states for the homogeneous electron gas},
   volume={110},
   ISSN={2469-9969},
   url={http://dx.doi.org/10.1103/PhysRevB.110.035108},
   DOI={10.1103/physrevb.110.035108},
   number={3},
   journal={Physical Review B},
   publisher={American Physical Society (APS)},
   author={Pescia, Gabriel and Nys, Jannes and Kim, Jane and Lovato, Alessandro and Carleo, Giuseppe},
   year={2024},
   month=jul }

@article{Cassella_2023,
   title={Discovering Quantum Phase Transitions with Fermionic Neural Networks},
   volume={130},
   ISSN={1079-7114},
   url={http://dx.doi.org/10.1103/PhysRevLett.130.036401},
   DOI={10.1103/physrevlett.130.036401},
   number={3},
   journal={Physical Review Letters},
   publisher={American Physical Society (APS)},
   author={Cassella, Gino and Sutterud, Halvard and Azadi, Sam and Drummond, N.D. and Pfau, David and Spencer, James S. and Foulkes, W.M.C.},
   year={2023},
   month=jan }

@article{Pescia_2022,
   title={Neural-network quantum states for periodic systems in continuous space},
   volume={4},
   ISSN={2643-1564},
   url={http://dx.doi.org/10.1103/PhysRevResearch.4.023138},
   DOI={10.1103/physrevresearch.4.023138},
   number={2},
   journal={Physical Review Research},
   publisher={American Physical Society (APS)},
   author={Pescia, Gabriel and Han, Jiequn and Lovato, Alessandro and Lu, Jianfeng and Carleo, Giuseppe},
   year={2022},
   month=may }

@article{Luo_2021,
   title={Gauge Equivariant Neural Networks for Quantum Lattice Gauge Theories},
   volume={127},
   ISSN={1079-7114},
   url={http://dx.doi.org/10.1103/PhysRevLett.127.276402},
   DOI={10.1103/physrevlett.127.276402},
   number={27},
   journal={Physical Review Letters},
   publisher={American Physical Society (APS)},
   author={Luo, Di and Carleo, Giuseppe and Clark, Bryan K. and Stokes, James},
   year={2021},
   month=dec }

@article{Martyn_2023,
   title={Variational Neural-Network Ansatz for Continuum Quantum Field Theory},
   volume={131},
   ISSN={1079-7114},
   url={http://dx.doi.org/10.1103/PhysRevLett.131.081601},
   DOI={10.1103/physrevlett.131.081601},
   number={8},
   journal={Physical Review Letters},
   publisher={American Physical Society (APS)},
   author={Martyn, John M. and Najafi, Khadijeh and Luo, Di},
   year={2023},
   month=aug }

@article{Nomura_2017,
   title={Restricted Boltzmann machine learning for solving strongly correlated quantum systems},
   volume={96},
   ISSN={2469-9969},
   url={http://dx.doi.org/10.1103/PhysRevB.96.205152},
   DOI={10.1103/physrevb.96.205152},
   number={20},
   journal={Physical Review B},
   publisher={American Physical Society (APS)},
   author={Nomura, Yusuke and Darmawan, Andrew S. and Yamaji, Youhei and Imada, Masatoshi},
   year={2017},
   month=nov }

@article{Romero_2025,
   title={Spectroscopy of two-dimensional interacting lattice electrons using symmetry-aware neural backflow transformations},
   volume={8},
   ISSN={2399-3650},
   url={http://dx.doi.org/10.1038/s42005-025-01955-z},
   DOI={10.1038/s42005-025-01955-z},
   number={1},
   journal={Communications Physics},
   publisher={Springer Science and Business Media LLC},
   author={Romero, Imelda and Nys, Jannes and Carleo, Giuseppe},
   year={2025},
   month=jan }

@article{Glasser_2018,
   title={Neural-Network Quantum States, String-Bond States, and Chiral Topological States},
   volume={8},
   ISSN={2160-3308},
   url={http://dx.doi.org/10.1103/PhysRevX.8.011006},
   DOI={10.1103/physrevx.8.011006},
   number={1},
   journal={Physical Review X},
   publisher={American Physical Society (APS)},
   author={Glasser, Ivan and Pancotti, Nicola and August, Moritz and Rodriguez, Ivan D. and Cirac, J. Ignacio},
   year={2018},
   month=jan }

@article{Hermann_2020,
   title={Deep-neural-network solution of the electronic Schrödinger equation},
   volume={12},
   ISSN={1755-4349},
   url={http://dx.doi.org/10.1038/s41557-020-0544-y},
   DOI={10.1038/s41557-020-0544-y},
   number={10},
   journal={Nature Chemistry},
   publisher={Springer Science and Business Media LLC},
   author={Hermann, Jan and Schätzle, Zeno and Noé, Frank},
   year={2020},
   month=sep, pages={891–897} }

@misc{chen2022simulating21dlatticequantum,
      title={Simulating 2+1D Lattice Quantum Electrodynamics at Finite Density with Neural Flow Wavefunctions}, 
      author={Zhuo Chen and Di Luo and Kaiwen Hu and Bryan K. Clark},
      year={2022},
      eprint={2212.06835},
      archivePrefix={arXiv},
      primaryClass={hep-lat},
      url={https://arxiv.org/abs/2212.06835}, 
}

@article{Robledo_Moreno_2022,
   title={Fermionic wave functions from neural-network constrained hidden states},
   volume={119},
   ISSN={1091-6490},
   url={http://dx.doi.org/10.1073/pnas.2122059119},
   DOI={10.1073/pnas.2122059119},
   number={32},
   journal={Proceedings of the National Academy of Sciences},
   publisher={Proceedings of the National Academy of Sciences},
   author={Robledo Moreno, Javier and Carleo, Giuseppe and Georges, Antoine and Stokes, James},
   year={2022},
   month=aug }

@misc{chen2024antnbridgingautoregressiveneural,
      title={ANTN: Bridging Autoregressive Neural Networks and Tensor Networks for Quantum Many-Body Simulation}, 
      author={Zhuo Chen and Laker Newhouse and Eddie Chen and Di Luo and Marin Soljačić},
      year={2024},
      eprint={2304.01996},
      archivePrefix={arXiv},
      primaryClass={quant-ph},
      url={https://arxiv.org/abs/2304.01996}, 
}

@misc{ma2025transformerbasedneuralnetworksbackflow,
      title={Transformer-Based Neural Networks Backflow for Strongly Correlated Electronic Structure}, 
      author={Huan Ma and Bowen Kan and Honghui Shang and Jinlong Yang},
      year={2025},
      eprint={2509.25720},
      archivePrefix={arXiv},
      primaryClass={quant-ph},
      url={https://arxiv.org/abs/2509.25720}, 
}

@article{PhysRevA.102.052819,
  title = {Dual-cone variational calculation of the two-electron reduced density matrix},
  author = {Mazziotti, David A.},
  journal = {Phys. Rev. A},
  volume = {102},
  issue = {5},
  pages = {052819},
  numpages = {8},
  year = {2020},
  month = {Nov},
  publisher = {American Physical Society},
  doi = {10.1103/PhysRevA.102.052819},
  url = {https://link.aps.org/doi/10.1103/PhysRevA.102.052819}
}

@article{Slavnov1989,
  author  = {Slavnov, N. A.},
  title   = {Calculation of scalar products of wave functions and form factors in the framework of the algebraic Bethe ansatz},
  journal = {Theoretical and Mathematical Physics},
  year    = {1989},
  volume  = {79},
  number  = {2},
  pages   = {502--508},
  doi     = {10.1007/BF01016531},
}

@article{Zhou_2002,
   title={Superconducting correlations in metallic nanoparticles: Exact solution of the BCS model by the algebraic Bethe ansatz},
   volume={65},
   ISSN={1095-3795},
   url={http://dx.doi.org/10.1103/PhysRevB.65.060502},
   DOI={10.1103/physrevb.65.060502},
   number={6},
   journal={Physical Review B},
   publisher={American Physical Society (APS)},
   author={Zhou, Huan-Qiang and Links, Jon and McKenzie, Ross H. and Gould, Mark D.},
   year={2002},
   month=jan 
}

@article{Faribault_2008,
   title={Exact mesoscopic correlation functions of the Richardson pairing model},
   volume={77},
   ISSN={1550-235X},
   url={http://dx.doi.org/10.1103/PhysRevB.77.064503},
   DOI={10.1103/physrevb.77.064503},
   number={6},
   journal={Physical Review B},
   publisher={American Physical Society (APS)},
   author={Faribault, Alexandre and Calabrese, Pasquale and Caux, Jean-Sébastien},
   year={2008},
   month=feb 
}

@article{Gorohovsky_2011,
   title={Exact expectation values within Richardson’s approach for the pairing Hamiltonian in a macroscopic system},
   volume={84},
   ISSN={1550-235X},
   url={http://dx.doi.org/10.1103/PhysRevB.84.224503},
   DOI={10.1103/physrevb.84.224503},
   number={22},
   journal={Physical Review B},
   publisher={American Physical Society (APS)},
   author={Gorohovsky, G. and Bettelheim, E.},
   year={2011},
   month=dec }

@article{PhysRev.56.340,
  title = {Forces in Molecules},
  author = {Feynman, R. P.},
  journal = {Phys. Rev.},
  volume = {56},
  issue = {4},
  pages = {340--343},
  numpages = {0},
  year = {1939},
  month = {Aug},
  publisher = {American Physical Society},
  doi = {10.1103/PhysRev.56.340},
  url = {https://link.aps.org/doi/10.1103/PhysRev.56.340}
}

@article{cybenko1989approximation,
  title={Approximation by superpositions of a sigmoidal function},
  author={Cybenko, George},
  journal={Mathematics of Control, Signals, and Systems},
  volume={2},
  pages={303–314},
  year={1989}
}

@article{hornik1989multilayer,
  title={Multilayer feedforward networks are universal approximators},
  author={Hornik, Kurt and Stinchcombe, Maxwell and White, Halbert},
  journal={Neural Networks},
  volume={2},
  number={5},
  pages={359–366},
  year={1989}
}

@article{pinkus1999approximation,
  title={Approximation theory of the MLP model in neural networks},
  author={Pinkus, Allan},
  journal={Acta Numerica},
  volume={8},
  pages={143–195},
  year={1999}
}

@article{PhysRevLett.130.153001,
  title = {Quantum Many-Body Theory from a Solution of the $N$-Representability Problem},
  author = {Mazziotti, David A.},
  journal = {Phys. Rev. Lett.},
  volume = {130},
  issue = {15},
  pages = {153001},
  numpages = {7},
  year = {2023},
  month = {Apr},
  publisher = {American Physical Society},
  doi = {10.1103/PhysRevLett.130.153001},
  url = {https://link.aps.org/doi/10.1103/PhysRevLett.130.153001}
}

@misc{chen2025neuralnetworkaugmentedpfaffianwavefunctions,
      title={Neural Network-Augmented Pfaffian Wave-functions for Scalable Simulations of Interacting Fermions}, 
      author={Ao Chen and Zhou-Quan Wan and Anirvan Sengupta and Antoine Georges and Christopher Roth},
      year={2025},
      eprint={2507.10705},
      archivePrefix={arXiv},
      primaryClass={cond-mat.str-el},
      url={https://arxiv.org/abs/2507.10705}, 
}

@misc{roth2025superconductivitytwodimensionalhubbardmodel,
      title={Superconductivity in the two-dimensional Hubbard model revealed by neural quantum states}, 
      author={Christopher Roth and Ao Chen and Anirvan Sengupta and Antoine Georges},
      year={2025},
      eprint={2511.07566},
      archivePrefix={arXiv},
      primaryClass={cond-mat.supr-con},
      url={https://arxiv.org/abs/2511.07566}, 
}

@misc{lange2026simulatingsuperconductivitymixeddimensionaltparalleljparalleljperp,
      title={Simulating superconductivity in mixed-dimensional $t_\parallel$-${J}_\parallel$-${J}_\perp$ bilayers with neural quantum states}, 
      author={Hannah Lange and Ao Chen and Antoine Georges and Fabian Grusdt and Annabelle Bohrdt and Christopher Roth},
      year={2026},
      eprint={2602.10091},
      archivePrefix={arXiv},
      primaryClass={cond-mat.str-el},
      url={https://arxiv.org/abs/2602.10091}, 
}

@article{tJ_NQS_model,
  title = {Simulating the Two-Dimensional $t\text{\ensuremath{-}}J$ Model at Finite Doping with Neural Quantum States},
  author = {Lange, Hannah and B\"ohler, Annika and Roth, Christopher and Bohrdt, Annabelle},
  journal = {Phys. Rev. Lett.},
  volume = {135},
  issue = {13},
  pages = {136504},
  numpages = {7},
  year = {2025},
  month = {Sep},
  publisher = {American Physical Society},
  doi = {10.1103/rc31-5hl9},
  url = {https://link.aps.org/doi/10.1103/rc31-5hl9}
}

@misc{roth2021groupconvolutionalneuralnetworks,
      title={Group Convolutional Neural Networks Improve Quantum State Accuracy}, 
      author={Christopher Roth and Allan H. MacDonald},
      year={2021},
      eprint={2104.05085},
      archivePrefix={arXiv},
      primaryClass={quant-ph},
      url={https://arxiv.org/abs/2104.05085}, 
}

@book{Mazziotti2007RDM,
  editor    = {David A. Mazziotti},
  title     = {Reduced-Density-Matrix Mechanics: With Application to Many-Electron Atoms and Molecules},
  series    = {Advances in Chemical Physics},
  volume    = {134},
  publisher = {John Wiley \& Sons, Inc.},
  address   = {New York},
  year      = {2007}
}

@book{ColemanYukalov2000,
  author    = {A. J. Coleman and V. I. Yukalov},
  title     = {Reduced Density Matrices: Coulson’s Challenge},
  publisher = {Springer},
  address   = {Berlin, Heidelberg},
  year      = {2000}
}

@article{Kummer1967,
  author  = {H. Kummer},
  title   = {N-Representability Problem for Reduced Density Matrices},
  journal = {Journal of Mathematical Physics},
  volume  = {8},
  number  = {10},
  pages   = {2063--2081},
  year    = {1967},
  doi     = {10.1063/1.1705122}
}

@misc{castillo2021effectivesolutionconvex1body,
      title={An effective solution to convex $1$-body $N$-representability}, 
      author={Federico Castillo and Jean-Philippe Labbé and Julia Liebert and Arnau Padrol and Eva Philippe and Christian Schilling},
      year={2021},
      eprint={2105.06459},
      archivePrefix={arXiv},
      primaryClass={quant-ph},
      url={https://arxiv.org/abs/2105.06459}, 
}

@article{PhysRevA.65.062511,
  title = {Variational minimization of atomic and molecular ground-state energies via the two-particle reduced density matrix},
  author = {Mazziotti, David A.},
  journal = {Phys. Rev. A},
  volume = {65},
  issue = {6},
  pages = {062511},
  numpages = {14},
  year = {2002},
  month = {Jun},
  publisher = {American Physical Society},
  doi = {10.1103/PhysRevA.65.062511},
  url = {https://link.aps.org/doi/10.1103/PhysRevA.65.062511}
}

@article{Cances2006RDM,
  author  = {Eric Canc{\`e}s and Gabriel Stoltz and Mathieu Lewin},
  title   = {The Electronic Ground-State Energy Problem: A New Reduced Density Matrix Approach},
  journal = {The Journal of Chemical Physics},
  volume  = {125},
  number  = {6},
  pages   = {064101},
  year    = {2006},
  doi     = {10.1063/1.2218332}
}

@incollection{Erdahl2007LowerBound,
  author    = {R. M. Erdahl},
  title     = {The Lower Bound Method for Density Matrices and Semidefinite Programming},
  booktitle = {Reduced-Density-Matrix Mechanics: With Application to Many-Electron Atoms and Molecules},
  editor    = {David A. Mazziotti},
  publisher = {John Wiley \& Sons, Inc.},
  address   = {New York},
  year      = {2007},
  pages     = {61--91}
}

@article{PhysRevLett.105.213003,
  title = {Active-Space $N$-Representability Constraints for Variational Two-Particle Reduced Density Matrix Calculations},
  author = {Shenvi, Neil and Izmaylov, Artur F.},
  journal = {Phys. Rev. Lett.},
  volume = {105},
  issue = {21},
  pages = {213003},
  numpages = {4},
  year = {2010},
  month = {Nov},
  publisher = {American Physical Society},
  doi = {10.1103/PhysRevLett.105.213003},
  url = {https://link.aps.org/doi/10.1103/PhysRevLett.105.213003}
}

@article{Verstichel2012Hubbard,
  author  = {Brecht Verstichel and Hannes {van Aggelen} and Wouter Poelmans and Dimitri {Van Neck}},
  title   = {Variational Two-Particle Density Matrix Calculation for the Hubbard Model Below Half Filling Using Spin-Adapted Lifting Conditions},
  journal = {Physical Review Letters},
  volume  = {108},
  number  = {21},
  pages   = {213001},
  year    = {2012},
  doi     = {10.1103/PhysRevLett.108.213001}
}

@article{PhysRevLett.117.153001,
  title = {Enhanced Constraints for Accurate Lower Bounds on Many-Electron Quantum Energies from Variational Two-Electron Reduced Density Matrix Theory},
  author = {Mazziotti, David A.},
  journal = {Phys. Rev. Lett.},
  volume = {117},
  issue = {15},
  pages = {153001},
  numpages = {5},
  year = {2016},
  month = {Oct},
  publisher = {American Physical Society},
  doi = {10.1103/PhysRevLett.117.153001},
  url = {https://link.aps.org/doi/10.1103/PhysRevLett.117.153001}
}

@article{RubioGarcia2019XXZ,
  author  = {A. Rubio-Garc{\'\i}a and J. Dukelsky and D. R. Alcoba and P. Capuzzi and O. B. O{\~n}a and E. R{\'\i}os and A. Torre and L. Lain},
  title   = {Variational Reduced Density Matrix Method in the Doubly-Occupied Configuration Interaction Space Using Four-Particle N-Representability Conditions: Application to the XXZ Model of Quantum Magnetism},
  journal = {The Journal of Chemical Physics},
  volume  = {151},
  number  = {15},
  pages   = {154104},
  year    = {2019},
  doi     = {10.1063/1.5120359}
}

@article{active_space_2RDM_4_strong_corr,
author = {Head-Marsden, Kade and Mazziotti, David A.},
title = {Active-Space Pair Two-Electron Reduced Density Matrix Theory for Strong Correlation},
journal = {The Journal of Physical Chemistry A},
volume = {124},
number = {23},
pages = {4848-4854},
year = {2020},
doi = {10.1021/acs.jpca.0c01937},
note ={PMID: 32469523},
URL = { https://doi.org/10.1021/acs.jpca.0c01937},
eprint = { https://doi.org/10.1021/acs.jpca.0c01937}
}

@article{Li2021ThreeParticle,
  author  = {R. R. Li and M. D. Liebenthal and A. E. DePrince},
  title   = {Challenges for Variational Reduced-Density-Matrix Theory with Three-Particle N-Representability Conditions},
  journal = {The Journal of Chemical Physics},
  volume  = {155},
  number  = {17},
  pages   = {174110},
  year    = {2021},
  doi     = {10.1063/5.0060247}
}

@article{Knight2022Ultracold,
  author  = {M. J. Knight and H. M. Quiney and A. M. Martin},
  title   = {Reduced Density Matrix Approach to Ultracold Few-Fermion Systems in One Dimension},
  journal = {New Journal of Physics},
  volume  = {24},
  number  = {5},
  pages   = {053004},
  year    = {2022},
  doi     = {10.1088/1367-2630/ac6b6d}
}

@article{PhysRevLett.127.023001,
  title = {Ensemble Reduced Density Matrix Functional Theory for Excited States and Hierarchical Generalization of Pauli's Exclusion Principle},
  author = {Schilling, Christian and Pittalis, Stefano},
  journal = {Phys. Rev. Lett.},
  volume = {127},
  issue = {2},
  pages = {023001},
  numpages = {7},
  year = {2021},
  month = {Jul},
  publisher = {American Physical Society},
  doi = {10.1103/PhysRevLett.127.023001},
  url = {https://link.aps.org/doi/10.1103/PhysRevLett.127.023001}
}

@article{PhysRevLett.124.180603,
  title = {Reduced Density Matrix Functional Theory for Bosons},
  author = {Benavides-Riveros, Carlos L. and Wolff, Jakob and Marques, Miguel A. L. and Schilling, Christian},
  journal = {Phys. Rev. Lett.},
  volume = {124},
  issue = {18},
  pages = {180603},
  numpages = {6},
  year = {2020},
  month = {May},
  publisher = {American Physical Society},
  doi = {10.1103/PhysRevLett.124.180603},
  url = {https://link.aps.org/doi/10.1103/PhysRevLett.124.180603}
}

@article{PhysRevLett.122.013001,
  title = {Diverging Exchange Force and Form of the Exact Density Matrix Functional},
  author = {Schilling, Christian and Schilling, Rolf},
  journal = {Phys. Rev. Lett.},
  volume = {122},
  issue = {1},
  pages = {013001},
  numpages = {7},
  year = {2019},
  month = {Jan},
  publisher = {American Physical Society},
  doi = {10.1103/PhysRevLett.122.013001},
  url = {https://link.aps.org/doi/10.1103/PhysRevLett.122.013001}
}

@article{semidefinite_programming,
author = {Vandenberghe, Lieven and Boyd, Stephen},
title = {Semidefinite Programming},
journal = {SIAM Review},
volume = {38},
number = {1},
pages = {49-95},
year = {1996},
doi = {10.1137/1038003},

URL = { 
        https://doi.org/10.1137/1038003
},
eprint = {
        https://doi.org/10.1137/1038003
}
}

@article{ERDAHL1979147,
title = {Two algorithms for the lower bound method of reduced density matrix theory},
journal = {Reports on Mathematical Physics},
volume = {15},
number = {2},
pages = {147-162},
year = {1979},
issn = {0034-4877},
doi = {https://doi.org/10.1016/0034-4877(79)90015-6},
url = {https://www.sciencedirect.com/science/article/pii/0034487779900156},
author = {R.M. Erdahl}
}

@article{PhysRevA.74.012501,
  title = {Computation of quantum phase transitions by reduced-density-matrix mechanics},
  author = {Gidofalvi, Gergely and Mazziotti, David A.},
  journal = {Phys. Rev. A},
  volume = {74},
  issue = {1},
  pages = {012501},
  numpages = {5},
  year = {2006},
  month = {Jul},
  publisher = {American Physical Society},
  doi = {10.1103/PhysRevA.74.012501},
  url = {https://link.aps.org/doi/10.1103/PhysRevA.74.012501}
}

@article{Schwerdtfeger2009Ising,
  author  = {C. A. Schwerdtfeger and D. A. Mazziotti},
  title   = {Convex-Set Description of Quantum Phase Transitions in the Transverse Ising Model Using Reduced-Density-Matrix Theory},
  journal = {The Journal of Chemical Physics},
  volume  = {130},
  number  = {22},
  pages   = {224102},
  year    = {2009},
  doi     = {10.1063/1.3149785}
}

@article{PhysRevA.69.042511,
  title = {Boson correlation energies via variational minimization with the two-particle reduced density matrix: Exact $N$-representability conditions for harmonic interactions},
  author = {Gidofalvi, Gergely and Mazziotti, David A.},
  journal = {Phys. Rev. A},
  volume = {69},
  issue = {4},
  pages = {042511},
  numpages = {8},
  year = {2004},
  month = {Apr},
  publisher = {American Physical Society},
  doi = {10.1103/PhysRevA.69.042511},
  url = {https://link.aps.org/doi/10.1103/PhysRevA.69.042511}
}

@article{PhysRevA.57.4219,
  title = {Contracted Schr\"odinger equation: Determining quantum energies and two-particle density matrices without wave functions},
  author = {Mazziotti, David A.},
  journal = {Phys. Rev. A},
  volume = {57},
  issue = {6},
  pages = {4219--4234},
  numpages = {0},
  year = {1998},
  month = {Jun},
  publisher = {American Physical Society},
  doi = {10.1103/PhysRevA.57.4219},
  url = {https://link.aps.org/doi/10.1103/PhysRevA.57.4219}
}

@article{Vogiatzis2023DataDriven,
  author  = {Boyn, Jan-Niklas and DePrince, A. Eugene and Vogiatzis, Konstantinos D.},
  title   = {Data-Driven Refinement of Electronic Energies from Two-Electron Reduced-Density-Matrix Theory},
  journal = {The Journal of Physical Chemistry Letters},
  volume  = {14},
  number  = {28},
  pages   = {6413--6419},
  year    = {2023},
  doi     = {10.1021/acs.jpclett.3c01382},
  url     = {https://doi.org/10.1021/acs.jpclett.3c01382}
}

\maketitle

\iftrue

\appendix
\onecolumngrid
\allowdisplaybreaks[3]  

\newpage

\section*{Supplementary Information}   

\section{NN Architecture}\label{sec:NN_arch_appendix}

We train NNs on $n$-RDMs computed on small $L \times L$ lattices and use them to predict $n$-RDMs for larger systems, exploiting their smooth, periodic structure over the BZ. We use two complementary architectures: (i) a self-attention NN that treats momenta as tokens and learns n-RDMs via self--attention \cite{NIPS2017_3f5ee243,bahdanau2016neuralmachinetranslationjointly,shaw2018selfattentionrelativepositionrepresentations}, and (ii) an implicit neural representation (INR) in the style of NeRF/SIREN that models n-RDMs as a continuous periodic function evaluable at arbitrary system sizes \cite{NeRF, sitzmann2020implicitneuralrepresentationsperiodic}.

\subsection{Self-Attention NN Architecture for the 1-RDM} \label{subsec:transformer_arch}

We start with the self-attention NN for 1-RDM.

\subsubsection{physical set-up} 
\label{subsubsec:physical set-up for attNN}

The self-attention NN is applied on a 1-RDM in two spatial dimensions (2D) for the examples in \cref{sec:toymodelpart} and \cref{app:hubbard}.
Therefore, in this part, we will first discuss 1-RDM that will be fed into the NN.
The 1-RDM has the general form 
\eq{
O_{\bsl k\alpha,\,\bsl k'\alpha'}=\langle c^\dagger_{\bsl k\alpha} c_{\bsl k'\alpha'}\rangle\,
}
where $\bsl{k}$ and $\bsl{k}'$ ranges over the first BZ, $\alpha$ represents intra-unit-cell degrees of freedom (sublattice, orbital, spin, etc.), taking $\nint$ different values of them per $\bsl k$. $c^\dagger_{\bsl   {k}\alpha}$ is the creation operator for a state with Bloch momentum $\bsl{k}$ and index $\alpha$, and $\left\langle ... \right\rangle$ is the ground state average.
$\bsl{k}$ takes $L\times L$ different values:
\eq{
\label{eq:k_expression}
\bsl{k} = (l_1/L)\bsl{b}_1 + (l_2/L)\bsl{b}_2\ ,
}
where $l_1,l_2 = 0,1,2,...,L-1$, and $\bsl{b}_1$ and $\bsl{b}_2$ are two primitive reciprocal lattice vectors.
When the translation symmetry is not spontaneously broken, the 1-RDM is reduced to 
\eq{
O_{\bsl k,\alpha\alpha'}=\langle c^\dagger_{\bsl k\alpha} c_{\bsl k\alpha'}\rangle,
\quad
O_{\bsl k}\in\mathbb C^{\nint\times\nint},
}
so the tensor has shape \(L\times L\times\nint\times\nint\).
This is the case for \cref{sec:toymodelpart}. When the translation symmetry is spontaneously broken, we will look at $O_{\bsl{k}\alpha,\bsl{k}'\alpha'}$ with fixed $\bsl{q} \equiv \bsl{k}'-\bsl{k}$, \ie,
\eq{
O_{\bsl{k},\alpha\alpha'}^{\bsl{q}} = O_{\bsl{k}\alpha,\bsl{k}+\bsl{q}\alpha'}\ .
}
For a fixed $\bsl{q}$, $O^{\bsl{q}}$ is also a tensor of dimension $L\times L \times \nint \times \nint$.
In the notation of $O_{\bsl{k},\alpha\alpha'}^{\bsl{q}}$, the translational invariant case is just to focus on $O_{\bsl{k},\alpha\alpha'}^{\bsl{q}=0}$.
Therefore, regardless of the translation symmetry,
$O_{\bsl{k},\alpha\alpha'}^{\bsl{q}}$ has dimension $L\times L \times \nint \times \nint$. 

\subsubsection{Preprocessing (transform 1-RDM to NN input)} \label{subsubsec:flatten_standardize_1RDM}

With the 1-RDM $O^{\bsl{q}}$ defined above, we now show how they are transformed into fixed-length vectors and standardized to form the input for our self-attention NN, which is preprocessing part of \cref{fig:transformer_drawing}(a).
From \cref{eq:k_expression}, the BZ grid has $L\times L$ momenta with the $i$th momentum labelled as $\bsl{k}_i$ ($i=1,2,...,L^2$).

For each $\bsl{k}_i$, we consider the corresponding 1-RDM block, $O^{\bsl q}(\bsl{k}_i)$.
We transform (flatten) this matrix into a real vector $\bsl{x}(\bsl{k}_i) \in \mathbb{R}^{\,2\nint^{2}}$ in the following way. For the translation-invariant case, $O^{\bsl q}(\bsl{k}_i)\in\mathbb{C}^{\nint\times \nint}$ and $\bsl q = 0$ (\cref{sec:toymodelpart}), 
\newcommand{\vecop}{\operatorname{vec}}
\begin{equation}
\bsl{x}(\bsl{k}_i) \equiv
\begin{bmatrix}
vec\big(\operatorname{Re} O^{\bsl q}(\bsl{k}_i)\big)\\
vec\big(\operatorname{Im} O^{\bsl q}(\bsl{k}_i)\big)
\end{bmatrix}
\in \mathbb{R}^{\,2\nint^{2}}.
\label{eq:flattenO}
\end{equation}
Here the operation \textit{vec(...)} simply turns a matrix into a long column vector by transposing and then stacking its rows from top to bottom. Namely,
\begin{equation}
\big[vec(M)\big]_{(\alpha-1)\xi+\alpha'}\equiv M_{\alpha\alpha'},
\qquad \alpha,\alpha'=1,\ldots,\xi.
\label{eq:vec()}
\end{equation}
For the specific translational-symmetry-breaking case in Hubbard model discussed in \cref{app:hubbard}, we have
\(O^{\bsl q}(\bsl k_i)\) real.
Furthermore, $\bsl{q}$ only takes two values , \( \bsl q\in\{0,\bsl Q\} \) with \(\bsl Q=(\pi,\pi)\). 
Thus, in this case, we do not split the 1-RDM into  real/imaginary parts; instead we stack the two \(q\)-blocks:
\eq{\label{eq:flattenO2}
\bsl x(\bsl k_i)
\equiv
\begin{bmatrix}
vec\big(O^{0}(\bsl k_i)\big)\\
vec\big(O^{\bsl Q}(\bsl k_i)\big)
\end{bmatrix}
\in \mathbb{R}^{\,2\xi^{2}} .}
Therefore, both cases (\cref{eq:flattenO,eq:flattenO2}) leads to $\bsl{x}(\bsl{k}_i)$ with the same shape, thus the components of $\bsl{x}(\bsl{k}_i)$ can be labeled by $f\in\{1,\dots,2\nint^2\}$. See \cref{sec:toymodelpart} and \cref{app:hubbard} for more details about the preprocessing steps for each model.
For one BZ mesh, stack the $L^2$ vectors $\bsl x(\bsl k_i)$ row-wise to form
\begin{equation}
X \equiv
\begin{bmatrix}
\bsl x(\bsl k_1)^\top\\\vdots\\\bsl x(\bsl k_{L^2})^\top
\end{bmatrix}
\in \mathbb{R}^{L^2\times 2\xi^2}.
\label{eq:stack-generic}
\end{equation}
A single training sample consists of two 1-RDMs over the entire BZ,
an initial $O^{\bsl q}_{\text{init}}$ and a final (target)
$O^{\bsl q}_{\text{tgt}}$. Here, “\(\mathrm{init}\)” means the 1-RDM we used to start HF;
“\(\mathrm{tgt}\)” means the true final 1-RDM we obtained at zero temperature after self-consistent HF convergence. After flattening in \cref{eq:flattenO,eq:flattenO2}, they become:
\begin{equation}
X^{(\beta)} \equiv
\begin{bmatrix}
\bsl x^{(\beta)}(\bsl k_1)^\top\\
\vdots\\
\bsl x^{(\beta)}(\bsl k_{L^2})^\top
\end{bmatrix}
\in \mathbb R^{\,L^2\times 2\xi^2}, \qquad
\beta\in\{\mathrm{init},\mathrm{tgt}\}.
\end{equation}
Define means and standard deviations over all  $\bsl k_i$ and all $N_\text{sample}$ training samples with $\beta \in \{\text{init}, \text{tgt}\}$
, for each feature $f\in\{1,\dots,2\nint^2\}$:
\begin{equation}\label{eq:mean stand_dev}
    \mu_f^{(\beta)} \equiv \frac{1}{N_{\mathrm{sample}} L^2} \sum_{\text{sample}}\sum_{i=1}^{L^2} X_{i,f}^{(\beta)}, 
    \qquad
    \left(\sigma_f^{(\beta)}\right)^2 \equiv \frac{1}{N_{\mathrm{sample}} L^2 - 1} \sum_{\text{sample}} \sum_{i=1}^{L^2} \left( X_{i,f}^{(\beta)} - \mu_f^{(\beta)} \right)^2.
\end{equation}
Note that for the translation-invariant case (\cref{sec:toymodelpart}), we calculate the mean and standard deviation across all features (see \cref{eq:toyModel_standardize}). We then standardize $X_{\mathrm{init}}, X_{\mathrm{tgt}}$ (computing the $z$-score):
\begin{equation}  \widetilde{X}^{\mathrm{(init)}}_{i,f}
  \equiv \frac{X^{\mathrm{(init)}}_{i,f}-\mu^{\mathrm{(init)}}_{f}}{\sigma^{\mathrm{(init)}}_{f}},
  \qquad
  \widetilde{X}^{\mathrm{(tgt)}}_{i,f}
  \equiv \frac{X^{\mathrm{(tgt)}}_{i,f}-\mu^{\mathrm{(tgt)}}_{f}}{\sigma^{\mathrm{(tgt)}}_{f}},
\label{eq:standardize}
\end{equation}
Equivalently, for each $\bsl{k}_i$
\begin{equation}
\label{eq:standardized_x_i}
\tilde{\bsl{x}}^{(\beta)}(\bsl k_i)
\equiv
\big(\widetilde{X}^{(\beta)}_{i,1},\ldots,\widetilde{X}^{(\beta)}_{i,2\xi^2}\big)^{\top}\in\mathbb{R}^{2\xi^2}.
\end{equation}
In what follows, the NN input is the standardized
$\tilde{\bsl{x}}^{\mathrm{(init)}}(\bsl k_i)$; for brevity we write
$\tilde{\bsl{x}}_i \equiv \tilde{\bsl{x}}^{\mathrm{(init)}}(\bsl k_i)$ unless stated otherwise. Standardization ensures that both the input and output data in the training dataset have a mean of $0$ and a standard deviation of $1$, which improves numerical conditioning and speeds training \cite{Goodfellow-et-al-2016,LeCun1998EfficientBackprop}. This combined set of transformations (flattening in \cref{eq:flattenO} and standardization in \cref{eq:standardize}) is denoted in the preprocessing part of \cref{fig:transformer_drawing}. Note that since these transformations are invertible, we can easily reverse the process to recover the predicted final 1-RDMs from the NN outputs. The matrices $\widetilde X^\text{(init)}$ and $\widetilde X^\text{(tgt)}$, containing exactly the information from $O^{\bsl q}_{\text{init}}$ and $O^{\bsl q}_{\text{tgt}}$, are thus the data we use to train the NN. During the training process, the self-attention NN parameters are optimized so that, starting from $O^{\bsl q}_\text{init}$ defined on all $L^{2}$ momenta, it produces $O^{\bsl q}_{\text{out}}$ whose values at each $\bsl k_i$ closely matches the $O^{\bsl q}_\text{tgt}$. After the training, we would expect the NN to be able to make accurate predictions for $O^{\bsl q}_\text{tgt}$ not only for the training system size but also for larger $L$.

\subsubsection{Overall Structure of Attention Layers} \label{subsubsec: Overall Structure consisting Attention Layers}
With each local $O^{\bsl q}(\bsl k_i)$ now flattened and standardized (\cref{eq:flattenO}-\cref{eq:flattenO2}) into token features $\tilde{\bsl x_i}$, we now describe the overall structure of how our model processes them through stacked attention layers to produce the predicted final 1-RDM, as shown in \cref{fig:transformer_drawing}(b).

\textbf{Input layer.} As shown in \cref{eq:flattenO,eq:flattenO2}, at each momentum \(\bsl k_i\) we have a normalized real feature vector 
\(\tilde{\bsl{x}}_i\in\mathbb{R}^{2\nint^2}\) (flattened entries of the 1-RDM at some $\bsl k_i$).
We first map it to a vector of dimension \(D\) (where $D$ is the ``model width"):
\begin{equation}  \bsl{h}^{(0)}_i = W_{\mathrm{in}}\tilde{\bsl{x}}_i + \bsl{b}_{\mathrm{in}}
  \in \mathbb{R}^{D},
\end{equation}
where $W_{\text{in}}, \bsl{b}_{\text{in}}$ are learnable parameters. Stacking all $\bsl h_i^{(0)}$ gives us a matrix $H^{(0)}$,
\begin{equation}
H^{(0)}
=\begin{bmatrix}
(\bm{h}_1^{(0)})^{\top}\\
\vdots\\
(\bm{h}_{L^{2}}^{(0)})^{\top}
\end{bmatrix}
\in \mathbb{R}^{L^{2}\times D}.
\end{equation}

\textbf{Attention layers.}
We then apply $N$ attention layers (\cref{fig:transformer_drawing}(a)) to $H^{(0)}$ in sequence in order to produce $H^{(N)} \in \mathbb{R}^{L^2\times D}$, where the $i$-th row is $h^{(N)}_i \in \mathbb{R}^{D}$.
In general, we denote the matrix $H^{(\ell-1)} \in \mathbb{R}^{L^2 \times D}$ as the input to the $\ell^{\text{th}}$ layer of our model, $\ell \in \{1, \dots N\}$.
The structure of one attention layer consists of three components: (a) one attention block discussed in  \cref{subsubsec:attentionblock_multipleheads}, (b) Dropout and LayerNorm discussed in \cref{subsubsec:dropout+layernorm} and (c) one FFN, also discussed in \cref{subsubsec:dropout+layernorm}. 
As we will see in \cref{subsubsec:dropout+layernorm}, the first FFN layer (see construction in \cref{subsubsec:dropout+layernorm}) maps $\mathbb{R}^D \rightarrow \mathbb{R}^{D_{\mathrm{ff}}}$ 
and the second maps back $\mathbb{R}^{D_{\mathrm{ff}}} \rightarrow \mathbb{R}^D$.

\textbf{Output head.} Finally, we define a linear head with which learnable parameters
$W_{\mathrm{out}}\in\mathbb{R}^{2\nint^2\times D}$, $\bsl b_{\mathrm{out}}\in\mathbb{R}^{2\nint^2}$ that maps each $\bsl h_i$ back to a vector $\widehat{\bsl x}_i$:
\begin{equation}
  \widehat{\bsl x}_i= W_{\mathrm{out}}\,\bsl h^{(N)}_i + \bsl b_{\mathrm{out}}\ .
\end{equation}
Stacking all  $\widehat{\bsl x}_i$ gives
\eq{
\widetilde X_{\mathrm{out}} = 
  \begin{bmatrix}
  \widehat{\bsl x}_1^{\top}\\
  \vdots\\
  \widehat{\bsl x}_{L^2}^{\top}
  \end{bmatrix}
  \in \mathbb{R}^{L^2\times 2\nint^2}.}
To recover the physics of 1-RDM, we de-standardize elementwise (reverse of \cref{eq:standardize}):
\begin{equation}
  (X_{\mathrm{out}})_{i,f} 
  \equiv(\widetilde X_{\mathrm{out}})_{i,f} \sigma^{\mathrm{tgt}}_{f} + \mu^{\mathrm{tgt}}_{f}.
\label{eq:destandardize1}
\end{equation}
Note that when testing the trained NN for larger $L$, we use the same $\mu^{\mathrm{tgt}}_{f}$ and $\sigma^{\mathrm{tgt}}_{f}$ computed from the smaller system size, since the assumption is that they possess similar distributions. We then reshape $X_{\mathrm{out}}$ to $L\times L\times \nint\times\nint$ and recombine
the real/imag halves to invert the flattening in \cref{eq:flattenO} - \cref{eq:flattenO2}, yielding the final predicted 1-RDM:

\newcommand{\veci}{\vecop^{-1}}
\begin{equation}
\bsl x_{\text{out}}(\bsl k_i)
\equiv
\bigl(\bsl {\widetilde{x}}_{\mathrm{out}}(\bsl k_i)\bigr)_f \, \sigma^{\mathrm{tgt}}_{f}
+ \mu^{\mathrm{tgt}}_{f},
\qquad f=1,\dots,2\nint^2 .
\end{equation}

\begin{equation}
\Re\, O^{\bsl q}_{\mathrm{out}}(\bsl k_i)
\equiv
\veci\left(\, \left(\bsl x_{\text{out}}(\bsl k_i)\right)_{1:\nint^2} \right),
\qquad
\Im\, O^{\bsl q}_{\mathrm{out}}(\bsl k_i)
\equiv
\veci\left( \left(\bsl x_{\text{out}}(\bsl k_i)\right)_{\nint^2+1:2\nint^2} \right),
\end{equation}
\begin{equation}
O^{\bsl q}_{\mathrm{out}}(\bsl k_i)
=
\Re O^{\bsl q}_{\mathrm{out}}(\bsl k_i) + i\Im O^{\bsl q}_{\mathrm{out}}(\bsl k_i),
\label{eq:destandardize2}
\end{equation}
where $\mu_f$, $\sigma_f$ are defined in \cref{eq:mean stand_dev}. And $\vecop^{-1}$ is the inverse of the vec in \cref{eq:flattenO}, which takes a length-$\nint^{2}$ vector and reshapes it back into a $\nint \times \nint$ matrix using the same stacking convention from $\mathrm{vec}$ (see \cref{eq:flattenO}). The process from \cref{eq:destandardize1} to \cref{eq:destandardize2} are shown as the decoding step in \cref{fig:transformer_drawing}.

\textbf{Training.} For computational efficiency, we divide the training dataset into several ``minibatches" (according to standard practice), where each minibatch consists of $N_b$ ($\widetilde X_\text{init}, \widetilde X_\text{tgt}$) pairs. For our loss function, we use the MSE between $\widetilde X_\text{out}$ and $\widetilde X_\text{tgt}$ (both of which have a total of $L^2$ elements):
\begin{equation}\label{eq:MSE_def}
\mathcal{L}
=\frac{1}{N_b L^{2}\,(2\nint^{2})}
\sum_{b =1}^{N_b}
\sum_{i=1}^{L^{2}}
\sum_{f=1}^{2\nint^{2}}
\big(\widetilde{X}^{(b)}_{\text{out},i, f}-\widetilde{X}^{(b)}_{\mathrm{tgt}, i, f}\big)^{2}.
\end{equation}
as indicated in \cref{fig:transformer_drawing}, then take the gradient of the loss $\mathcal{L}$ with respect to the trainable parameters (defined in \cref{subsubsec:dropout+layernorm}) and update the values of parameters using the AdamW optimizer \cite{loshchilov2019decoupledweightdecayregularization}.

\subsubsection{Attention Block (multiple attention heads)} \label{subsubsec:attentionblock_multipleheads}

Here we describe the detailed structure of the attention block we first mentioned in \cref{subsubsec: Overall Structure consisting Attention Layers} as part of each attention layer. Each attention block includes $N_{\text{head}}$ attention heads (\cref{fig:transformer_drawing}(c)).
The structure of one attention head is shown in \cref{fig:transformer_drawing}(d). We first introduce the \textit{Bias} block in \cref{fig:transformer_drawing}(d).
The bias block is constructed based on the geometry-aware relative positional bias that will be used in attention heads.
Intuitively speaking, attention \cite{NIPS2017_3f5ee243} decides “which $\bsl k_j$ should influence $\bsl k_i$.” We want this decision to respect BZ geometry and periodicity. Precisely, for each ordered pair of momenta $(\bsl k_i,\bsl k_j)$ with
$\bsl k_i=\tfrac{l_{i,1}}{L}\bsl b_1+\tfrac{l_{i,2}}{L}\bsl b_2$ and
$\bsl k_j=\tfrac{l_{j,1}}{L}\bsl b_1+\tfrac{l_{j,2}}{L}\bsl b_2$
defined in \cref{eq:k_expression}, we stack them to form the vector
\begin{equation}
\bigl(\tfrac{l_{i,1}}{L},\tfrac{l_{i,2}}{L},\tfrac{l_{j,1}}{L},\tfrac{l_{j,2}}{L}\bigr)\in[0,1)^4 .
\end{equation} In each attention head, we then define a learnable continuous function $f_{\theta}: \mathbb{R}^4 \rightarrow \mathbb{R}$ that maps each such pair of momenta to a scalar bias that represents how closely correlated they are. In general, this $f_{\theta}$ can be parameterized as a standard NN, as in \cref{sec:toymodelpart}. However, we can also explicitly incorporate our physical assumptions by using a carefully-chosen ansatz with a periodic basis, as is used in \cref{app:hubbard} and described below. Let \(W_r \in \mathbb{R}^{m\times 2}\) be a learnable matrix, where $m$ is a hyperparameter that determines the number of frequencies. Project each $\bsl k_i, \bsl k_j$,  separately:
\begin{equation}\label{eq:W_r_freqs}
  \bsl{u}_i = W_r
  \begin{bmatrix}
    \tfrac{l_{i,1}}{L} \\ \tfrac{l_{i,2}}{L}
  \end{bmatrix}
  \in \mathbb{R}^m,
  \qquad
  \bsl{u}_j = W_r
  \begin{bmatrix}
    \tfrac{l_{j,1}}{L} \\ \tfrac{l_{j,2}}{L}
  \end{bmatrix}
  \in \mathbb{R}^m.
\end{equation}
We build periodic features
\begin{equation}
  \boldsymbol{\phi}(\bsl k) \equiv \cos \bsl{u} + \sin \bsl{u}
  \in \mathbb{R}^m \quad (\text{applied elementwise}),
\end{equation}
and take the difference
\begin{equation}  \boldsymbol{\Delta}_{ij} = \boldsymbol{\phi}(\bsl k_i) - \boldsymbol{\phi}(\bsl k_j)
  \in \mathbb{R}^m .
\end{equation}
With a learnable vector of coefficients \(\bsl C\in\mathbb{R}^m\) (which is initially set to all $1$'s before being optimized during training), we define a scalar, nonnegative “periodic distance’’ $\sigma_{ij}$,
\begin{equation}\label{eq:inner_product_C}
  \sigma_{ij} \equiv \bigl| \langle \boldsymbol{\Delta}_{ij},\bsl C \rangle \bigr|
  \ge 0,
\end{equation}
(here $<,>$ means standard dot-product) and convert it to a decaying bias~\cite{shaw2018selfattentionrelativepositionrepresentations} $f_{\theta}$:
\begin{equation}
  f_{\theta}(\bsl k_i,\bsl k_j)
  \equiv \exp\left[-\bigl(\sigma_{ij}+\varepsilon\bigr)^{p}\right],
  \qquad \varepsilon = 10^{-6}\ ,
\label{eq:f_theta}
\end{equation}
where \(p\in[0.5,3]\) is an exponent that serves as a practical stability constraint \cite{Rasmussen2006GPML,Gneiting2002Compact,Buhmann2003RBF}.
Repeating this process for each pair of momenta, we obtain a matrix $B \in \mathbb{R}^{L^2 \times L^2}$, where we define
\begin{equation}
    B_{ij} \equiv f_{\theta}(\bsl k_i,\bsl k_j), \quad i, j \in \{1, \dots L^2\}.
    \label{eq:B_ij}
\end{equation}
Thus $B_{ij}$ acts as a smooth, periodic bias on the 1BZ \cite{Li_2024_FIRE,zhou2024transformersachievelengthgeneralization}, which is shown as the \textit{Bias} block in \cref{fig:transformer_drawing}(d).

Besides the bias block, the attention head also involves the query $Q$ matrix, key $K$ matrix, value $V$ matrix, and Scalable Softmax.
For each attention head \(h\in\{1,\dots,N_{\text{head}}\}\) with  $N_{\text{head}}$ the number of heads, we make matrices
$Q^{(h)}, K^{(h)}, V^{(h)} \in \mathbb{R}^{L^2 \times d_k}$
\cite{NIPS2017_3f5ee243} by multiplying $H^{(\ell-1)}$ (the current layer's input) with learnable matrices
$W_Q^{(h)},\, W_K^{(h)},\, W_V^{(h)} \in \mathbb{R}^{D \times d_k}$, where \(d_k \equiv D/N_{\text{head}}\) is the width of each head and $D$ is the model width as defined in \cref{subsubsec: Overall Structure consisting Attention Layers}. Then we have:
\begin{equation}
  Q^{(h)} = H^{(\ell-1)} W_Q^{(h)},\qquad
  K^{(h)} = H^{(\ell-1)} W_K^{(h)},\qquad
  V^{(h)} = H^{(\ell-1)} W_V^{(h)}.
\label{eq:Q_K_V}
\end{equation}
With $B_{ij}$ from \cref{eq:B_ij}, define the pre-softmax logits \cite{press2022trainshorttestlong,NIPS2017_3f5ee243} as:
\begin{equation}
\Lambda^{(h)}_{ij}
= \tau \left( \frac{\big\langle Q^{(h)}_{i}, K^{(h)}_{j}\big\rangle + B_{ij}} {\sqrt{d_k}} \right),
\qquad \text{where } \tau \equiv s\,\log\left(L^{2}\right)\ ,
\label{eq:scaled_logits}
\end{equation}
and $s$ is a learnable parameter. (We use the notation of $Q_i,K_i,$ and $V_i$ for the $i^{\text{th}}$ row of $Q,K,$ and $V$, respectively, with $i \in \{1, \dots L^2\}$). Notice that $B_{ij}\in(0,1]$ is large when $\bsl k_i$ and $\bsl k_j$ are close on 1BZ, so it adds extra preference to mix nearby momenta. Meanwhile, the multiplication by $s\,\log\left(L^{2}\right)$ implements Scalable Softmax \cite{nakanishi2025scalablesoftmaxsuperiorattention} to avoid the phenomenon of ``attention fading''; that is, it ensures that for a large number of $\bsl{k}$ points \cite{press2022trainshorttestlong}, the maximum value of the attention weights does not drop to $0$, which could otherwise hinder the NN's capacity for length generalization. The attention weights and head outputs are
\begin{equation}
A_{ij}^{(h)} = 
\frac{e^{\tau \Lambda_{ij}^{(h)}}}{\sum_{m=1}^{L^{2}} e^{\tau \Lambda_{im}^{(h)}}},
\qquad
\sum_{j} A_{ij}^{(h)} = 1,\quad A_{ij}^{(h)} > 0.
\label{eq:A_ij_z_ij}
\end{equation}
\ie, Softmax \cite{Bridle1989ProbabilisticIO} is applied here (row-wise normalization that turns attention
scores into positive weights summing to \(1\)). Then,
\eq{\bsl{z}^{(h)}_{i}=\sum_{j=1}^{L^2} A^{(h)}_{ij}\,V^{(h)}_{j}\in\mathbb{R}^{d_k}.}
We can stack all the $\bsl{z}_i's$ constructed in this way to form a matrix:
\eq{
Z^{(h)} = 
  \begin{bmatrix}
  \bsl{z}_1^{(h)} \\
  \vdots\\
  \bsl{z}_{L^2}^{(h)} \\
  \end{bmatrix}
  \in \mathbb{R}^{L^2\times d_k}.}
Therefore, a summary of the attention head we just discussed in \cref{eq:Q_K_V}-\cref{eq:A_ij_z_ij} is the following (see the red box in \cref{fig:transformer_drawing}):
\eq{
\mathrm{AttentionHead}^{(h)}\big(H^{(\ell -1)}\big)\coloneqq Z^{(h)}
\quad\text{with}\quad
\begin{cases}
Q^{(h)} = H^{(\ell-1)} W_Q^{(h)},\quad
K^{(h)} = H^{(\ell-1)} W_K^{(h)}, \quad
V^{(h)} = H^{(\ell-1)} W_V^{(h)},\\
\Lambda^{(h)}_{ij}
= \tau \left( \frac{\big\langle Q^{(h)}_{i},\, K^{(h)}_{j}\big\rangle + B_{ij}}{\sqrt{d_k}} \right), \text{where } \tau \equiv s\log\left(L^{2}\right),
\\
A_{ij}^{(h)} = 
\frac{e^{\tau \Lambda_{ij}^{(h)}}}{\sum_{m=1}^{L^{2}} e^{\tau \Lambda_{im}^{(h)}}},
\quad
\sum_{j} A_{ij}^{(h)} = 1,\quad A_{ij}^{(h)} > 0,\\
Z^{(h)} = A^{(h)} V^{(h)}.
\end{cases}
}
Finally, the $N_{\text{head}}\geq1$ attention heads form an attention block. Parameters are not shared between different attention heads (even if they are in the same attention block), (\ie, each individual attention head has its own set of learnable parameters which are optimized independently from the others).
For all $i \in \{1, \dots L^2\}$, we concatenate (\ie ``stack'') the output vectors $\bsl{z}^{(h)}_{i} \in \mathbb{R}^{d_k}$ from each attention head into a single vector of size $N_{\text{head}} d_k = D$:
\begin{equation}
\mathrm{Concat}\big( Z^{(1)},\ldots, Z^{(N_{\text{head}})}\big)
= \big[ Z^{(1)}\ \ Z^{(2)}\ \ \cdots\ \ Z^{(N_{\text{head}})} \,\big]
\in \mathbb{R}^{\,L^2 \times D} .
\end{equation}
and then apply an output linear map with learnable parameters $W_O\in\mathbb R^{D\times D}, \bsl{b}_O\in \mathbb R^{D}$:
\eq{ \label{eq:concat}
{Z} = \operatorname{Concat}\big({Z}^{(1)},\ldots,{Z}^{(N_{\text{head}})}\big)\, W_O + \bsl 1 \bsl{b}_O^\top,
}
and here $\bsl 1$ is a length-$L^2$ column of ones. This mixes information across heads and returns to the model width $D$.

\subsubsection{Dropout, Layer Normalization, and FFN}\label{subsubsec:dropout+layernorm}

We now discuss the dropout, layer normalization, and FFN in each each attention layer (we first introduced them in \cref{subsubsec: Overall Structure consisting Attention Layers}) as shown in \cref{fig:transformer_drawing}(b).

Applying $\operatorname{Dropout}$ to a vector or matrix means that we randomly set each element of the vector or matrix to zero with a probability $p$ or keep it with a probability of $1-p$~\cite{JMLR:v15:srivastava14a}.
We apply dropout to the attention output $\bsl{Z}$ (\cref{eq:Dropout}), add the result to the layer input $H^{(\ell-1)}$ (\cref{eq:Ui}), and then apply LayerNorm per $\bsl k_i$:
\begin{equation}\label{eq:Dropout}
\hat{{Z}} = \operatorname{Dropout}({Z}),
\qquad
\text{where }\operatorname{Dropout}({Z})_{ij} = \frac{m_{ij}{Z}_{ij}}{1-p},
\quad
m_{ij} \sim \mathrm{Bernoulli}(1-p),
\end{equation}
\begin{equation}
\bsl U_i = \operatorname{LayerNorm}\big(H_i^{(\ell-1)} + \hat{\bsl{Z}}_i\big), 
\quad {U} \in \mathbb{R}^{L^2 \times D}.
\label{eq:Ui}
\end{equation}
Here LayerNorm is applied independently to each \(\bsl U_i\in\mathbb{R}^D\), and is defined as follows:
define the mean and variance over \(d=1,\dots,D\):
\eq{
\mu_i \equiv \frac{1}{D}\sum_{d=1}^{D} U_{i,d}, 
\qquad
\sigma_i^{2} \equiv \frac{1}{D}\sum_{d=1}^{D}\big(U_{i,d}-\mu_i\big)^{2},
}
then, with learnable parameters \(\bsl \gamma, \bsl{\beta} \in\mathbb{R}^{D}\) and a small \(\epsilon>0\), we set
\eq{\label{eq:layernorm}
\mathrm{LayerNorm}(\bsl U_i)
\equiv
\gamma \odot \frac{\bsl U_i-\mu_i \bsl{1}}{\sqrt{\sigma_i^{2}+\epsilon}}
+
\beta
\in\mathbb{R}^{D},
}
where $\odot$ is the element-wise product. Namely,
\eq{
\big[\mathrm{LayerNorm}(U_i)\big]_d
=
\gamma_d \frac{U_{i,d}-\mu_i}{\sqrt{\sigma_i^{2}+\epsilon}} + \beta_d,
\qquad d=1,\ldots,D.
}
For every $\bsl k_i$, take its current vector $\bsl U_i \in \mathbb{R}^{D}$ and apply a learned linear map which increases the dimension to a larger width $D_{\mathrm{ff}}$, to get $\bsl F_i = W_1 \bsl U_i + \bsl{b}_1$ with $W_1\in\mathbb{R}^{D\times D_{\mathrm{ff}}}$ and $\bsl{b}_1\in\mathbb{R}^{D_{\mathrm{ff}}}$:
\begin{equation}
    F = U W_1 + \textbf{1}\bsl{b}_1^\top\in \mathbb{R}^{L^2 \times D_\text{ff}}.
\end{equation}
Apply the GELU (``Gaussian Error Linear Unit''\cite{hendrycks2023gaussianerrorlinearunits}) element-wise to $F$, which smoothly gates each component of $F$:
small negative values are damped, large positive values pass through almost linearly. It is written using the error function $\operatorname{erf}$, then mapped back from the hidden width to the model width:
\begin{equation}\label{eq:FFN}
\widehat{U} = W_2\,\mathrm{GELU}(F) + \textbf{1}\bsl{b}_2^\top \in \mathbb{R}^{L^2\times D} \quad
\text{where} \quad \mathrm{GELU}(F)\equiv \frac{F}{2}\left(\bsl{1}+\operatorname{erf}\left(\frac{ F}{\sqrt{2}}\right)\right),
\end{equation}
with $W_2\in\mathbb{R}^{D_{\mathrm{ff}}\times D}$, $\bsl{b}_2\in\mathbb{R}^{D}$, and $\bsl{1}\in \mathbb R^{L^2}$ (column of ones). Then,
\eq{H^{(\ell)} = \mathrm{LayerNorm}\left(\bsl U + \mathrm{Dropout}(\widehat{\bsl U})\right).}
\cref{eq:FFN} is labeled as FFN in \cref{fig:transformer_drawing}.

\subsection{SIREN with Custom Loss Function}
\label{SIREN_arch}
Alternatively, rather than learning a map from a random initial n-RDM to a final n-RDM, we may treat the task purely as interpolation: we learn a continuous function \(\Phi\) (the interpolant) represented by a NN that takes momentum–space coordinates \(\bsl{k}\) (or normalized coordinates \(\bsl v\in[-1,1]^2\), see \cref{eq:transformedk}) as input, and outputs a matrix $O^{\bsl q}(\bsl k_i)$ representing the n-RDM, e.g.\ \(\Phi(\bsl{k}) \approx O^{\bsl q}_{\alpha\alpha'}(\bsl{k})\) (or \(\Re/\Im\,O^{\bsl q}_{\alpha\alpha'}(\bsl{k})\) when training the real/imaginary parts separately). The explicit structure is in \cref{subsubsec:explicit_structure_SIREN}. One way to achieve this interpolation is the SIREN architecture.
In the following, we will discuss it using the example of fixed-$\bsl{q}$ $1$-RDM that only depends on one $\bsl{k}$.
Nevertheless, the discussion holds for a matrix that depends on more than one Bloch momenta, as shown in \cref{sec:RichardsonAppendix}.

\subsubsection{physical set-up}

We keep the BZ discretization of \cref{eq:k_expression}. All $n$-RDM are defined as in \cref{subsubsec:physical set-up for attNN}. As discussed in \cref{subsubsec:physical set-up for attNN}, we focus on considering the fixed-$\bsl{q}$ 1-RDMs ($O^{\bsl q}_{\bsl{k},\alpha\alpha'}$) over BZ with a shape of $L\times L\times \nint\times\nint$.
The overall goal of the SIREN \cite{sitzmann2020implicitneuralrepresentationsperiodic} approach is to train one continuous function \(\Phi(k_x,k_y)\) that predicts each 1-RDM block over the BZ. In particular, we train a function \(\Phi(k_x,k_y)\) per $(\alpha,\alpha')$
and per real/imaginary part  of $O^{\bsl q}_{\bsl{k},\alpha\alpha'}$ if needed, such that
\begin{equation}
    \Phi(\bsl k)\approx \Re\left[ O^{\bsl q}_{\bsl{k},\alpha\alpha'}\right] \quad \text{or} \quad \Im\left[ O^{\bsl q}_{\bsl{k},\alpha\alpha'}\right].
\end{equation}
(In practice, both the Hubbard (\cref{app:hubbard}) and the Richardson (\cref{sec:RichardsonAppendix}) model exhibit a vanishing imaginary part, so we only consider the real part for training in those cases.) Stacking all different $\Phi$ for all different values of  $(\alpha,\alpha')$ reconstructs $O^{\bsl q}(\bsl k)$.
We will train on smaller BZ grids, and then evaluate \(\Phi\) on denser BZ grids.

\subsubsection{NN input}
\label{subsubsec:Coordinate domain and periodic wrap and symmetry}
For an $L\times L$ BZ with $\bsl k_i$ from \cref{eq:k_expression}, we define the normalized coordinate
\begin{equation}
\boldsymbol{v}(\bsl{k}) \equiv \Big(2\,\tfrac{\ell_1}{L}-1, 2\,\tfrac{\ell_2}{L}-1\Big)\in[-1,1]^2,
\quad\text{for }\bsl{k}=\tfrac{\ell_1}{L}\bsl{b}_1+\tfrac{\ell_2}{L}\bsl{b}_2.
\label{eq:transformedk}
\end{equation}
For each fixed $(\alpha,\alpha')$, evaluate
$\Re[ O^{\bsl q}(\bsl k)]\text{ and } \Im [O^{\bsl q}(\bsl k)]\ \text{ for all } \bsl k.$
To enforce periodicity at the boundaries, extend each to a wrapped array
\eq{
\Re[\widetilde{O}^{\bsl q}],\ \Im[\widetilde{O}^{\bsl q}] \in \mathbb{R}^{(L+1)\times(L+1)}
\label{eq:Otilt}
}
by repeating the first row and column:
\begin{equation}
    \tilde O^{\bsl q}_{m,n}\equiv O^{\bsl q}_{\,m\bmod L, n\bmod L},\qquad m,n\in\{0,1,\dots,L\}
\label{eq:wrap_def}
\end{equation}
So $\widetilde O^{\bsl q}$ is just $O^{\bsl q}$ with its first row/column copied to the end—this “closes” the matrix at finite $L$. 

\subsubsection{SIREN architecture}\label{subsubsec:explicit_structure_SIREN}

Given input to our NN in the form of \cref{eq:transformedk}, we build one SIREN \cite{sitzmann2020implicitneuralrepresentationsperiodic} (depicted in \cref{fig:SIREN_drawing}) to predict each block of the 1-RDM respectively (e.g., $\Re [O(\bsl k)]$ or $\Im [O(\bsl k)]$). Each SIREN has 4 hidden sine layers: \(\bsl h_1,\bsl h_2, \bsl h_3, \bsl h_4\), followed by a final linear layer. Explicitly,
\eq{
\begin{aligned}
\bsl h_1(\bsl v) &= \sin\big(6\,(W_1\bsl v+\bsl b_1)\big) \quad \in \mathbb{R}^{d_H},\\
\bsl h_2(\bsl v) &= \sin\big(30\,(W_2 \bsl h_1(\bsl v)+\bsl b_2)\big) \quad \in \mathbb{R}^{d_H},\\
\bsl h_3(\bsl v) &= \sin\big(30\,(W_3 \bsl h_2(\bsl v)+\bsl b_3)\big) \quad \in \mathbb{R}^{d_H},\\
\bsl h_4(\bsl v) &= \sin\big(30\,(W_4 \bsl h_3(\bsl v)+\bsl b_4)\big) \quad \in \mathbb{R}^{d_H},\\
\Phi_\theta(\bsl v) &= \bsl w^\top \bsl h_4(\bsl v)+b_{\text{out}} \quad \in \mathbb{R}.
\label{eq:ourMLP}
\end{aligned}
}
where
\eq{
\begin{aligned}
& \bsl v \in \mathbb{R}^{2},\quad
W_1\in\mathbb{R}^{d_H\times 2},
W_2,W_3,W_4\in\mathbb{R}^{d_H\times d_H}, \bsl b_1, \bsl b_2, \bsl b_3,\bsl b_4\in\mathbb{R}^{d_H}, \bsl w\in\mathbb{R}^{d_H}, b_{\text{out}}\in\mathbb{R}.
\end{aligned}
}
\cref{eq:ourMLP} maps a BZ coordinate from \cref{eq:transformedk},
$\bsl v=(v_x,v_y)\in[-1,1]^2$, to a scalar
$\Phi_{\theta}(\bsl v)\in\mathbb{R}$ that predicts the corresponding element of
either $\Re [O^{\bsl q}](\bsl k_i)$ or $\Im[O^{\bsl q}](\bsl k_i)$. The layer widths are:
\(d_0=2\) for the input vector $\bsl{v}$ (from \cref{eq:transformedk}),
and \(d_1=d_2=d_3=d_4=d_H\) for the hidden layers ($h_{1,2,3,4}$ in \cref{eq:ourMLP}) for which we usually select \(d_H\in[256,800]\).
We choose sines because they naturally model smooth, periodic patterns across the BZ. Additionally, to keep gradients healthy, we incorporate a special weight initialization scheme \cite{sitzmann2020implicitneuralrepresentationsperiodic}:
define $W \sim \mathcal{U}(a,b)$ to mean ``sample uniformly at random between $a$ and $b$.'' Here \(W\sim\mathcal U\) denotes element-wise initialization:
each entry \(W_{ij}\) is drawn independently from \(\mathcal U(a,b)\). Then we initialize the weights in \cref{eq:ourMLP} with:

\begin{equation}
W_1 \sim \mathcal{U}\Bigl(-\tfrac{1}{d_0},\tfrac{1}{d_0}\Bigr), 
\qquad
\text{and }
w^\top, W_\ell \sim \mathcal{U}\Bigl(-\tfrac{\sqrt{6/d_H}}{30},\tfrac{\sqrt{6/d_H}}{30}\Bigr)
\text{ elementwise, }\quad
2 \leq \ell \leq 4.
\end{equation}
The full model can then be written as $\Phi_\theta$, where $\theta$ represents all the tunable parameters mentioned above.
\subsubsection{Datasets, standardization, and evaluation grids}\label{app: Datasetsstandardization}

For $\Re[\tilde{O}^{\bsl{q}}] \in \mathbb{R}^{(L+1)\times(L+1)}$ defined in \cref{eq:Otilt}
, we compute the mean and standard deviation over all entries:
\eq{\label{eq:SIREN section mu_mean}
\mu = \mathrm{mean}(\Re[\tilde{O}^{\bsl{q}}]), \qquad \sigma = \mathrm{std}(\Re[\tilde{O}^{\bsl{q}}]),}
and form the standardized version of $\widetilde{O}^{\bsl q}:$
\begin{equation}
    Z_L \equiv \frac{\Re[\tilde{O}^{\bsl{q}}] - \mu}{\sigma}\in \mathbb{R}^{(L+1)\times(L+1)}.
\label{eq:standardizeZ_L}
\end{equation}
A similar standardization procedure applies for $\Im[\tilde{O}^{\bsl{q}}]$.
For evaluation on larger $L$, we use the same $\mu$ and $\sigma$ (computed from the $O^{\bsl{q}}$ used for training) to normalize the data. Also notice that the unstandardized SIREN predictions for the target channel can be recovered after training by reversing \cref{eq:standardizeZ_L}. Each input point is a normalized BZ coordinate \(\nu\in[-1,1]^2\) (see \cref{eq:transformedk}). Define the coordinate grid $X_L \equiv \{\bsl{v}(\bsl{k}_{ij}) : i,j=0,\ldots,L\}
\subset [-1,1]^{(L+1)\times(L+1)\times 2},
\quad \bsl{k}_{ij}\equiv \tfrac{i}{L}\bsl{b}_1+\tfrac{j}{L}\bsl{b}_2.$
Our training data thus consists of the set of input coordinates $X_L \in [-1,1]^{(L+1)\times(L+1)\times 2}$ and the normalized values $Z_L$. 

When calculating the loss, we also evaluate SIREN on a $L'\times L'$ mesh with $L' > L$ and output $Z_{L'}\in [-1,1]^{(L'+1)\times(L'+1)\times 2}$, as seen in \cref{subsubsec:SIREN_loss_function}. 
However, we will eventually downsample the $Z_{L'}$ back to $L\times L$ mesh to compare to the training data.
In this process, we do \emph{not} use any true final RDMs on the $L'\times L'$ mesh as training data, since doing so would defeat our entire purpose of generalizing to larger system sizes.

\subsubsection{Loss function}\label{subsubsec:SIREN_loss_function}
Define $\Phi_{\theta} : [-1,1]^2 \to \mathbb{R}$ to be the SIREN we constructed in \cref{subsubsec:explicit_structure_SIREN} with tunable parameters denoted as $\theta$,
which takes a normalized BZ coordinate $\bsl v(\bsl{k}) \in [-1,1]^2$ as input and outputs a scalar prediction for the corresponding element of the chosen 1-RDM block:
\eq{\widehat{Z}(\bsl{k}) \equiv \Phi_{\theta}\bigl(\bsl v(\bsl{k})\bigr),}

For small-mesh MSE,
\eq{\mathcal{L}_{\mathrm{small}}(\theta)
= \frac{1}{(L+1)^2}
\sum_{\bsl v\in X_L}
\Big(\widehat{Z}(X_L)[\bsl v] - Z_L[\bsl v]\Big)^2 .}

Let \(G\) be any finite symmetry subgroup of the symmetry group of the system acting on normalized BZ coordinates
\(\bsl{v}\in[-1,1]^2\) (\eg, rotations/reflections of the square). 
Define the symmetrization operator \(S_G\) by averaging over the group action.
For example, for a square lattice we take \(G=D_4\) (order \(8\), consisting of four rotations \(0^\circ, 90^\circ, 180^\circ, 270^\circ\) and four reflections across the two axes and two diagonals).
Let \(\zeta\) be a fixed downsampler (\eg\ bilinear or bicubic interpolation~\cite{keys1981cubic,mitchell1988reconstruction}), which is effectively a classical algorithm used to interpolate the SIREN predictions on the denser coordinate mesh $X_{L'}$ back to the training system size. This allows us to control behavior at finer scales, without the need for additional training data (e.g. $Z_{L'}$) from the larger system. As such, we evaluate the SIREN on the dense grid $X_{L'}$, symmetrize, and then downsample back down to $X_L$:
\eq{\widehat{Z}^{\downarrow}(X_L) \equiv \zeta\big(\,\mathcal{S}_G[\widehat{Z}(X_{L'})] \to X_L \big),}
\eq{\mathcal{L}_{\mathrm{cons}}(\theta)
= \frac{1}{(L+1)^2}
\sum_{\bsl v\in X_L}
\Big(\widehat{Z}^{\downarrow}(X_L)[\bsl v] - Z_L[\bsl v]\Big)^2 .}
(Note that for the Richardson model in \cref{sec:richardson_ML}, we employ a slightly different technique to learn the symmetry group during training; see \cref{eq:L_symmetry} for details.) Randomly sample \(N\) points \(U=\{u_i\}_{i=1}^N \subset [-1,1]^2\).
With \(\nabla_u\) the gradient and \(\mathrm{Hess}_u\) the Hessian w.r.t.\ \(u\),
and with \(\|\cdot\|_1\) the $\ell^1$ norm and \(\|\cdot\|_F\) the Frobenius norm:
\eq{ \label{eq:FrobeniusNorms}
\mathcal{P}_{\mathrm{TV}}(\theta)
\equiv \frac{1}{N}\sum_{i=1}^N \big\|\nabla_u \Phi_\theta(u_i)\big\|_{1},
\qquad
\mathcal{P}_{\mathrm{FH}}(\theta)
\equiv \frac{1}{N}\sum_{i=1}^N \big\|\mathrm{Hess}_u \Phi_\theta(u_i)\big\|_{F}\ .
}
The total loss is:
\begin{equation}\label{eq:total_loss_function}
\mathcal{L}(\theta)
=
\mathcal{L}_{\mathrm{small}}(\theta)
+
\mathcal{L}_{\mathrm{cons}}(\theta)
+
\lambda\,\mathcal{P}(\theta),
\qquad
\mathcal{P}\in\{\mathcal{P}_{\mathrm{TV}},\,\mathcal{P}_{\mathrm{FH}},\,0\},
\end{equation}
with \(\lambda\in[10^{-7},\,10^{-5}]\) in practice.  We then take the gradient of $\mathcal{L}$ w.r.t.\ \(\theta\) and train the SIREN using the Adam optimizer \cite{kingma2017adammethodstochasticoptimization}. For all trainings using a SIREN-based NN in this paper, we set the constant learning rate for the optimizer equal to $1 \times 10^{-4}$.

\section{Richardson Model}\label{sec:RichardsonAppendix}
\subsection{Model Formulation}
We start by reviewing the Richardson model for superconductivity ~\cite{Richardson1963Restricted, RichardsonSherman1964, RevModPhys.76.643}.
\subsubsection{Analytic Solution via the Richardson Equations}

We begin with deriving the Richarson model for spinful fermions on a Bravis lattice with one atom per unit cell.
We label the hopping by $t_{\bsl r\bsl r'}$ and introduce an attraction $V(\bsl r - \bsl r')\le 0$ that depends only on separation $\bsl r-\bsl r'$ ~\cite{PhysRev.108.1175}:
\begin{equation}
H = \sum_{\bsl r,\bsl r',s} t_{\bsl r\bsl r'}\, c^\dagger_{\bsl rs} c_{\bsl r's}
+ \frac{1}{2}\sum_{\bsl r, \bsl r'} V(\bsl r{-}\bsl r') c^\dagger_{\bsl r\uparrow} c^\dagger_{\bsl r'\downarrow}\, c_{\bsl r'\downarrow} c_{\bsl r\uparrow}\ ,
\label{eq:realspaceBCS}
\end{equation}
Introduce lattice Fourier modes
\begin{equation}
c_{\bsl r,s}=\frac{1}{L}\sum_{\bsl k\in\mathrm{BZ}} e^{\,i\bsl k\cdot\bsl r}\,c_{\bsl k, s},
\qquad
c_{\bsl k, s}=\frac{1}{L}\sum_{\bsl r} e^{-i\bsl k\cdot\bsl r}\,c_{\bsl r, s},
\quad
\sum_{\bsl r} e^{\,i(\bsl k-\bsl k')\cdot \bsl r}=L^2\delta_{\bsl k,\bsl k'}.
\label{eq:Fouriertransform_lattice}
\end{equation}
so that the kinetic term diagonalizes as
\begin{align}
H_0
&= \sum_{\bsl r,\bsl r',\,s} t_{\bsl r-\bsl r'} c^\dagger_{\bsl r s}\, c_{\bsl r' s} \\
&= \sum_{\bsl r,\bsl r',\,s} t_{\bsl r-\bsl r'}
   \Big[\frac{1}{L}\sum_{\bsl k} e^{-i\bsl k\cdot \bsl r}\, c^\dagger_{\bsl k s}\Big]
   \Big[\frac{1}{L}\sum_{\bsl k'} e^{\,i\bsl k'\cdot \bsl r'}\, c_{\bsl k' s}\Big] \\
&= \frac{1}{L^2}\sum_{\bsl k,\bsl k',\,s}
   \Big[\sum_{\bsl r,\bsl r'} t_{\bsl r-\bsl r'}\,
   e^{-i\bsl k\cdot \bsl r}\, e^{\,i\bsl k'\cdot \bsl r'}\Big]\,
   c^\dagger_{\bsl k s} c_{\bsl k' s} \\
&= \frac{1}{L^2}\sum_{\bsl k,\bsl k',\,s}
   \Big[\sum_{\bsl r-\bsl r'} t_{\bsl r-\bsl r'} \sum_{\bsl r'} e^{-i\bsl k\cdot (\bsl r'+(\bsl r-\bsl r'))} e^{\,i\bsl k'\cdot \bsl r'}\Big]\,
   c^\dagger_{\bsl k s} c_{\bsl k' s} \\
&= \frac{1}{L^2}\sum_{\bsl k,\bsl k',\,s}
   \Big[\sum_{\bsl r-\bsl r'} t_{\bsl r-\bsl r'}\, e^{-i\bsl k\cdot (\bsl r-\bsl r'))}\Big]
   \Big[\sum_{\bsl r'} e^{\,i(\bsl k'-\bsl k)\cdot \bsl r'}\Big]\,
   c^\dagger_{\bsl k s} c_{\bsl k' s} \\
&= \sum_{\bsl k,\,s}
   \underbrace{\Big[\sum_{\bsl r-\bsl r'} t_{\bsl r-\bsl r'}\, e^{\,i\bsl k\cdot (\bsl r-\bsl r')}\Big]}_{\displaystyle \varepsilon_{\bsl k}}
   c^\dagger_{\bsl k s} c_{\bsl k s}
   \qquad \Big(\sum_{\bsl r'} e^{\,i(\bsl k'-\bsl k)\cdot \bsl r'} = L^2 \delta_{\bsl k,\bsl k'}\Big) \\
&= \sum_{\bsl k,\,s} \varepsilon_{\bsl k}\, c^\dagger_{\bsl k s} c_{\bsl k s},
\end{align}
The interaction becomes a pair–pair scattering in momentum space,
\allowdisplaybreaks
\begin{align}
H_{\text{int}}
&= \frac{1}{2}\sum_{\bsl r,\bsl r'} V(\bsl r-\bsl r')\,
   c^{\dagger}_{\bsl r\uparrow}\, c^{\dagger}_{\bsl r'\downarrow}\,
   c_{\bsl r'\downarrow}\, c_{\bsl r\uparrow} \\
&= \frac{1}{2}\sum_{\bsl r,\bsl r'} V(\bsl r-\bsl r')\,
   \Big[\frac{1}{L}\sum_{\bsl k}   e^{-i\bsl k\cdot\bsl r} \, c^{\dagger}_{\bsl k\uparrow}\Big]
   \Big[\frac{1}{L}\sum_{\bsl p}   e^{-i\bsl p\cdot\bsl r'}\, c^{\dagger}_{\bsl p\downarrow}\Big]
   \Big[\frac{1}{L}\sum_{\bsl p'} e^{\,i\bsl p'\cdot\bsl r'} c_{\bsl p'\downarrow}\Big]
   \Big[\frac{1}{L}\sum_{\bsl k'} e^{\,i\bsl k'\cdot\bsl r}  c_{\bsl k'\uparrow}\Big] \\
&= \frac{1}{2L^{4}}\sum_{\bsl k,\bsl k',\bsl p,\bsl p'}
   \Big[\sum_{\bsl r,\bsl r'} V(\bsl r-\bsl r')\,
        e^{-i\bsl k\cdot\bsl r}\, e^{-i\bsl p\cdot\bsl r'}\,
        e^{\,i\bsl p'\cdot\bsl r'} e^{\,i\bsl k'\cdot\bsl r}\Big]\,
   c^{\dagger}_{\bsl k\uparrow} c^{\dagger}_{\bsl p\downarrow}
   c_{\bsl p'\downarrow} c_{\bsl k'\uparrow} \\
&= \frac{1}{2L^{4}}\sum_{\bsl k,\bsl k',\bsl p,\bsl p'}
   \Big[\sum_{\bsl r - \bsl r'} V(\bsl r - \bsl r')\sum_{\bsl r'}
        e^{-i\bsl k\cdot(\bsl r'+(\bsl r - \bsl r'))} e^{-i\bsl p\cdot\bsl r'}
        e^{\,i\bsl p'\cdot\bsl r'} e^{\,i\bsl k'\cdot(\bsl r'+(\bsl r - \bsl r')))}\Big]\,
   c^{\dagger}_{\bsl k\uparrow} c^{\dagger}_{\bsl p\downarrow}
   c_{\bsl p'\downarrow} c_{\bsl k'\uparrow} \\
&= \frac{1}{2L^{4}}\sum_{\bsl k,\bsl k',\bsl p,\bsl p'}
   \Big[\sum_{\bsl r - \bsl r'} V(\bsl r - \bsl r')\sum_{\bsl r'}
        e^{\,i(\bsl k'-\bsl k)\cdot(\bsl r - \bsl r'))} 
        e^{-i \bsl k \cdot \bsl r' + i \bsl k' \cdot \bsl r' - i \bsl p \cdot \bsl r' + i \bsl p' \cdot \bsl r'}\Big]\,
   c^{\dagger}_{\bsl k\uparrow} c^{\dagger}_{\bsl p\downarrow}
   c_{\bsl p'\downarrow} c_{\bsl k'\uparrow} \\
&= \frac{1}{2L^{2}}\sum_{\bsl k,\bsl k',\bsl p,\bsl p'}
   \Big[\sum_{\bsl r - \bsl r'} V(\bsl r - \bsl r')\, e^{\,i(\bsl k'-\bsl k)\cdot(\bsl r - \bsl r')}\Big]
   \delta_{\bsl k+\bsl p,\ \bsl k'+\bsl p'}
   c^{\dagger}_{\bsl k\uparrow} c^{\dagger}_{\bsl p\downarrow}
   c_{\bsl p'\downarrow} c_{\bsl k'\uparrow}
\label{eq:H_int_lastline}
\end{align}
Define
\begin{equation}
V(\bsl q)\equiv\sum_{\bsl r - \bsl r'} V(\bsl r - \bsl r')e^{i\bsl q\cdot(\bsl r - \bsl r')}, \qquad \bsl q\equiv \bsl k'-\bsl k
\end{equation}
and use  $\bsl{p}'=\bsl{k}+\bsl{p}-\bsl{k}'$ from the Kronecker delta to obtain,
\begin{equation}
H_{\text{int}}
= \frac{1}{2L^{2}}\sum_{\bsl k,\bsl p,\bsl q}
  V(\bsl q)
  c^{\dagger}_{\bsl k\uparrow} c^{\dagger}_{\bsl p\downarrow}
  c_{\bsl p-\bsl q,\downarrow} c_{\bsl k+\bsl q,\uparrow}.
\label{eq:H_int_Vq}
\end{equation}
To obtain the Richardson model from \cref{eq:realspaceBCS}, we select the subspace of
time-reversed pairs with the following constraints:
\begin{equation}
\bsl k+\bsl p=\bsl 0,\qquad \bsl k'+\bsl p'=\bsl 0
\quad\Longleftrightarrow\quad
\bsl p=-\bsl k,\ \ \bsl p'=-\bsl k'.
\end{equation}
where $\bsl k$ and $\bsl p$ are the crystal momenta of the two electrons that form a pair, we set their total momentum to be zero:
\begin{equation}
H_{\text{int}}^{\text{red}}
= \frac{1}{2L^{2}}\sum_{\bsl k,\bsl k'} V(\bsl k'-\bsl k)
  c^{\dagger}_{\bsl k\uparrow} c^{\dagger}_{-\bsl k\,\downarrow}\,
  c_{-\bsl k'\,\downarrow} c_{\bsl k'\uparrow}
= \frac{1}{2L^{2}}\sum_{\bsl k,\bsl k'} V(\bsl k'-\bsl k) A_{\bsl k}^{\dagger}A_{\bsl k'},
\end{equation}
with
\begin{equation}
    A^{\dagger}_{\bsl k} \equiv c^{\dagger}_{\bsl k\uparrow}c^{\dagger}_{-\bsl k\downarrow}, \text{ and } A_{\bsl k} \equiv c_{-\bsl k\downarrow}c_{\bsl k\uparrow}.
\label{eq:AAdagger}
\end{equation}
Therefore, the Richardson model is
\begin{equation}
H = \sum_{\bsl k,s}\varepsilon_{\bsl k} c^\dagger_{\bsl k s} c_{\bsl k s}
+ g \Big(\sum_{\bsl k} A_{\bsl k}^\dagger\Big)\Big(\sum_{\bsl k'} A_{\bsl k'}\Big),
\label{eq:RichardsonH}
\end{equation}
where $u = \frac{1}{2}V(\bsl k' - \bsl k)$ and $g=u/L^2$ with $g<0$ denoting attractive pairing, and we choose $\epsilon_{-\bsl{k}} = \epsilon_{\bsl{k}}$. Define $N_{\bsl k} \equiv n_{\bsl k\uparrow}+n_{-\bsl k\downarrow} = c^\dagger_{\bsl k\uparrow}c_{\bsl k\uparrow}+ c^\dagger_{-\bsl k\downarrow}c_{-\bsl k\downarrow}$. The quasi-spin algebra reads
\begin{equation}
[A_{\bsl k}, A_{\bsl p}^\dagger] = \delta_{\bsl k \bsl p}\,(1-N_{\bsl k}),\quad
[N_{\bsl k}, A_{\bsl p}^\dagger] = 2\delta_{\bsl k \bsl p}\,A_{\bsl k}^\dagger,\quad
[A_{\bsl k}^\dagger, A_{\bsl p}^\dagger]=0.
\label{eq:commutators}
\end{equation}
In the seniority-zero (all particles paired) sector with \(M\) pairs, eigenstates take the factorized form \cite{RevModPhys.76.643},
\begin{equation}
\ket{\Psi_{2M}} = \prod_{\alpha=1}^{M} B_\alpha^\dagger \ket{0},
\qquad
B_\alpha^\dagger = \sum_{\bsl k} \frac{A_{\bsl k}^\dagger}{2\varepsilon_{\bsl k} - E_\alpha},
\label{eq:ansatz}
\end{equation}
where the unknown ``rapidities'' \(\{E_\alpha\}_{\alpha=1}^M\) are determined by the eigenequation. Demanding \(H\ket{\Psi_{2M}} = E\ket{\Psi_{2M}}\) and using \cref{eq:commutators} yields the set of coupled equations
\begin{equation}
\frac{1}{g}
+\sum_ {\bsl k} \frac{1}{\,2\varepsilon_ {\bsl k} - E_\alpha\,}
-2\sum_{\beta(\neq \alpha)} \frac{1}{E_\beta - E_\alpha} = 0,
\quad \alpha=1,\dots,M 
\label{eq:Richardson}
\end{equation}
and the total many-body energy
\begin{equation}
 E = \sum_{\alpha=1}^{M} E_\alpha .
\label{eq:Energy}
\end{equation}
Here, we will derive \cref{eq:Richardson} explicitly. Use $\{c_{i},c^\dagger_{j}\}=\delta_{ij}$ and $\{c_{i},c_{j}\}=\{c^\dagger_i,c^\dagger_j\}=0$, we can show~\cref{eq:commutators}:
\begin{align}
[A_{\bsl k}, A_{\bsl k}^\dagger]
&= A_{\bsl k} A_{\bsl k}^\dagger - A_{\bsl k}^\dagger A_{\bsl k} \nonumber = c_{-{\bsl k}\downarrow} c_{{\bsl k}\uparrow} c_{{\bsl k}\uparrow}^\dagger c_{-{\bsl k}\downarrow}^\dagger
   - c_{{\bsl k}\uparrow}^\dagger c_{-{\bsl k}\downarrow}^\dagger c_{-{\bsl k}\downarrow} c_{{\bsl k}\uparrow} \nonumber = c_{-{\bsl k}\downarrow} (1 - c_{{\bsl k}\uparrow}^\dagger c_{{\bsl k}\uparrow}) c_{-{\bsl k}\downarrow}^\dagger
   - c_{{\bsl k}\uparrow}^\dagger (c_{-{\bsl k}\downarrow}^\dagger c_{-{\bsl k}\downarrow}) c_{{\bsl k}\uparrow} \nonumber\\
&= c_{-{\bsl k}\downarrow} c_{-{\bsl k}\downarrow}^\dagger
   - c_{-{\bsl k}\downarrow} c_{{\bsl k}\uparrow}^\dagger c_{{\bsl k}\uparrow} c_{-{\bsl k}\downarrow}^\dagger
   - c_{{\bsl k}\uparrow}^\dagger c_{-{\bsl k}\downarrow}^\dagger c_{-{\bsl k}\downarrow} c_{{\bsl k}\uparrow} \nonumber = (1 - c_{-{\bsl k}\downarrow}^\dagger c_{-{\bsl k}\downarrow})
   - c_{{\bsl k}\uparrow}^\dagger c_{{\bsl k}\uparrow}(1 - c_{-{\bsl k}\downarrow}^\dagger c_{-{\bsl k}\downarrow})
   - c_{{\bsl k}\uparrow}^\dagger c_{-{\bsl k}\downarrow}^\dagger c_{-{\bsl k}\downarrow} c_{{\bsl k}\uparrow} \nonumber\\
&= 1 - n_{-{\bsl k}\downarrow} - n_{{\bsl k}\uparrow}
   + n_{{\bsl k}\uparrow} n_{-{\bsl k}\downarrow} - n_{{\bsl k}\uparrow} n_{-{\bsl k}\downarrow} = 1 - n_{{\bsl k}\uparrow} - n_{-{\bsl k}\downarrow} = 1 - N_{\bsl k};
\end{align}

\begin{equation}
[A_{\bsl k}, A_{\bsl q}^\dagger] = c_{-{\bsl k}\downarrow} c_{{\bsl k}\uparrow} c_{\bsl q\uparrow}^\dagger c_{-\bsl q\downarrow}^\dagger
   - c_{\bsl q\uparrow}^\dagger c_{-\bsl q\downarrow}^\dagger c_{-\bsl k\downarrow} c_{\bsl k\uparrow} = c_{-\bsl k\downarrow} c_{\bsl k\uparrow} c_{\bsl q\uparrow}^\dagger c_{-\bsl q\downarrow}^\dagger
   - c_{-\bsl k\downarrow} c_{\bsl k\uparrow} c_{\bsl q\uparrow}^\dagger c_{-\bsl q\downarrow}^\dagger = 0\quad (\bsl k\neq \bsl q);
\end{equation}

\begin{align}
[N_{\bsl k}, A_{\bsl p}^\dagger]
&= [n_{{\bsl k}\uparrow} + n_{-{\bsl k}\downarrow},\, c_{{\bsl p}\uparrow}^\dagger c_{-{\bsl p}\downarrow}^\dagger] \nonumber = [n_{{\bsl k}\uparrow}, c_{{\bsl p}\uparrow}^\dagger c_{-{\bsl p}\downarrow}^\dagger]
   + [n_{-{\bsl k}\downarrow}, c_{{\bsl p}\uparrow}^\dagger c_{-{\bsl p}\downarrow}^\dagger] \nonumber\\
&= [n_{{\bsl k}\uparrow}, c_{{\bsl p}\uparrow}^\dagger]\,c_{-{\bsl p}\downarrow}^\dagger
   + c_{{\bsl p}\uparrow}^\dagger [n_{{\bsl k}\uparrow}, c_{-{\bsl p}\downarrow}^\dagger]
   + [n_{-{\bsl k}\downarrow}, c_{{\bsl p}\uparrow}^\dagger]\,c_{-{\bsl p}\downarrow}^\dagger
   + c_{{\bsl p}\uparrow}^\dagger [n_{-{\bsl k}\downarrow}, c_{-{\bsl p}\downarrow}^\dagger] \nonumber\\
&= [n_{{\bsl k}\uparrow}, c_{{\bsl p}\uparrow}^\dagger]\,c_{-{\bsl p}\downarrow}^\dagger
   + c_{{\bsl p}\uparrow}^\dagger [n_{-{\bsl k}\downarrow}, c_{-{\bsl p}\downarrow}^\dagger] 
   \quad\text{(two terms vanish)} \nonumber\\
&= \delta_{{\bsl k}{\bsl p}}\,c_{{\bsl p}\uparrow}^\dagger c_{-{\bsl p}\downarrow}^\dagger
   + \delta_{-{\bsl k},-{\bsl p}}\,c_{{\bsl p}\uparrow}^\dagger c_{-{\bsl p}\downarrow}^\dagger
   \quad\text{(since } [n_i, c_j^\dagger]=\delta_{ij}c_j^\dagger\text{)} \nonumber\\
&= (\delta_{{\bsl k}{\bsl p}} + \delta_{{\bsl k}{\bsl p}}) c_{{\bsl p}\uparrow}^\dagger c_{-{\bsl p}\downarrow}^\dagger  = 2\,\delta_{{\bsl k}{\bsl p}} A_{\bsl p}^\dagger
 = 2\delta_{{\bsl k}{\bsl p}} A_k^\dagger;
\end{align}
\eq{
[A_{\bsl k}^\dagger,A_{\bsl p}^\dagger]
=c^\dagger_{\bsl k\uparrow}c^\dagger_{-\bsl k\downarrow}c^\dagger_{\bsl p\uparrow}c^\dagger_{-\bsl p\downarrow}
-c^\dagger_{\bsl p\uparrow}c^\dagger_{-\bsl p\downarrow}c^\dagger_{\bsl k\uparrow}c^\dagger_{-\bsl k\downarrow}=0.}
We introduce the shorthand for \cref{eq:ansatz}
\begin{equation}
\phi_{\alpha}(\bsl k) \equiv \frac{1}{2\varepsilon_{\bsl k} - E_\alpha}.
\label{eq:phi}
\end{equation}
Thus $B_\alpha^\dagger=\sum_{\bsl k} \phi_\alpha(\bsl k)\,A_{\bsl k}^\dagger$. For later convenience define the collective pair operators
\begin{equation}
A^\dagger \equiv \sum_{\bsl k} A_{\bsl k}^\dagger, 
\qquad 
K_\alpha \equiv \sum_{\bsl k} \phi_\alpha(\bsl k)\,\big(1-N_{\bsl k}\big),
\label{eq:AandK}
\end{equation}
i.e. $K_\alpha$ is obtained from $B_\alpha^\dagger$ expression by replacing $A_{\bsl k}^\dagger$ with $(1-N_{\bsl k})$.
We split the Hamiltonian as
\begin{equation}
H = H_0 + V, 
\qquad 
H_0=\sum_{\bsl k,s}\varepsilon_{\bsl k}c^\dagger_{\bsl k s}c_{\bsl k s},
\qquad 
V=g \sum_{\bsl k,\bsl k'} A_{\bsl k}^\dagger A_{\bsl k'} = gA^\dagger A .
\label{eq:split}
\end{equation}
Evaluate the following commutators using~\cref{eq:commutators}:
\begin{align}
[H_0, A_{\bsl p}^\dagger] 
&= \left[ \sum_{\bsl k} \varepsilon_{\bsl k} N_{\bsl k}, A_{\bsl p}^\dagger \right] = \sum_{\bsl k} \varepsilon_{\bsl k} [N_{\bsl k}, A_{\bsl p}^\dagger] = \sum_{\bsl k} \varepsilon_{\bsl k} 2 \delta_{{\bsl k}{\bsl p}} A_{\bsl k}^\dagger = 2 \varepsilon_{\bsl p} A_{\bsl p}^\dagger.
\end{align}

\begin{equation}
[V,A_{\bsl p}^\dagger] = g\sum_{\bsl k,\bsl k'}A_{\bsl k}^\dagger[A_{\bsl k'},A_{\bsl p}^\dagger]
= g\sum_{\bsl k} A_{\bsl k}^\dagger(1-N_{\bsl p})
= g\,A^\dagger(1-N_{\bsl p})
\end{equation}
Combine them:
\begin{equation}
[H,A_{\bsl p}^\dagger]=2\varepsilon_{\bsl p} A_{\bsl p}^\dagger + g\,A^\dagger(1-N_{\bsl p}).
\end{equation}
Using~\cref{eq:ansatz},
\begin{equation}
\begin{aligned}
[H,B_\alpha^\dagger]
&=\sum_{\bsl p} \phi_\alpha(\bsl p)\,[H,A_{\bsl p}^\dagger]=\sum_{\bsl p} \phi_\alpha(\bsl p)\,2\varepsilon_{\bsl p} A_{\bsl p}^\dagger + g\,A^\dagger\sum_{\bsl p} \phi_\alpha(\bsl p)(1-N_{\bsl p})
\end{aligned}
\end{equation}
We have the identity
\eq{\frac{2\varepsilon_{\bsl p}}{2\varepsilon_{\bsl p}-E_\alpha}
=\frac{(2\varepsilon_{\bsl p}-E_\alpha)+E_\alpha}{2\varepsilon_{\bsl p}-E_\alpha}
=1+\frac{E_\alpha}{2\varepsilon_{\bsl p}-E_\alpha}}
Hence the first sum becomes
\begin{equation}
\begin{aligned}
\sum_{\bsl p} \phi_\alpha(\bsl p)\,2\varepsilon_{\bsl p}\,A_{\bsl p}^\dagger
&= \sum_{\bsl p} \left[ 1 + \frac{E_\alpha}{2\varepsilon_{\bsl p}-E_\alpha} \right] A_{\bsl p}^\dagger
= A^\dagger + E_\alpha \sum_{\bsl p} \frac{A_{\bsl p}^\dagger}{2\varepsilon_{\bsl p}-E_\alpha} = A^\dagger + E_\alpha B_\alpha^\dagger .
\end{aligned}
\end{equation}
Therefore,
\begin{equation}
[H,B_\alpha^\dagger]
= E_\alpha B_\alpha^\dagger + A^\dagger + g\,A^\dagger K_\alpha,
\qquad
K_\alpha \equiv \sum_{\bsl p} \phi_\alpha(\bsl p)\,(1-N_{\bsl p}).
\label{eq:HB-final}
\end{equation}
We will need how $K_\alpha$ commutes with $B_\beta^\dagger$. Using \cref{eq:AandK} with \cref{eq:phi},
\begin{align}
[K_\alpha, B_\beta^\dagger]
&= \left[ \sum_{\bsl k} \phi_\alpha(\bsl k)(1-N_{\bsl k}), \sum_{\bsl p} \phi_\beta(\bsl p) A_{\bsl p}^\dagger \right] = \sum_{\bsl k,\bsl p} \phi_\alpha(\bsl k) \phi_\beta(\bsl p) [1-N_{\bsl k}, A_{\bsl p}^\dagger] = \sum_{\bsl k,\bsl p} \phi_\alpha(\bsl k) \phi_\beta(\bsl p) (-2 \delta_{\bsl k \bsl p} A_{\bsl k}^\dagger) \\
&= -2 \sum_{\bsl k} \phi_\alpha(\bsl k)\phi_\beta(\bsl k) A_{\bsl k}^\dagger = -2 \sum_{\bsl k} \frac{1}{(2\varepsilon_{\bsl k} - E_\alpha)(2\varepsilon_{\bsl k} - E_\beta)} A_{\bsl k}^\dagger .
\label{eq:K_Bcom}
\end{align}
Using the identity $\dfrac{1}{x-a}-\dfrac{1}{x-b}=\dfrac{a-b}{(x-a)(x-b)}$
with $x\equiv 2\varepsilon_{\bsl k}$, $a\equiv E_\alpha$, $b\equiv E_\beta$, we obtain
\eq{\frac{1}{(2\varepsilon_{\bsl k}-E_\alpha)(2\varepsilon_{\bsl k}-E_\beta)}
=\frac{1}{E_\alpha-E_\beta}\left[\frac{1}{2\varepsilon_{\bsl k}-E_\alpha}
-\frac{1}{2\varepsilon_{\bsl k}-E_\beta}\right].}
Multiplying by $A_{\bsl k}^\dagger$ and summing over $k$ gives
\begin{equation}
\begin{aligned}
\sum_{\bsl k} \frac{A_{\bsl k}^\dagger}{(2\varepsilon_{\bsl k}-E_\alpha)(2\varepsilon_{\bsl k}-E_\beta)}
&=\frac{1}{E_\alpha-E_\beta}
\sum_{\bsl k}\left[\frac{A_{\bsl k}^\dagger}{2\varepsilon_{\bsl k}-E_\alpha}
-\frac{A_{\bsl k}^\dagger}{2\varepsilon_{\bsl k}-E_\beta}\right] =\frac{1}{E_\alpha-E_\beta}\Big(B_\alpha^\dagger-B_\beta^\dagger\Big)
=\frac{B_\alpha^\dagger-B_\beta^\dagger}{E_\alpha-E_\beta}.
\end{aligned}
\end{equation}
Insert it into \cref{eq:K_Bcom}:
\begin{equation}
[\,K_\alpha,B_\beta^\dagger\,]
=-2\,\frac{B_\alpha^\dagger-B_\beta^\dagger}{E_\alpha-E_\beta}.
\label{eq:KB_commutators}
\end{equation}
Equivalently, $K_\alpha B_\beta^\dagger
= B_\beta^\dagger K_\alpha
-2\,\frac{B_\alpha^\dagger-B_\beta^\dagger}{E_\alpha-E_\beta}$. Now act with $H$ on the ansatz state $|\Psi_{2M}\rangle=\prod_{\lambda=1}^{M}B_\lambda^\dagger|0\rangle$.
Since $H|0\rangle=0$, the commutator trick gives
\begin{equation}
\begin{aligned}
H\big|\Psi_{2M}\big\rangle
&=H\left(\prod_{\alpha=1}^{M} B_\alpha^\dagger\right)|0\rangle
= \Big([H,\prod_{\alpha=1}^{M} B_\alpha^\dagger]+\prod_{\alpha=1}^{M} B_\alpha^\dagger H\Big)|0\rangle \\
&=\,[H,\prod_{\alpha=1}^{M} B_\alpha^\dagger]\,|0\rangle 
\qquad(\text{because }H|0\rangle=0) \\
&=\Big(\sum_{\alpha=1}^{M} 
\Big(\prod_{\gamma<\alpha} B_\gamma^\dagger\Big)[H,B_\alpha^\dagger]
\Big(\prod_{\gamma>\alpha} B_\gamma^\dagger\Big)\Big)|0\rangle
\quad\big(\text{using } [H,XY]=[H,X]Y+X[H,Y]\big).
\end{aligned}
\end{equation}
Insert \cref{eq:HB-final},
\begin{equation}
\begin{aligned}
H \lvert \Psi_{2M} \rangle
&= \sum_{\alpha} \left( \prod_{\gamma < \alpha} B_\gamma^\dagger \right)
\left( E_\alpha B_\alpha^\dagger + A^\dagger + g A^\dagger K_\alpha \right)
\left( \prod_{\gamma > \alpha} B_\gamma^\dagger \right) \lvert 0 \rangle\\
&= \sum_{\alpha} E_\alpha \left( \prod_{\gamma} B_\gamma^\dagger \right) \lvert 0 \rangle
+ A^\dagger \sum_{\alpha} \left( \prod_{\gamma \neq \alpha} B_\gamma^\dagger \right) \lvert 0 \rangle
+ g A^\dagger \sum_{\alpha} \left( \prod_{\gamma < \alpha} B_\gamma^\dagger \right)
K_\alpha \left( \prod_{\gamma > \alpha} B_\gamma^\dagger \right) \lvert 0 \rangle\\
&=\sum_{\alpha} E_{\alpha} \lvert \Psi_{2M} \rangle
+ A^{\dagger} \sum_{\alpha} \left( \prod_{\gamma \neq \alpha} B_{\gamma}^{\dagger} \right) \lvert 0 \rangle  \\
&\quad + g A^{\dagger} \sum_{\alpha} \left( \prod_{\gamma < \alpha} B_{\gamma}^{\dagger} \right)
\Bigg[
\left( \prod_{\gamma > \alpha} B_{\gamma}^{\dagger} \right) K_{\alpha}
+ \sum_{\beta > \alpha} \left( \prod_{\alpha < \gamma < \beta} B_{\gamma}^{\dagger} \right)
\big[ K_{\alpha}, B_{\beta}^{\dagger} \big]
\left( \prod_{\gamma > \beta} B_{\gamma}^{\dagger} \right)
\Bigg] \lvert 0 \rangle\\
&= \sum_{\alpha} E_{\alpha} \lvert \Psi_{2M} \rangle
+ A^{\dagger} \sum_{\alpha} \left( \prod_{\gamma \neq \alpha} B_{\gamma}^{\dagger} \right) \lvert 0 \rangle
+ g A^{\dagger} \sum_{\alpha} \left( \prod_{\gamma \neq \alpha} B_{\gamma}^{\dagger} \right) K_{\alpha} \lvert 0 \rangle
+ g A^{\dagger} \sum_{\alpha} \sum_{\beta > \alpha}
\left( \prod_{\gamma \neq \alpha, \beta} B_{\gamma}^{\dagger} \right)
[ K_{\alpha}, B_{\beta}^{\dagger} ] \lvert 0 \rangle
\end{aligned}
\end{equation}
\eqa{
H\ket{\Psi_{2M}}
&=\Bigl(\sum_\alpha E_\alpha\Bigr)\ket{\Psi_{2M}}
+A^\dagger\sum_\alpha 
\Bigl[1+g\sum_{\bsl k}\phi_\alpha(\bsl k)\Bigr]
\prod_{\gamma\neq\alpha} B_\gamma^\dagger \ket{0} -g\,A^\dagger\sum_{\alpha}\sum_{\beta(\neq\alpha)}
\frac{B_\alpha^\dagger-B_\beta^\dagger}{E_\alpha-E_\beta}
\prod_{\substack{\gamma\neq\alpha,\beta}} B_\gamma^\dagger\ket{0}.
}
Then, 
\eqa{
& \sum_{\alpha}\sum_{\beta(\neq\alpha)}
\frac{B_\alpha^\dagger-B_\beta^\dagger}{E_\alpha-E_\beta}
\prod_{\substack{\gamma\neq\alpha\\\gamma\neq\beta}} B_\gamma^\dagger  = \sum_{\alpha}\sum_{\beta(\neq\alpha)}
\frac{1}{E_\alpha-E_\beta}
\prod_{\substack{\gamma\neq\beta}} B_\gamma^\dagger -  \sum_{\alpha}\sum_{\beta(\neq\alpha)}
\frac{1}{E_\alpha-E_\beta}
\prod_{\substack{\gamma\neq\alpha}} B_\gamma^\dagger \\
&  = \sum_{\alpha}\sum_{\beta(\neq\alpha)}
\frac{1}{E_\beta-E_\alpha}
\prod_{\substack{\gamma\neq\alpha}} B_\gamma^\dagger -  \sum_{\alpha}\sum_{\beta(\neq\alpha)}
\frac{1}{E_\alpha-E_\beta}
\prod_{\substack{\gamma\neq\alpha}} B_\gamma^\dagger \quad (\text{swapped dummy indices in the first term})\\
& = 2 \sum_{\alpha}\sum_{\beta(\neq\alpha)}
\frac{1}{E_\beta-E_\alpha}
\prod_{\substack{\gamma\neq\alpha}} B_\gamma^\dagger \ ,
}
which leads to
\eqa{
H\ket{\Psi_{2M}}
&=\Bigl(\sum_\alpha E_\alpha\Bigr)\ket{\Psi_{2M}}
+A^\dagger\sum_\alpha 
\Bigl[\,1+g\sum_{\bsl k}\phi_\alpha(\bsl k)\Bigr]
\prod_{\gamma\neq\alpha} B_\gamma^\dagger \ket{0} -2 g\,A^\dagger\sum_{\alpha}\sum_{\beta(\neq\alpha)}
\frac{1}{E_\beta-E_\alpha}
\prod_{\substack{\gamma\neq\alpha}} B_\gamma^\dagger \ket{0} \\
& = \Bigl(\sum_\alpha E_\alpha\Bigr)\ket{\Psi_{2M}}
+A^\dagger\sum_\alpha 
\Bigl[\,1+g\sum_{\bsl k}\phi_\alpha(\bsl k)-2 g\sum_{\beta(\neq\alpha)}
\frac{1}{E_\beta-E_\alpha}\Bigr]
\prod_{\gamma\neq\alpha} B_\gamma^\dagger \ket{0}\ , 
}
which gives the following equation:

\begin{equation}
\frac{1}{g}
+\sum_{\bsl k}\frac{1}{2\varepsilon_{\bsl k}-E_\mu}
-\sum_{\nu\neq\mu}\frac{2}{E_\nu-E_\mu}=0,
\qquad \mu=1,\dots,M.\
\label{eq:Richardson_eqn}
\end{equation}
as claimed~\cite{RevModPhys.76.643}.

After obtaining the Richardson equations, it is convenient to introduce the Jacobian matrix with respect to the rapidities:
\begin{equation}
G_{\mu\nu}=\frac{\partial R_\mu}{\partial E_\nu},
\label{eq:Gaudin_def_mu}
\end{equation}
where $R_\mu(\{E\})
\equiv\frac{1}{g}
+\sum_{k}\frac{1}{2\varepsilon_k-E_\mu}
-\sum_{\substack{ \nu\neq\mu}}^{M}\frac{2}{E_\nu-E_\mu}
=0.$ For off-diagonal elements ($\nu\neq\mu$),
only the third term depends on $E_\nu$, and only through the $\nu$th term:
\begin{equation}
\begin{aligned}
G_{\mu\nu}
&=\frac{\partial}{\partial E_\nu}\left(
-\frac{2}{E_\nu-E_\mu}
\right)
=\frac{2}{(E_\nu-E_\mu)^2},
\qquad \nu\neq\mu.
\end{aligned}
\label{eq:Gaudin_offdiag_mu}
\end{equation}
For diagonal elements ($\nu=\mu$), differentiating with respect to $E_\mu$ yields
\begin{align}
G_{\mu\mu}
&=\frac{\partial R_\mu}{\partial E_\mu}=\sum_{\bsl k}\frac{\partial}{\partial E_\mu}(2\varepsilon_{\bsl k}-E_\mu)^{-1}
-\sum_{\substack{\nu\neq \mu}}2\frac{\partial}{\partial E_\mu}(E_\nu-E_\mu)^{-1}\\
&=\sum_{\bsl k}\left[-(2\varepsilon_{\bsl k}-E_\mu)^{-2}\cdot(-1)\right]
-2\sum_{\substack{\nu\neq \mu}}^{M}\left[-(E_\nu-E_\mu)^{-2}\cdot(-1)\right] =\sum_{\bsl k}\frac{1}{(2\varepsilon_{\bsl k}-E_\mu)^2}
-2\sum_{\substack{\nu\neq \mu}}^{M}\frac{1}{(E_\nu-E_\mu)^2}.
\end{align}
Together,
\begin{equation}
G_{\mu\nu}=
\begin{cases}
\displaystyle \sum_{\bsl k}\frac{1}{(2\varepsilon_{\bsl k}-E_\mu)^2}
-2\sum_{\substack{\lambda\neq\mu}}\frac{1}{(E_\lambda-E_\mu)^2}, & \nu=\mu,\\
\displaystyle \frac{2}{(E_\nu-E_\mu)^2}, & \nu\neq\mu.
\end{cases}
\label{eq:Gaudin_summary_mu}
\end{equation}
This is called Gaudin matrix \cite{RevModPhys.76.643}, which will be useful later in \cref{app:correlator_derivation} when expressing correlation functions.

\subsubsection{Pair–Pair Correlators: Exact Formula}\label{app:correlator_derivation}

With the Cooper-pair operator $A_k^\dagger$ and $A_k$ defined in \cref{eq:AAdagger}, for a general many‑body state, the pair–pair correlator is
\begin{equation}
\widetilde C_{\bsl k \bsl k'} \equiv \big\langle A_{\bsl k}^\dagger A_{\bsl k'} \big\rangle
- \big\langle A_{\bsl k}^\dagger \big\rangle \big\langle A_{\bsl k'} \big\rangle \ ,
\label{eq:pairpair_connected}
\end{equation}
where $\big\langle ... \big\rangle$ is the average with respect to the state of interest.
In the Richardson model we work with number‑conserving eigenstates
$\ket{\Psi_{2M}}$ (fixed particle number $2M$). Since
$A_{\bsl k}$ lowers particle number by two, $A_{\bsl k}\ket{\Psi_{2M}}$ lies in the
$2M-2$ sector and is orthogonal to $\ket{\Psi_{2M}}$, hence
\begin{equation}
\big\langle A_{\bsl k} \big\rangle = \bra{\Psi_{2M}} A_{\bsl k} \ket{\Psi_{2M}} = 0.
\label{eq:Ak_zero}
\end{equation}
Therefore the connected and full correlators coincide,
\begin{equation}
\widetilde C_{\bsl k \bsl k'} = \big\langle A_{\bsl k}^\dagger A_{\bsl k'} \big\rangle .
\label{eq:pairpair_equals_full}
\end{equation}
Writing out $A_k^\dagger$ and $A_{k'}$,
\begin{equation}
\big\langle A_{\bsl k}^\dagger A_{\bsl k'} \big\rangle
= \big\langle c^\dagger_{\bsl k\uparrow}c^\dagger_{-\bsl k\downarrow}
c_{-\bsl k'\downarrow}c_{\bsl k'\uparrow} \big\rangle,
\label{eq:pairpair_cccc}
\end{equation}
which is exactly the Cooper‑channel combination of the 2-RDM
defined in \cref{eq:nRDM}, i.e.
\begin{equation}
C_{\bsl k \bsl k'} \equiv \widetilde C_{\bsl k \bsl k'}
= O_{\bsl k\uparrow,-\bsl k\downarrow,\bsl k'\uparrow,-\bsl k'\downarrow}.
\label{eq:pairpair_is_2rdm}
\end{equation}
(For U(1)–breaking mean‑field states we may have $\langle A_{\bsl k}\rangle \neq 0$,
in which case the connected correlator \eqref{eq:pairpair_connected} differs
from the full $2$‑RDM matrix element by
$\langle A_{\bsl k}^\dagger\rangle\langle A_{\bsl k'}\rangle$; in our number‑conserving
case, \eqref{eq:Ak_zero} ensures they are identical.)

Now we derive a general expression of 
$\bra{\Psi_{2M}} A_{\bsl k}^\dagger A_{\bsl k'} \ket{\Psi_{2M}} $.
Let's start from
\eq{A_{\bsl k'} \lvert \Psi_{2M} \rangle
=
A_{\bsl k'} \left( \prod_{\epsilon=1}^{M} B_\epsilon^\dagger \right) \lvert 0 \rangle .}
Because \( A_{\bsl k'} \lvert 0 \rangle = 0 \), we can commute \( A_{\bsl k'} \) through the product and keep only commutators:
\begin{align}
A_{\bsl k'} \left( \prod_{\epsilon=1}^{M} B_\epsilon^\dagger \right) \lvert 0 \rangle 
&=\prod_{\epsilon=1}^{M} B_\epsilon^\dagger A_{\bsl k'} \lvert 0 \rangle + \sum_{\mu=1}^{M}
\left( \prod_{\epsilon<\mu} B_\epsilon^\dagger \right)
\bigl[ A_{\bsl k'}, B_\mu^\dagger \bigr]
\left( \prod_{\epsilon>\mu} B_\epsilon^\dagger \right)
\lvert 0 \rangle\\
&=
\sum_{\mu=1}^{M}
\left( \prod_{\epsilon<\mu} B_\epsilon^\dagger \right)
\bigl[ A_{\bsl k'}, B_\mu^\dagger \bigr]
\left( \prod_{\epsilon>\mu} B_\epsilon^\dagger \right)
\lvert 0 \rangle .
\end{align}
We can compute the commutator using
\( B_\mu^\dagger = \sum_{\bsl p} \phi_\mu(\bsl p) A_{\bsl p}^\dagger \)
and
\( [ A_{\bsl k'}, A_{\bsl p}^\dagger ] = \delta_{\bsl k' \bsl p} (1 - N_{\bsl k'}) \):
\begin{equation}
\bigl[ A_{\bsl k'}, B_\mu^\dagger \bigr] =
\sum_{\bsl p} \phi_\mu(\bsl p) \bigl[ A_{\bsl k'}, A_{\bsl p}^\dagger \bigr] =
\sum_{\bsl p} \phi_\mu(\bsl p) \delta_{\bsl k' \bsl p} (1 - N_{\bsl k'}) =
\phi_\mu(\bsl k') (1 - N_{\bsl k'}) .
\end{equation}
Therefore,
\begin{align}
A_{\bsl k'} \lvert \Psi_{2M} \rangle
&=
\sum_{\mu=1}^{M}
\phi_\mu(\bsl k')
\left( \prod_{\epsilon<\mu} B_\epsilon^\dagger \right)
(1 - N_{\bsl k'})
\left( \prod_{\epsilon>\mu} B_\epsilon^\dagger \right)
\lvert 0 \rangle\\
&= \sum_{\mu=1}^{M}
\phi_\mu(\bsl k')
\left( \prod_{\epsilon<\mu} B_\epsilon^\dagger \right)
\Biggl[
\left( \prod_{\epsilon>\mu} B_\epsilon^\dagger \right)
(1 - N_{\bsl k'}) \lvert 0 \rangle
+
\sum_{\nu>\mu}
\left( \prod_{\mu<\epsilon<\nu} B_\epsilon^\dagger \right)
\bigl[ 1 - N_{\bsl k'}, B_\nu^\dagger \bigr]
\left( \prod_{\epsilon>\nu} B_\epsilon^\dagger \right)
\lvert 0 \rangle
\Biggr],
\end{align}
where $\bigl[ N_{\bsl k'}, B_\nu^\dagger \bigr]
=
\sum_{\bsl p} \phi_\nu(\bsl p) \bigl[ N_{\bsl k'}, A_{\bsl p}^\dagger \bigr]
=
2\phi_\nu(\bsl k') A_{\bsl k'}^\dagger,$ using \cref{eq:commutators}, so $
\bigl[ 1 - N_{\bsl k'}, B_\nu^\dagger \bigr]
=
-2\phi_\nu(\bsl k') A_{\bsl k'}^\dagger$. Thus,
\begin{align}
A_{\bsl k'} \lvert \Psi_{2M} \rangle
&= \sum_{\mu=1}^{M}
\phi_\mu(\bsl k')
\left( \prod_{\epsilon<\mu} B_\epsilon^\dagger \right)
\Biggl[
\left( \prod_{\epsilon>\mu} B_\epsilon^\dagger \right) \lvert 0 \rangle
-
\sum_{\nu>\mu}
\left( \prod_{\mu<\epsilon<\nu} B_\epsilon^\dagger \right)
2\,\phi_\nu(k')\, A_{k'}^\dagger
\left( \prod_{\epsilon>\nu} B_\epsilon^\dagger \right)
\lvert 0 \rangle
\Biggr]\\
&= \sum_{\mu=1}^{M}
\phi_\mu(\bsl k')
\left( \prod_{\epsilon \neq \mu} B_\epsilon^\dagger \right)
\lvert 0 \rangle
-
\sum_{\mu=1}^{M}
\phi_\mu(\bsl k')
\left( \prod_{\epsilon<\mu} B_\epsilon^\dagger \right)
\sum_{\nu>\mu}
\left( \prod_{\mu<\epsilon<\nu} B_\epsilon^\dagger \right)
2\phi_\nu(\bsl k') A_{\bsl k'}^\dagger
\left( \prod_{\epsilon>\nu} B_\epsilon^\dagger \right)
\lvert 0 \rangle\\
&= \sum_{\mu=1}^{M}
\phi_\mu(\bsl k')
\left( \prod_{\epsilon \neq \mu} B_\epsilon^\dagger \right)
\lvert 0 \rangle - 2A_{\bsl k'}^\dagger
\sum_{\mu=1}^{M}
\sum_{\nu>\mu}
\phi_\mu(\bsl k')\, \phi_\nu(\bsl k')
\left( \prod_{\epsilon<\mu} B_\epsilon^\dagger \right)
\left( \prod_{\mu<\epsilon<\nu} B_\epsilon^\dagger \right)
\left( \prod_{\epsilon>\nu} B_\epsilon^\dagger \right)
\lvert 0 \rangle\\
&= \sum_{\mu=1}^{M}
\phi_\mu(\bsl k')
\left( \prod_{\epsilon \neq \mu} B_\epsilon^\dagger \right)
\lvert 0 \rangle
-
2A_{\bsl k'}^\dagger
\sum_{1 \le \mu < \nu \le M}
\phi_\mu(\bsl k') \phi_\nu(\bsl k')
\left( \prod_{\epsilon \neq \mu,\nu} B_\epsilon^\dagger \right)
\lvert 0 \rangle.
\end{align}
Define the shorthand:
\begin{equation}
\lvert \Psi^{(\mu)} \rangle
\equiv
\left( \prod_{\epsilon \neq \mu} B_\epsilon^\dagger \right) \lvert 0 \rangle,
\quad
\lvert \Psi^{(\mu\nu)} \rangle
\equiv
\left( \prod_{\epsilon \neq \mu,\nu} B_\epsilon^\dagger \right) \lvert 0 \rangle,
\quad
\langle \Psi^{(\rho)} \rvert
\equiv
\langle 0 \rvert
\left( \prod_{\omega \neq \rho} B_\omega \right),
\quad
\langle \Psi^{(\rho\sigma)} \rvert
\equiv
\langle 0 \rvert
\left( \prod_{\omega \neq \rho,\sigma} B_\omega \right).
\end{equation}
Therefore,
\begin{align}
    \bra{\Psi_{2M}} A_{\bsl k}^\dagger A_{\bsl k'} \ket{\Psi_{2M}} 
    &= \langle \Psi_{2M} \lvert A_{\bsl k}^\dagger \Bigl[ \sum_{\mu} \phi_\mu(\bsl k') \lvert \Psi^{(\mu)} \rangle - 2 A_{\bsl k'}^\dagger \sum_{\mu<\nu} \phi_\mu(\bsl k') \phi_\nu(\bsl k') \lvert \Psi^{(\mu\nu)} \rangle \Bigr]\\
    &= \sum_{\mu} \phi_\mu(\bsl k') \langle \Psi_{2M} \lvert A_{\bsl k}^\dagger \rvert \Psi^{(\mu)} \rangle - 2 \sum_{\mu<\nu} \phi_\mu(\bsl k') \phi_\nu(\bsl k') \langle \Psi_{2M} \lvert A_{\bsl k}^\dagger A_{\bsl k'}^\dagger \rvert \Psi^{(\mu\nu)} \rangle\\
    &= \sum_{\mu} \phi_\mu(\bsl k') F_{\bsl k}^{(\mu)} - 2 \sum_{\mu<\nu} \phi_\mu(\bsl k') \phi_\nu(\bsl k') F_{\bsl k \bsl k'}^{(\mu\nu)}.
\label{eq:correlatorInFormfactors}
\end{align}
where we define the form factors
\begin{equation}
F_{\bsl k}^{(\mu)}\equiv
\langle \Psi_{2M} \lvert A_{\bsl k}^\dagger \rvert \Psi^{(\mu)} \rangle,
\qquad
F_{\bsl k \bsl k'}^{(\mu\nu)}\equiv
\langle \Psi_{2M} \lvert A_{\bsl k}^\dagger A_{\bsl k'}^\dagger \rvert \Psi^{(\mu\nu)} \rangle.
\label{eq:form_factors}
\end{equation}
Define the general pair-creation operators
\begin{equation}
B^\dagger(u)\equiv \sum_{\bsl p}\frac{A_{\bsl p}^\dagger}{d_{\bsl p}-u},
\quad u\in\mathbb{C}.
\label{eq:Bdu}
\end{equation}
where $d_{\bsl p}=2\varepsilon_{\bsl p}$. For any choice of parameters $\{u_\mu\}_{\mu=1}^M$, this generates the general $M$-pair state
\begin{equation}
|\{u\}\rangle\equiv\prod_{\mu=1}^{M} B^\dagger(u_\mu)\,|0\rangle.
\label{eq:trial_state_u}
\end{equation}
In particular, taking $u_\mu=E_\mu$ equal to the Richardson rapidities, we recover the Richardson state
$|\Psi_{2M}\rangle
=\prod_{\mu=1}^{M} B^\dagger(E_\mu)|0\rangle,
\quad
B^\dagger(E_\mu)=\sum_{\bsl p=1}\frac{A_{\bsl p}^\dagger}{d_{\bsl p}-E_\mu},$ which matches the amplitudes $\phi_\mu(\bsl p)=1/(d_{\bsl p}-E_\mu)$ we discussed. Now we can show the residue identity for $A_{\bsl k}^\dagger$. From \cref{eq:Bdu},
\begin{align}
(d_{\bsl k} - u) B^\dagger(u)
&= (d_{\bsl k} - u) \sum_{\bsl p} \frac{A_{\bsl p}^\dagger}{d_{\bsl p} - u} = \sum_{\bsl p} (d_{\bsl k} - u) \frac{A_{\bsl p}^\dagger}{d_{\bsl p} - u} \\
&= (d_{\bsl k} - u) \frac{A_{\bsl k}^\dagger}{d_{\bsl k} - u}
+ \sum_{\bsl p \neq \bsl k} (d_{\bsl k} - u) \frac{A_{\bsl p}^\dagger}{d_{\bsl p} - u} = A_{\bsl k}^\dagger
+ \sum_{\bsl p \neq \bsl k} \frac{d_{\bsl k} - u}{d_{\bsl p} - u} A_{\bsl p}^\dagger \\
\lim_{u \to d_{\bsl k}} (d_{\bsl k} - u) B^\dagger(u)
&= \lim_{u \to d_{\bsl k}}
\left[
A_{\bsl k}^\dagger
+ \sum_{\bsl p \neq \bsl k} \frac{d_{\bsl k} - u}{d_{\bsl p} - u} A_{\bsl p}^\dagger
\right] = A_{\bsl k}^\dagger
+ \sum_{\bsl p \neq \bsl k}
\left( \lim_{u \to d_{\bsl k}} \frac{d_{\bsl k} - u}{d_{\bsl p} - u} \right)
A_{\bsl p}^\dagger = A_{\bsl k}^\dagger
+ \sum_{\bsl p \neq \bsl k} 0 \cdot A_{\bsl p}^\dagger = A_{\bsl k}^\dagger .
\end{align}
Therefore, 
\eq{
A_{\bsl k}^\dagger = \lim_{u \to d_{\bsl k}} (d_{\bsl k} - u) B^\dagger(u).
}
We can re-write the form factors in \cref{eq:form_factors}: 
\begin{align}
F_{\bsl k}^{(\mu)}
&= \langle \Psi \lvert A_{\bsl k}^\dagger \rvert \Psi^{(\mu)} \rangle = \langle \{E\} \lvert A_{\bsl k}^\dagger \prod_{\nu \neq \mu} B^\dagger(E_\nu) \rvert 0 \rangle = \langle \{E\} | \lim_{u \to d_{\bsl k}} (d_{\bsl k} - u) B^\dagger(u)
\prod_{\nu \neq \mu} B^\dagger(E_\nu)| 0 \rangle
\\
&= \lim_{u \to d_{\bsl k}}
(d_{\bsl k} - u)
S\left( \{E\}, \{F^{(\mu)}(u)\} \right),
\label{eq:formfactor1}
\end{align}
where we defined $S(\{E\}, \{ F^{(\mu)}(u) \}) \equiv
\langle \{E\} \mid \{ F^{(\mu)}(u) \} \rangle$, and $ \lvert \{ F^{(\mu)}(u) \} \rangle
\equiv
B^\dagger(u)
\prod_{\nu \neq \mu} B^\dagger(E_\nu)
\lvert 0 \rangle $. Notice that, here we denote $\{F^{(\mu)}(u)\} = \{E_1,\ldots,E_{\mu-1},u,E_{\mu+1},\ldots,E_M\}$, the set obtained from $\{E\}$ by replacing only the $\mu$-th element with $u$. Similarly,
\begin{equation}
F_{\bsl k \bsl k'}^{(\mu\nu)}
=
\lim_{u \to d_{\bsl k}}
\lim_{v \to d_{\bsl k'}}
(d_{\bsl k} - u)(d_{\bsl k'} - v)\,
S\left( \{E\}, \{ F^{(\mu\nu)}(u,v) \} \right),
\label{eq:formfactor2}
\end{equation}
where we defined
$\lvert \{ F^{(\mu\nu)}(u,v) \} \rangle \equiv
B^\dagger(u) B^\dagger(v)
\prod_{\lambda \neq \mu,\nu} B^\dagger(E_\lambda)
\lvert 0 \rangle .$ 

Now all we need is to evaluate the form factors from \cref{eq:formfactor1} and \cref{eq:formfactor2}, so that we can plug into the correlator expression in \cref{eq:correlatorInFormfactors}. To do so, we use the standard Slavnov's formula \cite{Slavnov1989,Zhou_2002,Faribault_2008,Gorohovsky_2011} that gives an exact determinant expression for $S(\{E\},\{F\})$ when $\{E\}$ are the solution from Richardson equations (which we will show explicitly in \cref{app:Slavnov's theorem}); we can derive:
\begin{equation}
F_{\bsl k}^{(\mu)}
=
(d_{\bsl k} - E_\mu)
\det\bigl( G(\mu \rightarrow C^{(\bsl k)}) \bigr),
\label{eq:formfactorsimplified1}
\end{equation}
\begin{equation}
F_{\bsl k \bsl k'}^{(\mu\nu)}
=
\frac{
(d_{\bsl k} - E_\mu)(d_{\bsl k} - E_\nu)(d_{\bsl k'} - E_\mu)(d_{\bsl k'} - E_\nu)
}{
(d_{\bsl k} - d_{\bsl k'})(E_\nu - E_\mu)
}\,
\det\bigl( G(\mu,\nu \rightarrow C^{(\bsl k)}, C^{(\bsl k')}) \bigr), \quad (\bsl k \neq \bsl k')
\label{eq:formfactorsimplified2}
\end{equation}
where $G$ is the Gaudin matrix derived in \cref{eq:Gaudin_summary_mu}, and $G(\mu \rightarrow C^{(\bsl k)})$ here denotes the matrix obtained from $G$ by replacing its $\mu$-th column with the vector $C^{(\bsl k)}$ whose elements are $C_\alpha^{(\bsl k)} = \dfrac{1}{(d_{\bsl k} - E_\alpha)^2}$ , while $G(\mu,\nu \rightarrow C^{(\bsl k)}, C^{(\bsl k')})$ denotes the matrix obtained by replacing the $\mu$-th and $\nu$-th columns of $G$ with $C^{(\bsl k)}$ and $C^{(\bsl k')}$, respectively. Now plug \cref{eq:formfactorsimplified1} and \cref{eq:formfactorsimplified2} into \cref{eq:correlatorInFormfactors},
\begin{align}
    \bra{\Psi_{2M}} A_{\bsl k}^\dagger A_{\bsl k'} \ket{\Psi_{2M}}
    &= \sum_{\mu} \phi_\mu(\bsl k') F_k^{(\mu)} - 2 \sum_{\mu<\nu} \phi_\mu(\bsl k') \phi_\nu(\bsl k') F_{\bsl k \bsl k'}^{(\mu\nu)}\\
    &= \sum_{\mu} \phi_\mu(\bsl k') (d_{\bsl k} - E_\mu) \det\bigl( G(\mu \rightarrow C^{(\bsl k)}) \bigr) \\
    &\quad - 2 \sum_{\mu<\nu} \phi_\mu(\bsl k') \phi_\nu(\bsl k') \frac{ (d_{\bsl k} - E_\mu)(d_{\bsl k} - E_\nu)(d_{\bsl k'} - E_\mu)(d_{\bsl k'} - E_\nu) }{(d_{\bsl k} - d_{\bsl k'})(E_\nu - E_\mu)} \det\bigl( G(\mu,\nu \rightarrow C^{(\bsl k)}, C^{(\bsl k')}) \bigr)\\
    &= \sum_{\mu} \frac{d_{\bsl k} - E_\mu}{d_{\bsl k'} - E_\mu} \det\bigl( G(\mu \rightarrow C^{(\bsl k)}) \bigr) - 2 \sum_{\mu<\nu}\frac{(d_{\bsl k} - E_\mu)(d_{\bsl k} - E_\nu)}{(d_{\bsl k} - d_{\bsl k'})(E_\nu - E_\mu)} \det\bigl( G(\mu,\nu \rightarrow C^{(\bsl k)}, C^{(\bsl k')}) \bigr).
\label{eq:correlatorInFormfactors_new}
\end{align}
We normalize it by $\langle \Psi_{2M} \mid \Psi_{2M} \rangle = \det G$:
\begin{align}
\frac{\big\langle\Psi_{2M}\big|A_{\bsl k}^{\dagger}A_{\bsl k'}\big|\Psi_{2M}\big\rangle}{\langle\Psi_{2M}|\Psi_{2M}\rangle}
&= \sum_{\mu} \frac{d_{\bsl k} - E_\mu}{d_{\bsl k'} - E_\mu} 
\frac{\det\bigl( G(\mu \rightarrow C^{(\bsl k)}) \bigr) }{\det G}
- 2 \sum_{\mu<\nu}\frac{(d_{\bsl k} - E_\mu)(d_{\bsl k} - E_\nu)}{(d_{\bsl k} - d_{\bsl k'})(E_\nu - E_\mu)}\frac{\det\bigl( G(\mu,\nu \rightarrow C^{(\bsl k)}, C^{(\bsl k')}) \bigr)}{\det G}
\label{eq:CorrelatorInDeterminantRatio}
\end{align}
Now all we need is to simplify the terms $\frac{\det\bigl( G(\mu \rightarrow C^{(\bsl k)}) \bigr) }{\det G}$ and $\frac{\det\bigl( G(\mu,\nu \rightarrow C^{(\bsl k)}, C^{(\bsl k')}) \bigr)}{\det G}$, here is how we derive an expression for them: 

We differentiate Richardson equations:

\begin{align}
0
&=\frac{\partial R_\alpha}{\partial d_{\bsl k}}=\frac{\partial}{\partial d_{\bsl k}}\left[
\frac{1}{g}
+\sum_{i}\frac{1}{d_i-E_\alpha}
-2\sum_{\beta\neq\alpha}\frac{1}{E_\beta-E_\alpha}
\right]\\
&=\frac{\partial}{\partial d_{\bsl k}}\left(\frac{1}{d_{\bsl k}-E_\alpha}\right)
+\sum_{\bsl i\neq \bsl k}\frac{\partial}{\partial d_{\bsl k}}\left(\frac{1}{d_{\bsl i}-E_\alpha}\right)
-2\sum_{\beta\neq\alpha}\frac{\partial}{\partial d_{\bsl k}}\left(\frac{1}{E_\beta-E_\alpha}\right)
\\
&=\frac{-1}{(d_{\bsl k}-E_\alpha)^2}\frac{\partial}{\partial d_{\bsl k}}(d_{\bsl k}-E_\alpha)
+\sum_{\bsl i\neq \bsl k}\left[\frac{-1}{(d_{\bsl i}-E_\alpha)^2}\frac{\partial}{\partial d_{\bsl k}}(d_{\bsl i}-E_\alpha)\right]
-2\sum_{\beta\neq\alpha}\left[
\frac{\partial}{\partial E_\beta}\left(\frac{1}{E_\beta-E_\alpha}\right)\frac{\partial E_\beta}{\partial d_{\bsl k}}
+\frac{\partial}{\partial E_\alpha}\left(\frac{1}{E_\beta-E_\alpha}\right)\frac{\partial E_\alpha}{\partial d_{\bsl k}}
\right]
\\
&=\frac{-1}{(d_{\bsl k}-E_\alpha)^2}\Big(1-\frac{\partial E_\alpha}{\partial d_{\bsl k}}\Big)
+\sum_{\bsl i\neq \bsl k}\left[-\frac{1}{(d_{\bsl i}-E_\alpha)^2}\Big(0-\frac{\partial E_\alpha}{\partial d_{\bsl k}}\Big)\right]-2\sum_{\beta\neq\alpha}\left[
\Big(-\frac{1}{(E_\beta-E_\alpha)^2}\Big)\frac{\partial E_\beta}{\partial d_{\bsl k}}
+\Big(\frac{1}{(E_\beta-E_\alpha)^2}\Big)\frac{\partial E_\alpha}{\partial d_{\bsl k}}
\right]
\\
&=\frac{-1}{(d_{\bsl k}-E_\alpha)^2}
+\frac{1}{(d_{\bsl k}-E_\alpha)^2}\frac{\partial E_\alpha}{\partial d_{\bsl k}}
+\sum_{\bsl i\neq \bsl k}\frac{1}{(d_{\bsl i}-E_\alpha)^2}\frac{\partial E_\alpha}{\partial d_{\bsl k}} +2\sum_{\beta\neq\alpha}\frac{1}{(E_\beta-E_\alpha)^2}\frac{\partial E_\beta}{\partial d_{\bsl k}}
-2\sum_{\beta\neq\alpha}\frac{1}{(E_\beta-E_\alpha)^2}\frac{\partial E_\alpha}{\partial d_{\bsl k}}
\\
&=-\frac{1}{(d_{\bsl k}-E_\alpha)^2}
+\left[\sum_{\bsl i}\frac{1}{(d_{\bsl i}-E_\alpha)^2}-2\sum_{\beta\neq\alpha}\frac{1}{(E_\beta-E_\alpha)^2}\right]\frac{\partial E_\alpha}{\partial d_{\bsl k}}
+\sum_{\beta\neq\alpha}\frac{2}{(E_\beta-E_\alpha)^2}\frac{\partial E_\beta}{\partial d_{\bsl k}}
\end{align}
We can identify the Gaudin matrix from \cref{eq:Gaudin_summary_mu}, then 
\begin{align}
0
&= -\frac{1}{(d_{\bsl k}-E_\alpha)^2}
+G_{\alpha\alpha}x_\alpha^{(\bsl k)}
+\sum_{\beta\neq\alpha}G_{\alpha\beta}x_\beta^{(\bsl k)},
\qquad
x_\beta^{(\bsl k)}\equiv \frac{\partial E_\beta}{\partial d_{\bsl k}},
\\
&G_{\alpha\alpha}x_\alpha^{(\bsl k)}+\sum_{\beta\neq\alpha}G_{\alpha\beta}x_\beta^{(\bsl k)}
=\frac{1}{(d_{\bsl k}-E_\alpha)^2},
\\
&\sum_{\beta=1}^{M}G_{\alpha\beta}x_\beta^{(\bsl k)}
=\frac{1}{(d_{\bsl k}-E_\alpha)^2}.
\end{align}
Namely,$\quad G \hat{x} = C^{(k)},$ where $C^{(k)}\equiv\frac{1}{(d_{\bsl k}-E_\alpha)^2}$. 

According to Cramer's rule, if a matrix $A \in \mathbb{C}^{n \times n}$ is invertible and $Ax = b$, then for each $i$,
\begin{equation}
x_i=\frac{\det A^{(i)}}{\det A}
\qquad (i=1,\dots,n).
\label{eq:cramer_basic}
\end{equation}
where $A^{(i)}$ is obtained from $A$ by replacing its $i$-th column with $b$. Applying \cref{eq:cramer_basic} with $A=G$ and $b=C^{(\bsl k)}$ gives
\begin{equation}
x^{(k)}_\mu
=
\frac{\det\big(G(\mu\to C^{(\bsl k)})\big)}{\det G},
\label{eq:cramer_onecol}
\end{equation}
where $G(\mu\to C^{(\bsl k)})$ denotes the matrix obtained by replacing the $\mu$th column of $G$ by the vector $C^{(k)}$.
Since $x^{(k)}_\mu=\partial E_\mu/\partial d_k$, we conclude
\begin{equation}
\frac{\det\big(G(\mu\to C^{(\bsl k)})\big)}{\det G}
=
\frac{\partial E_\mu}{\partial d_{\bsl k}}.
\label{eq:det_ratio_equals_derivative}
\end{equation}
Similarly, consider the matrix $G(\mu,\nu\to C^{(k)},C^{(k')})$ obtained from $G$ by replacing the $\mu$th column by $C^{(k)}$ and the $\nu$th column by $C^{(k')}$.
We can also get the standard identity
\begin{equation}
\frac{\det\big(G(\mu,\nu\to C^{(\bsl k)},C^{(\bsl k')})\big)}{\det G}
=
x^{(\bsl k)}_\mu x^{(\bsl k')}_\nu
-
x^{(\bsl k)}_\nu x^{(\bsl k')}_\mu .
\label{eq:two_col_identity_x}
\end{equation}
Substituting $x^{(\bsl k)}_\mu=\partial E_\mu/\partial d_{\bsl k}$ and $x^{(\bsl k')}_\mu=\partial E_\mu/\partial d_{\bsl k'}$ leads to
\begin{equation}
\frac{\det\big(G(\mu,\nu\to C^{(\bsl k)},C^{(\bsl k')})\big)}{\det G}
=
\frac{\partial E_\mu}{\partial d_{\bsl k}}\frac{\partial E_\nu}{\partial d_{\bsl k'}}
-
\frac{\partial E_\nu}{\partial d_{\bsl k}}\frac{\partial E_\mu}{\partial d_{\bsl k'}}.
\label{eq:two_col_identity_derivatives}
\end{equation}
We plug \cref{eq:det_ratio_equals_derivative} and \cref{eq:two_col_identity_derivatives} into \cref{eq:CorrelatorInDeterminantRatio}:
\begin{align}
\frac{\big\langle\Psi_{2M}\big|A_{\bsl k}^{\dagger}A_{\bsl k'}\big|\Psi_{2M}\big\rangle}{\langle\Psi_{2M}|\Psi_{2M}\rangle}
&= \sum_{\mu} \frac{d_{\bsl k} - E_\mu}{d_{\bsl k'} - E_\mu} 
\frac{\partial E_\mu}{\partial d_{\bsl k}}- 2 \sum_{\mu<\nu}\frac{(d_{\bsl k} - E_\mu)(d_{\bsl k} - E_\nu)}{(d_{\bsl k} - d_{\bsl k'})(E_\nu - E_\mu)}(\frac{\partial E_\mu}{\partial d_{\bsl k}}\frac{\partial E_\nu}{\partial d_{\bsl k'}}
-
\frac{\partial E_\nu}{\partial d_{\bsl k}}\frac{\partial E_\mu}{\partial d_{\bsl k'}}), \quad (\bsl k \neq \bsl k').
\label{eq:finalCorrelatorexpression}
\end{align}
We derived the correlator formula for $\bsl k\neq \bsl k'$ case, now for $\bsl k= \bsl k'$ we have:
\begin{align}
H
&= \sum_{\bsl k,s} \varepsilon_{\bsl k} c_{\bsl k s}^\dagger c_{\bsl k s}
  + g \sum_{\bsl k} A_{\bsl k}^\dagger \sum_{\bsl k'} A_{\bsl k'} = \sum_{\bsl k} \varepsilon_{\bsl k} \left( n_{\bsl k\uparrow} + n_{-\bsl k\downarrow} \right)
  + g \sum_{\bsl k,\bsl k'} A_{\bsl k}^\dagger A_{{\bsl k}'} \\
&= \frac{1}{2} \sum_{\bsl k} d_{\bsl k} N_{\bsl k}
  + g \sum_{\bsl k,\bsl k'} A_{\bsl k}^\dagger A_{\bsl k'}, \quad ( N_{\bsl k} \equiv n_{\bsl k\uparrow} + n_{-\bsl k\downarrow}).
\end{align}
Take derivative:
\begin{align}
\frac{\partial H}{\partial d_{\bsl k}}
&= \frac{1}{2} N_{\bsl k} = n_{{\bsl k}\uparrow} n_{-{\bsl k}\downarrow} = c_{{\bsl k}\uparrow}^\dagger c_{-{\bsl k}\downarrow}^\dagger
   c_{-{\bsl k}\downarrow} c_{{\bsl k}\uparrow} = A_{\bsl k}^\dagger A_{\bsl k} .
\end{align}
where we focus on the seniority-zero (fully paired) subspace. We apply the Hellmann--Feynman theorem \cite{PhysRev.56.340} to the exact eigenstate
$\lvert \Psi_{2M} \rangle$:
\begin{align}
\frac{\partial E}{\partial d_{\bsl k}}
&= \frac{\langle \Psi_{2M} \lvert \dfrac{\partial H}{\partial d_{\bsl k}} \rvert \Psi_{2M} \rangle}
        {\langle \Psi_{2M} \mid \Psi_{2M} \rangle} = \frac{\langle \Psi_{2M} \lvert A_{\bsl k}^\dagger A_{\bsl k} \rvert \Psi_{2M} \rangle}
        {\langle \Psi_{2M} \mid \Psi_{2M} \rangle} .
\end{align}
Therefore,
\begin{equation}
    \frac{\langle \Psi_{2M} \lvert A_{\bsl k}^\dagger A_{\bsl k} \rvert \Psi_{2M} \rangle} {\langle \Psi_{2M} \mid \Psi_{2M} \rangle} = \frac{\partial E}{\partial d_{\bsl k}} = \sum_{\mu=1}^{M} \frac{\partial E_\mu}{\partial d_{\bsl k}}, \quad (\bsl k=\bsl k')
\label{eq:finalCorrelatorexpression2}
\end{equation}
\cref{eq:finalCorrelatorexpression} and \cref{eq:finalCorrelatorexpression2} are the final expressions of the pair-pair correlator, where $d_{\bsl k} = 2 \varepsilon_{\bsl k}$, matching the result from \cite{Gorohovsky_2011}.

\subsubsection{Slavnov's Formula and Application}\label{app:Slavnov's theorem}

In this section, we will derive \cref{eq:formfactorsimplified1} and \cref{eq:formfactorsimplified2} used in Pair–Pair Correlators section (\cref{app:correlator_derivation}). Recall the $M$-pair states
\begin{equation}
\ket{\{u\}} \equiv \prod_{\alpha=1}^{M} B^\dagger(u_\alpha)\ket{0},
\qquad
B^\dagger(u)=\sum_{\bsl p}\frac{A_{\bsl p}^\dagger}{d_{\bsl p}-u},
\end{equation}
and we defined the overlap
\begin{equation}
S(\{E\},\{F\}) \equiv \braket{\{E\}|\{F\}}.
\end{equation}
In this subsection, we assume $\{E\}$ is on-shell (namely, $\{E\}$ solves the Richardson equations), while $\{F\}$ is arbitrary and it does not need to solve the Richardson equations (we call it off-shell).

According to Slavnov’s formula and its application in Richardson model \cite{Slavnov1989,Zhou_2002,Faribault_2008,Gorohovsky_2011}, for on-shell $\{E\}=\{E_1,\dots,E_M\}$ and arbitrary $\{F\}=\{F_1,\dots,F_M\}$,
\begin{equation}
S(\{E\},\{F\})
=
\frac{
\displaystyle
\prod_{b=1}^{M}
\prod_{\substack{a=1 \\ a\neq b}}^{M}
\left( E_a - F_b \right)
}{
\displaystyle
\prod_{1 \le a < b \le M}
\left( E_b - E_a \right)
\prod_{1 \le b < a \le M}
\left( F_b - F_a \right)
}
\det \mathcal{S}(\{E\},\{F\}) .
\label{eq:slavnov-master}
\end{equation}
and the $M\times M$ matrix $\mathcal{S}$ has entries (with $a,b = 1,...,M$)
\begin{equation}
\mathcal{S}_{ab}(\{E\},\{F\})
=
\frac{E_b-F_b}{E_a-F_b}
\left[
\sum_{i}\frac{1}{(E_a-d_i)(F_b-d_i)}
-
2\sum_{\substack{c=1\\ c\neq a}}^{M}\frac{1}{(E_a-E_c)(F_b-E_c)}
\right].
\label{eq:slavnov-matrix}
\end{equation}
We now apply \eqref{eq:slavnov-master}–\eqref{eq:slavnov-matrix} to the specific sets in our case
\begin{equation}
\{F^{(\mu)}(u)\}=\{E_1,\dots,E_{\mu-1},u,E_{\mu+1},\dots,E_M\},
\qquad
\{F^{(\mu\nu)}(u,v)\}=\{E_1,\dots,u,\dots,v,\dots,E_M\},
\label{eq:twosets}
\end{equation}
because they are exactly the sets in our form factors (the factors to compute correlators) in \cref{eq:formfactor1}
and \cref{eq:formfactor2}:
\begin{equation}
F_{\bsl k}^{(\mu)}=\lim_{u\to d_{\bsl k}}(d_{\bsl k}-u)S(\{E\},\{F^{(\mu)}(u)\}),
\qquad
F_{\bsl k \bsl k'}^{(\mu\nu)}=\lim_{u\to d_{\bsl k}}\lim_{v\to d_{\bsl k'}}(d_{\bsl k}-u)(d_{\bsl k'}-v)S(\{E\},\{F^{(\mu\nu)}(u,v)\}).
\end{equation}
For $\{F\}=\{F^{(\mu)}(u)\}$ we have $F_b=E_b$ for $b\neq\mu$ and $F_\mu=u$, so the prefactor in \cref{eq:slavnov-master} can be reduced to 1:
\begin{align}
\frac{\displaystyle\prod_{b}\ \prod_{\substack{a\neq b}}(E_a-F_b)}
{\displaystyle
\prod_{a<b}(E_b-E_a)\;
\prod_{b<a}(F_b-F_a)
}
&= \frac{
\displaystyle
\prod_{\substack{b\neq \mu, a\neq b}} (E_a-E_b)
\prod_{a<b=\mu}(E_a-u)\prod_{a>b=\mu}(E_a-u)
}{
\displaystyle
\prod_{a<b}(E_b-E_a)
\prod_{\substack{b<a\\ b,a\neq \mu}}(E_b-E_a)
\prod_{b<a=\mu}(E_b-u)
\prod_{a>b=\mu}(u-E_a)}\\
&= (\frac{\prod_{\substack{b\neq \mu, a\neq b}} (E_a-E_b)}{\prod_{a<b}(E_b-E_a)
\prod_{\substack{b<a\\ b,a\neq \mu}}(E_b-E_a)} )(\frac{\prod_{a<b=\mu}(E_a-u)}{\prod_{b<a=\mu}(E_b-u)})(\frac{\prod_{a>b=\mu}(E_a-u)}{\prod_{a>b=\mu}(u-E_a)})\\
&=(-1)^{M-\mu}(1)(-1)^{M-\mu} = (-1)^{2({M-\mu})} =1.
\end{align}
Hence \cref{eq:slavnov-master} becomes
\begin{equation}
S(\{E\},\{F^{(\mu)}(u)\})=\det \mathcal{S}\big(\{E\},\{F^{(\mu)}(u)\}\big).
\label{eq:overlap-one-repl-det}
\end{equation}

We now evaluate the limit $F_b\to E_b$ in \eqref{eq:slavnov-matrix}. First, for the columns $b\neq\mu$, according to \cref{eq:twosets}, we have $F_b=E_b$, then

\text{Off-diagonal ($a\neq b$):}
\begin{align}
\lim_{F_b\to E_b}\mathcal S_{ab}
&=\lim_{F_b\to E_b}\frac{E_b-F_b}{E_a-F_b}\cdot
\frac{-2}{(E_a-E_b)(F_b-E_b)} =\frac{-2}{(E_a-E_b)}
\lim_{F_b\to E_b}\frac{E_b-F_b}{(E_a-F_b)(F_b-E_b)} \\
&=\frac{-2}{(E_a-E_b)}\,
\lim_{F_b\to E_b}\frac{-(F_b-E_b)}{(E_a-F_b)(F_b-E_b)} =\frac{2}{(E_a-E_b)}
\lim_{F_b\to E_b}\frac{1}{E_a-F_b}
=\frac{2}{(E_a-E_b)^2}.
\label{eq:slavnov-limit-offdiag}
\end{align}

\text{Diagonal ($a=b$):}
\begin{align}
\lim_{F_b\to E_b}\mathcal S_{bb}
&=\lim_{F_b\to E_b}\frac{E_b-F_b}{E_b-F_b}
\left[
\sum_{i}\frac{1}{(E_b-d_i)(F_b-d_i)}
-2\sum_{\substack{c\neq b}}\frac{1}{(E_b-E_c)(F_b-E_c)}
\right] =\sum_{i}\frac{1}{(E_b-d_i)^2}
-2\sum_{\substack{c\neq b}}\frac{1}{(E_b-E_c)^2}.
\label{eq:slavnov-limit-diag}
\end{align}
Comparing \eqref{eq:slavnov-limit-offdiag}–\eqref{eq:slavnov-limit-diag} with the Gaudin matrix
$G$ in \cref{eq:Gaudin_summary_mu}, we conclude: for every $b\neq\mu$, the $b$-th column of $\mathcal{S}(\{E\},\{F^{(\mu)}(u)\})$
matches exactly the $b$-th column of $G$ in the limit $F_b\to E_b$.
Hence, inside the determinant \eqref{eq:overlap-one-repl-det}, all columns except the $\mu$-th
become Gaudin columns.

Second, for the special column $b=\mu$, we have $F_\mu=u$. Multiply \eqref{eq:slavnov-matrix} by $(d_{\bsl k}-u)$ and take $u\to d_{\bsl k}$. Only the $\bsl i=\bsl k$ term in the first sum is singular, therefore,
\begin{align}
\lim_{u\to d_{\bsl k}}(d_{\bsl k}-u)\mathcal{S}_{a\mu}
&=
\lim_{u\to d_{\bsl k}}(d_{\bsl k}-u)\frac{E_\mu-u}{E_a-u}
\frac{1}{(E_a-d_{\bsl k})(u-d_{\bsl k})} =
\frac{E_\mu-d_{\bsl k}}{E_a-d_{\bsl k}}
\frac{-1}{E_a-d_{\bsl k}}
=
\frac{d_{\bsl k}-E_\mu}{(d_{\bsl k}-E_a)^2}.
\label{eq:slavnov-residue-column}
\end{align}
Define the vector $C^{(\bsl k)}_a \equiv \frac{1}{(d_{\bsl k}-E_a)^2},$ with $a=1,\dots,M,$ then
\begin{equation}
    \lim_{u\to d_{\bsl k}}(d_{\bsl k}-u)\mathcal{S}_{a\mu}=(d_{\bsl k}-E_\mu)C^{(\bsl k)}_a.
\end{equation}
Namely, as a column vector statement:
\begin{equation}
\lim_{u \to d_{\bsl k}} (d_{\bsl k} - u) \mathcal S_\mu(u) = (d_{\bsl k} - E_\mu) C^{(\bsl k)},
\label{eq:column-limit}
\end{equation}
where $C^{(\bsl k)}$ is the column vector with components $C_a^{(\bsl k)}$.

Multiply \cref{eq:overlap-one-repl-det} by $(d_{\bsl k} - u)$ and take the limit:
\begin{equation}
\lim_{u \to d_{\bsl k}} (d_{\bsl k} - u)
S\bigl(\{E\}, \{F^{(\mu)}(u)\}\bigr)
=
\lim_{u \to d_{\bsl k}} (d_{\bsl k} - u)
\det\Bigl(
\mathcal S\bigl(\{E\}, \{F^{(\mu)}(u)\}\bigr)
\Bigr).
\end{equation}
Since only the $\mu$-th column of $\mathcal S\bigl(\{E\}, \{F^{(\mu)}(u)\}\bigr)$ contains a pole as we take the limit (having finite limit), multiplying its determinant
by $(d_{\bsl k}-u)$ and taking $u\to d_{\bsl k}$ is equivalent to replacing the $\mu$-th column by
$(d_{\bsl k}-E_\mu)C^{(\bsl k)}$ (see \cref{eq:column-limit}), while all other columns become Gaudin matrix's columns (see \eqref{eq:slavnov-limit-offdiag}–\eqref{eq:slavnov-limit-diag}).
Thus,
\begin{equation}
\lim_{u\to d_{\bsl k}}(d_{\bsl k}-u)S(\{E\},\{F^{(\mu)}(u)\})
= (d_{\bsl k}-E_\mu)\det G\big(\mu \rightarrow C^{(\bsl k)}\big),
\end{equation}
where $G(\mu\rightarrow C^{(k)})$ denotes the matrix obtained from $G$ by replacing its $\mu$-th column
with $C^{(\bsl k)}$.
Using \cref{eq:formfactor1} gives precisely \cref{eq:formfactorsimplified1} that we used in the previous subsection:
\begin{equation}
F_{\bsl k}^{(\mu)}
=
(d_{\bsl k} - E_\mu)
\det\bigl( G(\mu \rightarrow C^{(\bsl k)}) \bigr),
\end{equation}
Similarly, we can also show that
\begin{equation}
F_{\bsl k \bsl k'}^{(\mu\nu)}
=
\frac{
(d_{\bsl k} - E_\mu)(d_{\bsl k} - E_\nu)(d_{\bsl k'} - E_\mu)(d_{\bsl k'} - E_\nu)
}{
(d_{\bsl k} - d_{\bsl k'})(E_\nu - E_\mu)
}\,
\det\bigl( G(\mu,\nu \rightarrow C^{(\bsl k)}, C^{(\bsl k')}) \bigr), \quad (\bsl k \neq \bsl k')
\end{equation}
as claimed in \cref{eq:formfactorsimplified2}, where $G(\mu,\nu\rightarrow C^{(\bsl k)},C^{(\bsl k')})$ denotes the matrix obtained from $G$ by replacing
the $\mu$-th and $\nu$-th columns by $C^{(\bsl k)}$ and $C^{(\bsl k')}$, respectively.

\subsubsection{Numerical Exact Pair-Pair Correlator Data}\label{app:Numeric_CC}
\begin{figure*}[t]
  \centering \includegraphics[width=1.0\textwidth]{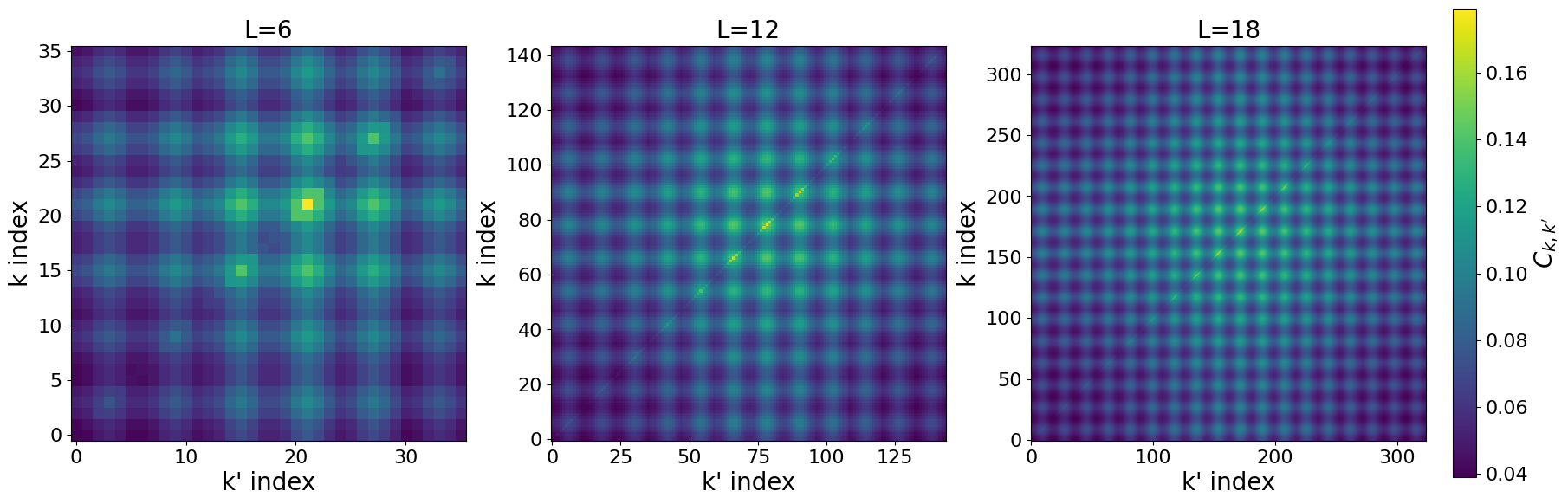}
  \caption{\textbf{Exact pair–pair correlators from the 2-dimensional Richardson model across mesh sizes.} Heat maps of the ground-state $C_{\bsl k,\bsl k'}$ on the BZ for $L=6,12,18$, computed from the exact solutions of Richardson equations.}
  \label{fig:ck0-grid}
\end{figure*}

We solve the Richardson equations \cref{eq:Richardson_eqn} on 2-dimensional Richardson models ~\cite{Richardson1963Restricted, RichardsonSherman1964, RevModPhys.76.643} on $L = 6, 12, 18$ respectively, with fixed electron filling $2M/L^2$ = 1/6.
On each $L\times L$ BZ mesh we sample
\begin{equation}
\bsl k=(k_x,k_y),\qquad
k_x=\frac{2\pi}{L}\,n_x,\quad
k_y=\frac{2\pi}{L}\,n_y,\qquad
n_x,n_y\in\{0,1,\ldots,L-1\},
\end{equation}
with single–particle energy
\begin{equation}
    \varepsilon_{\bsl k}=t(\cos k_x+\cos k_y).
\end{equation}
We choose hopping $t= 0.1$ in this work. For each of the meshes, we calculate the numerical pair-pair correlator data using \cref{eq:finalCorrelatorexpression} and \cref{eq:finalCorrelatorexpression2}.
\begin{table}[t]
\centering
\setlength{\tabcolsep}{10pt}
\renewcommand{\arraystretch}{1.2}
\begin{tabular}{c c c c c c}
\hline
$L$ & electron number & $g$ & $u$ & $ \sqrt{\sum_{\mu=1}^{M} |R_\mu|^2}.$\\
\hline
2  & 2 & $-0.250000$ & $-1$ &$2.22\times 10^{-16}$\\
4  & 4 & $-0.062500$ & $-1$ & $0.0$\\
6  & 6 & $-0.027778$ & $-1$ & $2.52\times 10^{-12}$\\
12 & 24 & $-0.006944$ & $-1$ & $1.80\times 10^{-10}$\\
18 & 54 & $-0.003086$ & $-1$ & $1.18\times 10^{-12}$\\
\hline
\end{tabular}
\caption{\textbf{Verification metrics across mesh sizes.}
For each $L$, we report the electron number, coupling $g\equiv u/L^{2}$ and Richardson-equation residual 2-norm $\sqrt{\sum_{\mu=1}^{M} |R_\mu|^2}$ where $R_\mu \equiv \frac{1}{g} +\sum_{\bsl k}\frac{1}{2\varepsilon_{\bsl k}-E_\mu}
-\sum_{\nu\neq\mu}\frac{2}{E_\nu-E_\mu}$. Here, $L=2$ and $L=4$ are used to benchmark against Exact-Diagonalization (ED) calculation, $L=6$ provides training data for our SIREN, and $L=12,18$ are reserved for SIREN benchmarking; for $L=6,12,18$ we fix the electron filling to $1/6$.}
\label{tab:rg-verify}
\end{table}

\subsubsection{Exact-Diagonalization (ED) Validation}

We validate our numerical Richardson-equation solver by benchmarking against ED calculation on small system sizes in the fully-paired sector. For each $L \times L$ momentum grid and coupling $g = u/L^2$ with $u=-1$, we solve the Richardson equations for the rapidities $\{E_\mu\}_{\mu=1}^M$ and compute the many-body energy $E_{\mathrm{numerical}} = \sum_{\mu=1}^M E_\mu$. For the cases $L = 2$, $N_e = 2$ and $L = 4$, $N_e = 4$, our $E_{\mathrm{numerical}}$ matches the ED ground-state energy exactly (with differences at most $\sim10^{-16}$), and the $\frac{1}{g}
+\sum_{\bsl k}\frac{1}{2\varepsilon_{\bsl k}-E_\mu}
-\sum_{\nu\neq\mu}\frac{2}{E_\nu-E_\mu}$  are negligible ($\sim10^{-16}$ and $\sim10^{-14}$ respectively), confirming that our solver converges to the exact solution.

We also further validate our pair--pair correlator evaluation by benchmarking against ED. Using the same momentum grid, electron filling, u, and g, we compute the correlator matrix in two independent ways: (i) from the Richardson rapidities $\{E_\mu\}$ via the closed-form expressions, \cref{eq:finalCorrelatorexpression} and \cref{eq:finalCorrelatorexpression2}, and (ii) directly from the ED ground-state wavefunction in the $M$-pair Hilbert space. For the small-size test cases (e.g., $L=2,N_e=2$ and $L=4,N_e=4$), the resulting $C_{\bsl{k}\bsl{k}'}$ agree exactly, confirming the correctness of our numerical correlator solver.

\subsection{ML Results}\label{sec:richardson_ML}

We wish to train a NN on the pair-pair correlation function (a special $2$-RDM) $C \in \mathbb{R}^{L^2 \times L^2}$, which is a symmetric positive semi-definite real matrix with $L^4$ elements in total. Since the self-attention mechanism has quadratic time and space complexity, the memory requirements are proportional to $L^8$, which can be extremely difficult for large $L$. Therefore, we chose to focus exclusively on the SIREN architecture for this model.

Numerically, the eigenvalue spectrum of $C$ is dominated by a single eigenvalue, with the remaining eigenvalues suppressed by several orders of magnitude. Consequently, the matrix has an effective rank close to 1. As such, we can diagonalize $C$ for each value of $L$ and choose $\bsl{A} \in \mathbb{R}^{L^2 \times 1}$ to be a normalized (\ie, $||\bsl{A}||_2=1$) eigenvector corresponding to the largest eigenvalue. Since the trace of the correlator is equal to $\frac{L^2}{12}$ (electron pair number), $C$ can then be approximately reconstructed as $C \approx \frac{L^2}{12} \bsl{A}\bsl{A}^T$.

Note that we can reshape $\bsl{A}$ into a matrix of shape $L\times L$, which matches the matrix dimensions in the 1-RDM case. Thus, we train a SIREN to output the elements of $\bsl{A}$ given 2D normalized coordinates, as described in \cref{SIREN_arch}. For inference, we enforce full $D_4$ symmetry by averaging the SIREN predictions for large $L$ with all of their rotations and reflections (i.e. the entire group of symmetries of the square). However, during training, we instead allow the SIREN to learn the symmetry by adding another term to the loss function in \cref{eq:total_loss_function}:

\begin{equation}\label{eq:L_symmetry}
    \mathcal{L}_{\text{symmetry}} = \lambda_{\text{symmetry}} \left(\frac{1}{|S|}\right) \sum_{g \in S}
    \text{MSE}\left(\widetilde{\bsl{A}}, g(\widetilde{\bsl{A}})\right),
\end{equation}

where $\widetilde{\bsl{A}} \in \mathbb{R}^{(L+1)\times (L+1)}$ is the ``wrapped around" version of the reshaped $\bsl{A}$ matrix for large L (see \cref{subsubsec:Coordinate domain and periodic wrap and symmetry} for details), $S$ is a generating subset of $D_4$ ($\ie, D_4 = \langle S\rangle$), which we choose to be $S=\{R_{90}, D\}$ ($R_{90}$ denotes rotation by $90^\circ$ and $D$ denotes a reflection across the main diagonal), $|S|$ is the cardinality of $S$ ($|S|=2$ in our case), and the $\lambda_{\text{symmetry}}$ coefficient (a hyperparameter) controls the contribution to the total loss function. This symmetry loss is added after a small number of warmup steps ($50$ steps in our case) to prevent gradients from misleading the training by chasing a moving target. For training on the Richardson model, we omit the optional statistical prior (TV/FH) described in the general SIREN architecture (see \cref{subsubsec:SIREN_loss_function} for details), so we choose $\mathcal{P}=0$ in \cref{eq:total_loss_function}.
Finally, we use the Tree-Structured Parzen Estimation (TPE) \cite{TPE_paper} algorithm to search for the optimal hyperparameter configuration that can enable the SIREN to generalize effectively to larger $L$, as intended. For each trial, the SIREN is allowed to be trained for a total of $500$ epochs, although we also save intermediate checkpoints in order to find the optimal parameter values.

Using the above method, we trained a SIREN on $\bsl{A}$ for $L=6$, and then used it to predict $\bsl{A}$ for $L=12$ and $L=18$. For each test size, $\bsl{A}$ is first normalized such that $||\bsl{A}||_2=1$ (where \(\|\cdot\|_2\) denotes the $\ell^2$ norm) before reconstructing the predicted final $2$-RDM via:
\begin{equation}\label{eq:C_SIREN_reconstruction}
    C^{\text{SIREN}} = \frac{L^2}{12} \bsl{A} \bsl{A}^T.
\end{equation}
For the Richardson model, the MSE is defined as:
\begin{equation}
    \text{MSE}(C, \,C^{\text{SIREN}}) = \frac{1}{L^4}\sum_{i=1}^{L^2}\sum_{j=1}^{L^2}\left(C^{\text{SIREN}}_{ij} - C_{ij}\right)^2,
\label{eq:MSE_def_for_richardson}
\end{equation}
where $C^{\text{SIREN}}$ and $C$ are the SIREN-predicted final values and the true final values for the $2$-RDM of pair-pair correlators, respectively.
In order to evaluate the trained SIREN's performance, we take the square root of \cref{eq:MSE_def_for_richardson}, $\ie\ \sqrt{\text{MSE}}$, in order to ensure that the power on the unit of the error is the same as that for $C$ and $C^{\text{SIREN}}$. To measure the relative accuracy $r_n$, we normalize the $\sqrt{\text{MSE}}$ by the total range of values of the original $C$, and subtract this fraction from unity. Results are shown in \cref{tab:richardson_results_combined}. We can also compare the predicted final $2$-RDM from the SIREN to the true final $2$-RDM (\ie\ the actual values of $C$) visually, as shown in \cref{fig:richardson_comparison_app} for a test size of $L=12$. (See \cref{fig:richardson_comparison_maintext} in the maintext for a comparison on $L=18$.)

\begin{table}[t]
  \centering
  \small
  \setlength{\tabcolsep}{6pt}
  \renewcommand{\arraystretch}{1.12}
  \begin{tabular}{@{}rcc@{}}
    \toprule
    $L$ & $\sqrt{\mathrm{MSE}}$ & $r_n$ \\
    \midrule
    12  & $0.0074$ & $94.56\%$ \\
    18  & $0.0077$ & $94.29\%$ \\
    \bottomrule
  \end{tabular}
  \caption{\textbf{Pair-Pair Correlation function of Richardson model interpolation with SIREN.} Metrics at evaluation mesh size $L\times L$ for a SIREN trained on $6\times 6$. Columns report root--mean--square error ($\sqrt{\mathrm{MSE}}$) and relative accuracy ($r_n$) between SIREN predictions and true final $2$-RDM on the evaluation mesh.}
  \label{tab:richardson_results_combined}
\end{table}

\begin{figure}[t]
    \centering
    \includegraphics[width=1.0\linewidth]{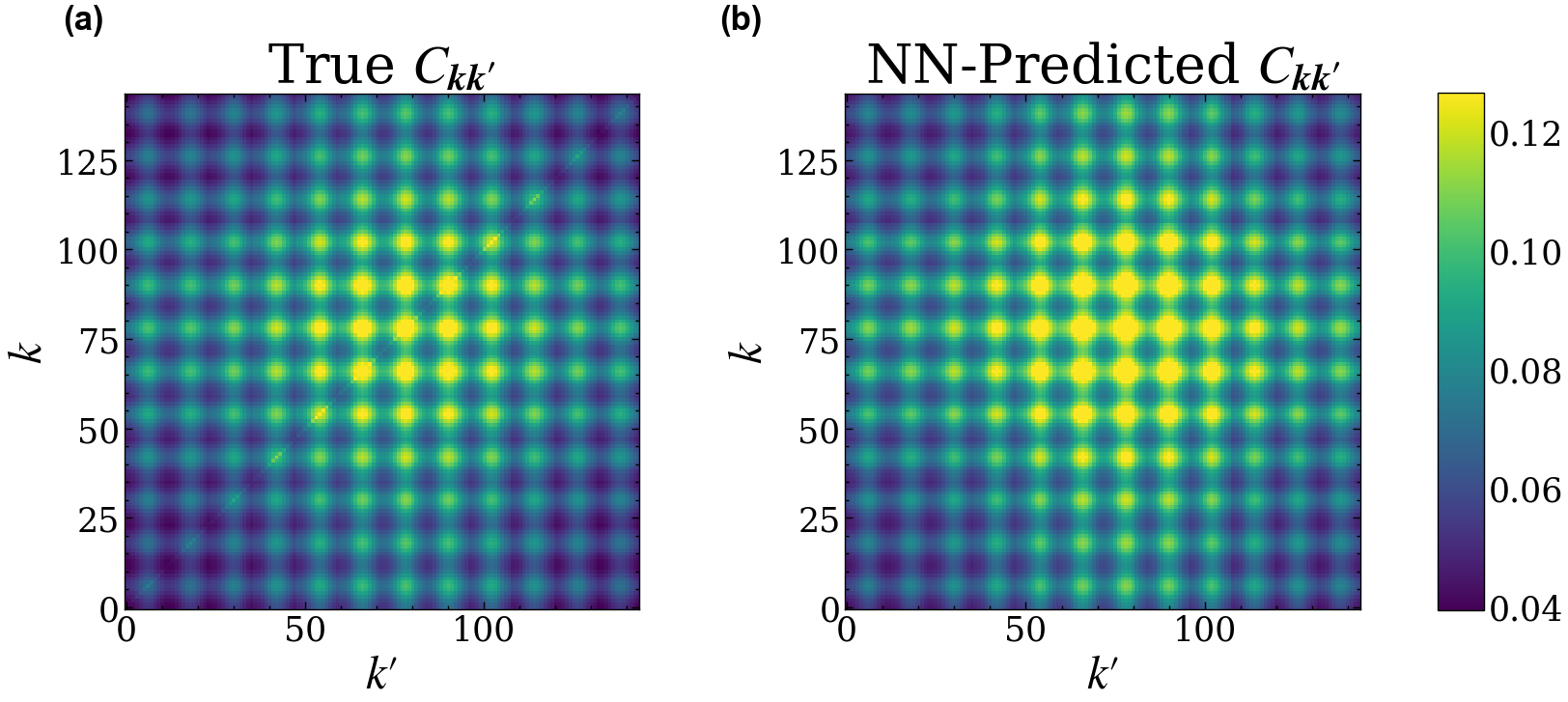}
    \caption{Comparison between (a) the true final $C_{\bsl{k}\bsl{k}'}$ for $L=12$ and (b) the predicted final $C_{\bsl{k}\bsl{k}'}$ for $L=12$ given by a SIREN trained on $L=6$. Indices \(k,k'\in\{0,\dots,L^{2}-1\}\) enumerate the \(L\times L\) $\bsl{k}$ points by flattening \((\ell_{1},\ell_{2})\) with \(\ell_{1,2}\in\{0,\dots,L-1\}\) via \(k=\ell_{1}+L\,\ell_{2}\) with its momentum vector in \cref{eq:k_expression}, and similarly for \(k'\).}
    \label{fig:richardson_comparison_app}
\end{figure}

\begin{figure}[t]
    \centering
    \includegraphics[width=1.0\linewidth]{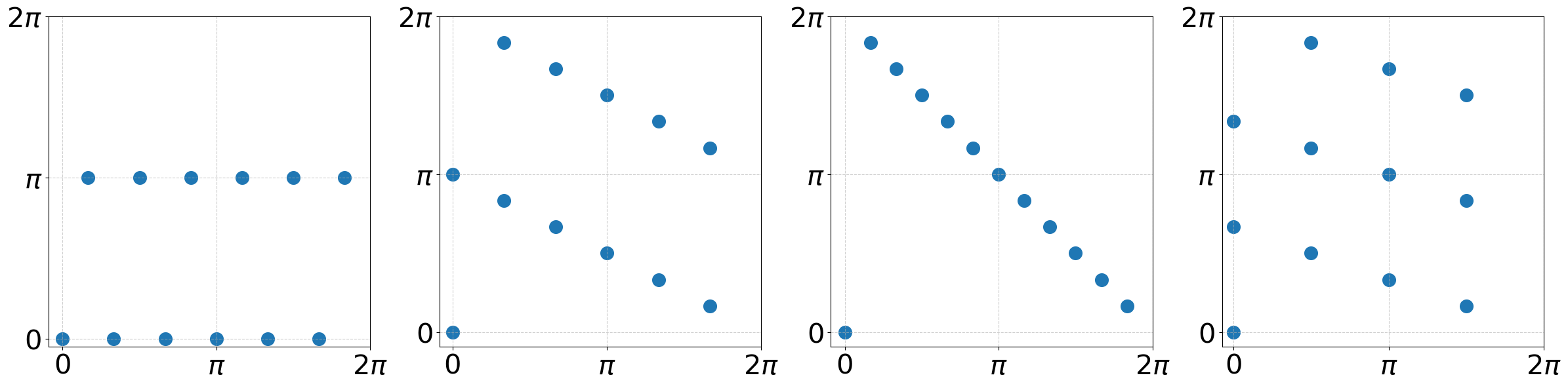}
    \caption{The $\bsl{k}$ points chosen for the tilted meshes the SIREN is trained on for the Richardson model. The coordinates are wrapped around modulo $2\pi$ in each direction.}
    \label{fig:tilted_mesh_kpts}
\end{figure}

Next, we train an additional SIREN on 4 $L_1 \times L_2$ tilted meshes with different side lengths $L_1, L_2\in\mathbb{Z}_{>0}$. Define $\bsl{b}_1=(2\pi, 0)$ and $\bsl{b}_2=(0, 2\pi)$ to be the primitive reciprocal lattice vectors. For each tilted mesh, we construct the transformed reciprocal lattice vectors $\bsl{f}_1$ and $\bsl{f}_2$ as a linear combination of $\bsl{b}_1$ and $\bsl{b}_2$, $\ie$ $\left(\bsl{f}_1, \bsl{f}_2\right)^T = M \left(\bsl{b}_1, \bsl{b}_2\right)^T$, where $M\in\mathbb{Z}^{2\times 2}$ is different for each tilted mesh. The side lengths and transformed reciprocal lattice vectors are chosen so that each tilted mesh has $L_1 L_2 = 12$ $\bsl{k}$ points, and so that the overlap between different meshes ($\ie$, the number of $\bsl{k}$ points in common) is minimized as shown in \cref{fig:tilted_mesh_kpts}.
We diagonalize $C_{\bsl{k}\bsl{k}'}$ for each tilted mesh, and choose the normalized eigenvector $\bsl{A}\in\mathbb{R}^{L_1 L_2}$ corresponding to the highest eigenvalue, which can be viewed as a function of $L_1 L_2$ $\bsl{k}$ points. By combining the eigenvectors for all four of the tilted meshes, we obtain the training dataset for the SIREN after normalizing the $\bsl{k}$ points to lie in $[-1, 1]^2$ and enforcing periodic boundary conditions for each reshaped eigenvector (see \cref{subsubsec:Coordinate domain and periodic wrap and symmetry}). We train for a total of $1000$ epochs, and apply the $D_4$ symmetry loss according to \cref{eq:L_symmetry} after the first $20$ epochs. Note that we do not use TPE \cite{TPE_paper} to optimize hyperparameters, as was done when training on $L=6$.
After training, the resulting SIREN is then used to predict a new eigenvector, which we divide by its $\ell^2$ norm to produce $\bsl{A}\in\mathbb{R}^{18^2}$ for $L=18$, where $||\bsl{A}||=1$. The corresponding $C_{\bsl{k}\bsl{k}'}$ is again reconstructed via \cref{eq:C_SIREN_reconstruction}, after which we compute the accuracy with respect to the true final $C_{\bsl{k}\bsl{k}'}$ using \cref{eq:MSE_def_for_richardson}. The result yields
a relative accuracy of $r_n=93.77\%$. A comparison between $C^{\text{SIREN}}$ and the true $C_{\bsl{k}\bsl{k}'}$ for this case can be seen in \cref{fig:tilted_mesh_comparison}.
\begin{figure}[t]
    \centering
    \includegraphics[width=1.0\linewidth]{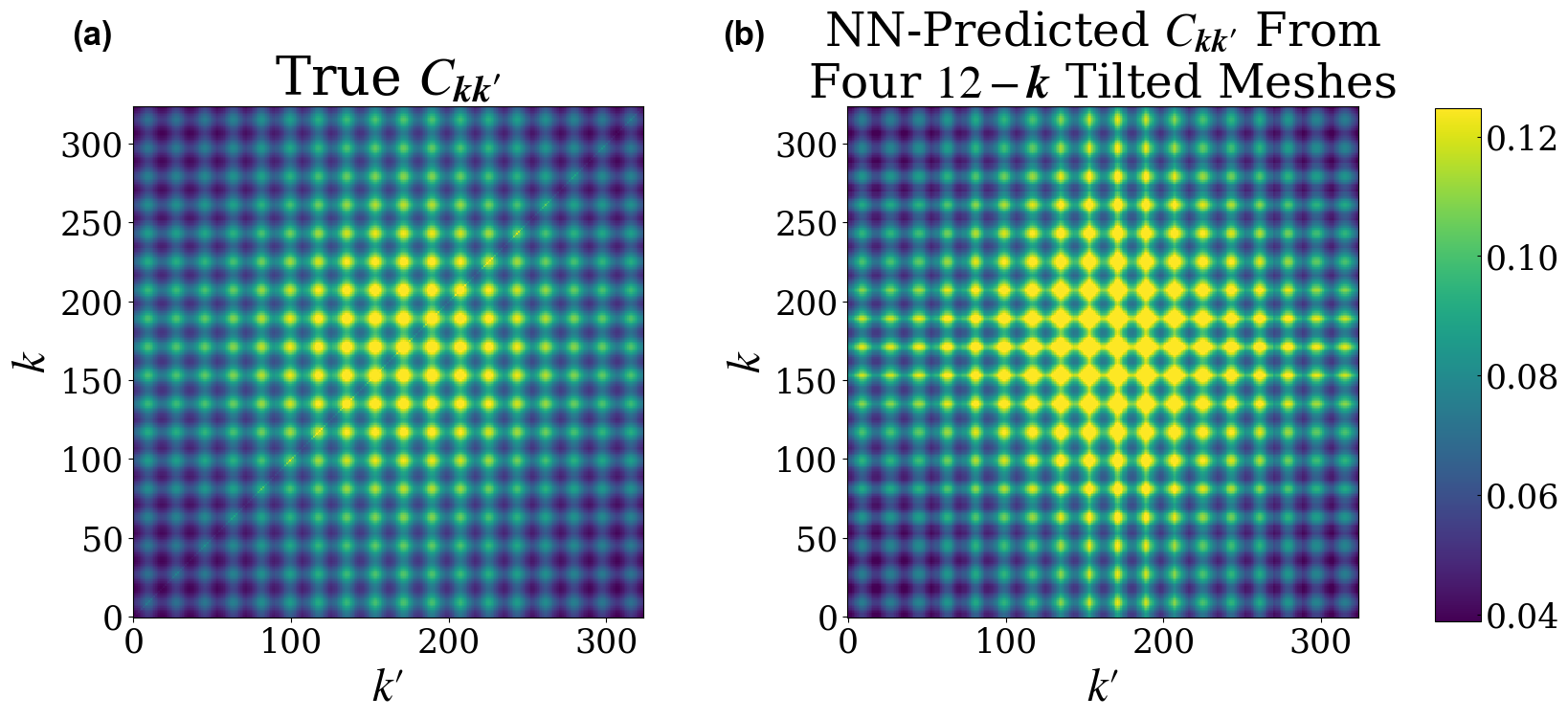}
    \caption{Comparison between (a) the true final $C_{\bsl{k}\bsl{k}'}$ for $L=18$ and (b) the predicted final $C_{\bsl{k}\bsl{k}'}$ for $L=18$ given by a SIREN trained on four tilted meshes with $12$ $\bsl{k}$ points each. Indices \(k,k'\in\{0,\dots,L^{2}-1\}\) enumerate the \(L\times L\) $\bsl{k}$ points by flattening \((\ell_{1},\ell_{2})\) with \(\ell_{1,2}\in\{0,\dots,L-1\}\) via \(k=\ell_{1}+L\,\ell_{2}\) with its momentum vector in \cref{eq:k_expression}, and similarly for \(k'\).}
    \label{fig:tilted_mesh_comparison}
\end{figure}

\section{Translation-invariant 1-RDM in a four-band model}
\label{sec:toymodelpart}

\subsection{Model Set-Up}\label{subsec:fourbandmodelsetup}
We work on an $L\times L$ Bravis lattice with a four-component orbital (or layer/sublattice) spinor
$\psi_{\bsl r}=(\psi_{\bsl r,1},\dots,\psi_{\bsl r,4})^{\mathsf T}$ at site $\bsl r\in\mathbb{Z}^2$.
The real-space single-particle Hamiltonian contains on-site and short-range hoppings
written with $4\times4$ matrices $h(\boldsymbol{\delta})$ that connect $\bsl r$ to $\bsl r+\boldsymbol{\delta}$:
\begin{equation}
  H_0 = \sum_{\bsl r}\psi_{\bsl r}^\dagger\,h(\bsl 0)\,\psi_{\bsl r}
  + \sum_{\bsl r}\sum_{\boldsymbol{\delta}\neq \bsl 0}\Big(\psi_{\bsl r+\boldsymbol{\delta}}^\dagger\,h(\boldsymbol{\delta})\,\psi_{\bsl r} + \text{h.c.}\Big).
  \label{eq:toy-real}
\end{equation}
We choose the five gamma matrices as
\begin{equation}
  \Gamma_1=\sigma_x\otimes\sigma_x,\quad
  \Gamma_2=\sigma_x\otimes\sigma_y,\quad
  \Gamma_3=\sigma_x\otimes\sigma_z,\quad
  \Gamma_4=\sigma_y\otimes\mathbb{I},\quad
  \Gamma_5=\sigma_z\otimes\mathbb{I},
  \label{eq:toy-gammas}
\end{equation}
where $\sigma_x$, $\sigma_y$ and $\sigma_z$ are Pauli matrices.
The hoppings have the following form:
\begin{equation}
  \begin{aligned}
    &\text{on-site:} && h(\bsl 0)=\Gamma_5,\\
    &\text{NN (odd, along $x,y$):} &&
      h(+\hat{\bsl x})=\tfrac{1}{2i}\Gamma_1,\quad h(-\hat{\bsl x})=-\tfrac{1}{2i}\Gamma_1,\qquad
      h(+\hat{\bsl y})=\tfrac{1}{2i}\Gamma_2,\quad h(-\hat{\bsl y})=-\tfrac{1}{2i}\Gamma_2,\\
    &\text{NNN(odd, along $x,y$):} &&
      h(+2\hat{\bsl x})=\tfrac{1}{2i}\Gamma_3,\quad h(-2\hat{\bsl x})=-\tfrac{1}{2i}\Gamma_3,\qquad
      h(+2\hat{\bsl y})=\tfrac{1}{2i}\Gamma_4,\quad h(-2\hat{\bsl y})=-\tfrac{1}{2i}\Gamma_4,\\
    &\text{NN (even, in the $\Gamma_5$ channel):} &&
      h(\pm\hat{\bsl x})=\tfrac{1}{2}\Gamma_5,\qquad h(\pm\hat{\bsl y})=\tfrac{1}{2}\Gamma_5.
  \end{aligned}
  \label{eq:toy-hops}
\end{equation}

Here “NN” denotes nearest–neighbor displacements $\boldsymbol{\delta}$ with
$|\boldsymbol{\delta}|=1$ ($\pm \hat{\bsl{x}},\, \pm \hat{\bsl{y}}$);
“NNN” denotes next–nearest neighbors along the axes
$|\boldsymbol{\delta}|=2$ ($\pm 2\hat{\bsl{x}},\, \pm 2\hat{\bsl{y}}$).
Unlisted $h(\boldsymbol{\delta})$ are zero. With the lattice Fourier transform
\begin{equation}
  \psi_{\bsl r}=\frac{1}{\sqrt{L^2}}\sum_{\bsl k} e^{i\bsl k\cdot r}\,\psi_{\bsl k},\qquad
  \bsl k=(k_x,k_y)\in[0,2\pi)^2,
  \label{eq:toy-fourier}
\end{equation}
\cref{eq:toy-real} becomes
\begin{equation}
    \begin{aligned}
    H_0
    &= \sum_{\bsl r}\psi_{\bsl r}^\dagger h(\bsl 0)\psi_{\bsl r}
      + \sum_{\bsl r}\sum_{\boldsymbol\delta\ne\bsl 0}
        \big(\psi_{\bsl r+\boldsymbol\delta}^\dagger h(\boldsymbol\delta)\psi_{\bsl r}+\text{h.c.}\big) \\
    &= \frac{1}{L^2}\sum_{\bsl r}\sum_{\bsl k,\bsl k'}
        e^{-i\bsl k'\cdot\bsl r}\psi_{\bsl k'}^\dagger\,
        h(\bsl 0)\,
        e^{i\bsl k\cdot\bsl r}\psi_{\bsl k} + \frac{1}{L^2}\sum_{\bsl r}\sum_{\boldsymbol\delta\ne\bsl 0}
        \sum_{\bsl k,\bsl k'}
        e^{-i\bsl k'\cdot(\bsl r+\boldsymbol\delta)}\psi_{\bsl k'}^\dagger\,
        h(\boldsymbol\delta)\,
        e^{i\bsl k\cdot\bsl r}\psi_{\bsl k}
        +\text{h.c.} \\
    &= \sum_{\bsl k,\bsl k'}
        \psi_{\bsl k'}^\dagger\,h(\bsl 0)\,\psi_{\bsl k}
        \frac{1}{L^2}\sum_{\bsl r}e^{i(\bsl k-\bsl k')\cdot\bsl r} + \sum_{\boldsymbol\delta\ne\bsl 0}\sum_{\bsl k,\bsl k'}
        \psi_{\bsl k'}^\dagger\,h(\boldsymbol\delta)\,\psi_{\bsl k}
        \frac{1}{L^2}\sum_{\bsl r}e^{i(\bsl k-\bsl k')\cdot\bsl r}\,
        e^{-i\bsl k'\cdot\boldsymbol\delta}
        +\text{h.c.} \\
    &= \sum_{\bsl k}\psi_{\bsl k}^\dagger h(\bsl 0)\psi_{\bsl k}
      + \sum_{\boldsymbol\delta\ne\bsl 0}\sum_{\bsl k}
        \psi_{\bsl k}^\dagger\big(e^{\,i\bsl k\cdot\boldsymbol\delta}h(\boldsymbol\delta)\big)\psi_{\bsl k}
        +\text{h.c.}
    \qquad\bigg(\frac{1}{L^2}\sum_{\bsl r}e^{i(\bsl k-\bsl k')\cdot\bsl r}
    = \delta_{\bsl k,\bsl k'}\bigg) \\
    &= \sum_{\bsl k}\psi_{\bsl k}^\dagger
        \Big[h(\bsl 0)+\sum_{\boldsymbol\delta\ne\bsl 0}
          \big(e^{\,i\bsl k\cdot\boldsymbol\delta}h(\boldsymbol\delta)
            + e^{-i\bsl k\cdot\boldsymbol\delta}h^\dagger(\boldsymbol\delta)\big)\Big]
        \psi_{\bsl k} \\
    &\equiv \sum_{\bsl k}\psi_{\bsl k}^\dagger H_0(\bsl k)\psi_{\bsl k},
    \qquad
    H_0(\bsl k)=h(\bsl 0)+\sum_{\boldsymbol\delta\ne\bsl 0}
    \big(e^{\,i\bsl k\cdot\boldsymbol\delta}h(\boldsymbol\delta)
      + e^{-i\bsl k\cdot\boldsymbol\delta}h^\dagger(\boldsymbol\delta)\big) = \sum_{a=1}^{5} d_a(\bsl k)\Gamma_a,
    \end{aligned}
\label{eq:toy-H0}
\end{equation}
and the form factors are
\begin{equation}
  d_1=\sin k_x,\quad d_2=\sin k_y,\quad d_3=\sin(2k_x),\quad d_4=\sin(2k_y),\quad
  d_5=1+\cos k_x+\cos k_y.
  \label{eq:toy-dvec}
\end{equation}
The odd/even harmonics follow from $e^{\pm i \bsl k\cdot \boldsymbol \delta}=(\cos \bsl k\cdot \boldsymbol \delta)\pm i(\sin \bsl k\cdot \boldsymbol\delta)$:
antisymmetric pairs $h(+\delta)=-h(-\delta)$ generate $\sin$, while symmetric pairs $h(+\boldsymbol\delta)=h(-\boldsymbol\delta)$ generate $\cos$.
By the Clifford algebra of the $\Gamma_a$, the spectrum is doubly degenerate with
\begin{equation}  E_\pm(\bsl k)=\pm\|d(\bsl k)\|=\pm\sqrt{\sum_{a=1}^5 d_a(\bsl k)^2}.
  \label{eq:toy-spectrum}
\end{equation}
At filling \(f=1\) we occupy the lower band.
We diagnose its topology via the Wilson loop \(W_x(k_y)\) of the occupied projector \(P(\bsl k)\) in \cref{eq:toy-projector},
yielding the \(\mathbb Z_2\) index
$\nu_{\mathbb Z_2}=1$.
This certifies that our training/benchmark data come from a $\dsZ_2$ non-trivial phase.
 For the interaction in second quantization and its lattice Fourier transform \cref{eq:Fouriertransform_lattice},
\begin{align}
H_{\text{int}}
&= \frac12 \sum_{\bsl r}\sum_{\bsl R} V(\bsl R)\, n_{\bsl r}\, n_{\bsl r+\bsl R}
\notag\\
&= \frac12 \sum_{\bsl r,\bsl R} V(\bsl R)
\left(\frac{1}{\sqrt N}\sum_{\bsl k} e^{i\bsl k\cdot \bsl r}\, \psi_{\bsl k}^\dagger\right)
\left(\frac{1}{\sqrt N}\sum_{\bsl k'} e^{-i\bsl k'\cdot \bsl r}\, \psi_{\bsl k'}\right)
\left(\frac{1}{\sqrt N}\sum_{\bsl p} e^{i\bsl p\cdot (\bsl r+\bsl R)}\, \psi_{\bsl p}^\dagger\right)
\left(\frac{1}{\sqrt N}\sum_{\bsl p'} e^{-i\bsl p'\cdot (\bsl r+\bsl R)}\, \psi_{\bsl p'}\right)
\notag\\
&= \frac{1}{2N^2}\sum_{\bsl R} V(\bsl R)
\sum_{\bsl k,\bsl k',\bsl p,\bsl p'}
\Big(\sum_{\bsl r} e^{i(\bsl k-\bsl k'+\bsl p-\bsl p')\cdot \bsl r}\Big)
e^{i(\bsl p-\bsl p')\cdot \bsl R}\,
\psi_{\bsl k}^\dagger \psi_{\bsl k'} \psi_{\bsl p}^\dagger \psi_{\bsl p'}
\notag\\
&= \frac{1}{2N}\sum_{\bsl R} V(\bsl R)
\sum_{\bsl k,\bsl k',\bsl q}
e^{-i\bsl q\cdot \bsl R}\,
\psi_{\bsl k+\bsl q}^\dagger \psi_{\bsl k}\,
\psi_{\bsl k'-\bsl q}^\dagger \psi_{\bsl k'}
\qquad (\bsl q\equiv \bsl p'-\bsl p)
\notag\\
&= \frac{1}{2N}\sum_{\bsl q}\underbrace{\Big(\sum_{\bsl R} V(\bsl R)e^{-i\bsl q\cdot \bsl R}\Big)}_{U(\bsl q)}
\sum_{\bsl k,\bsl k'}
\psi_{\bsl k+\bsl q}^\dagger \psi_{\bsl k}\,
\psi_{\bsl k'-\bsl q}^\dagger \psi_{\bsl k'}.
\label{eq:int-momentum}
\end{align}
Thus the translation-invariant interaction enters through the lattice Fourier transform,
\begin{equation}
U(\bsl q)=\sum_{\bsl R} V(\bsl R)\,e^{-i\bsl q\cdot \bsl R}.
\label{eq:lattice-FT}
\end{equation}
For our specific short-range choice
\begin{equation}
V(\bsl R) = U_0\left[\delta_{\bsl R,\bsl 0}
+\frac12\big(\delta_{\bsl R,\hat{\bsl x}}+\delta_{\bsl R,-\hat{\bsl x}}
+\delta_{\bsl R,\hat{\bsl y}}+\delta_{\bsl R,-\hat{\bsl y}}\big)\right]
\end{equation}
Therefore,
\begin{equation}
U(\bsl q) = \sum_{\bsl R} V(\bsl R)e^{-i\bsl q\cdot \bsl R}
= U_0\left[1+\tfrac12\big(e^{-iq_x}+e^{+iq_x}+e^{-iq_y}+e^{+iq_y}\big)\right]
= U_0\big(1+\cos q_x+\cos q_y\big).
\label{eq:Uq-final}
\end{equation}
So the momentum-space interaction is
\begin{equation}
H_{\text{int}}
= \frac{1}{2N}\sum_{\bsl q}U_0\big(1+\cos q_x+\cos q_y\big)
\sum_{\bsl k,\bsl k'}
\psi_{\bsl k+\bsl q}^\dagger \psi_{\bsl k}\,
\psi_{\bsl k'-\bsl q}^\dagger \psi_{\bsl k'}.
\end{equation}
The mean-field order parameter is the 1-RDM (projector onto occupied HF bands) at each momentum:
\begin{equation}
  P(\bsl k)=\sum_{n=1}^{f} \,|u_{n}(\bsl k)\rangle\langle u_{n}(\bsl k)|,
  \label{eq:toy-projector}
\end{equation}
where n is the band index, and \(f\in\{1,2,3,4\}\) is the filling (number of occupied bands per \(\bsl k\)). We choose $f=1$ for our ML training. The projector obeys
\begin{equation}
  P^\dagger(\bsl k)=P(\bsl k),\qquad P^2(\bsl k)=P(\bsl k),\qquad \operatorname{tr} P(\bsl k)=f.
  \label{eq:toy-projector-constraints}
\end{equation}

Normal ordering the interaction with respect to \(P(\bsl k)\) yields the self-consistent HF Hamiltonian
\begin{equation}
  H_{\mathrm{HF}}(\bsl k)= H_0(\bsl k)+ \Sigma_H+\Sigma_F(\bsl k;[P]),
  \label{eq:toy-HF-H}
\end{equation}
with a Hartree shift
\begin{equation}
  \Sigma_H = U(0)\,\bar{n}\mathbb{I}_4,\quad 
  \bar{n}\equiv \frac{1}{L^2}\sum_{\bsl k}\operatorname{tr}P(\bsl k),
  \label{eq:toy-HF-Hartree}
\end{equation}
and a Fock self-energy given by a BZ convolution
\begin{equation}
  \Sigma_F(\bsl k;[P]) = -\frac{1}{L^2}\sum_{\bsl q} U(\bsl q)\,P(\bsl k+\bsl q).
  \label{eq:toy-HF-Fock}
\end{equation}
Because \(\Sigma_H\propto \mathbb{I}_4\) and \(\operatorname{tr}P(\bsl k)=f\) is fixed by the filling, the Hartree term produces a uniform band shift and does not affect the projector; symmetry breaking (if any) is controlled by \(\Sigma_F\). We monitor the total energy density (per unit cell)
\begin{equation}
  \mathcal{E}[P]=\frac{1}{2 L^2}\sum_{\bsl k}\operatorname{tr}\Bigl[\bigl(H_{\mathrm{HF}}(\bsl k)+H_0(\bsl k)\bigr)\,P(\bsl k)\Bigr],
  \label{eq:toy-energy}
\end{equation}
where the factor \(1/2\) avoids double counting. The stopping criterion is
\(
  \big|\mathcal{E}_{n+1}-\mathcal{E}_{n}\big|<10^{-6}
\),
supplemented by the mean-squared change of the projector between iterations. 
Arrays \(H_0(\bsl k)\), \(U(\bsl q)\), and \(P(\bsl k)\) live on the BZ grid. Given \(P^{(n)}\), we build \(\Sigma_H[P^{(n)}]\) and \(\Sigma_F[P^{(n)}]\), diagonalize \(H_{\mathrm{HF}}[P^{(n)}](\bsl k)\) at each \(\bsl k\), and update
\begin{equation}
  P^{(n+1)}(\bsl k)=\sum_{m=1}^{f}\,|u^{(n)}_{m}(\bsl k)\rangle\langle u^{(n)}_{m}(\bsl k)|,
  \label{eq:toy-update}
\end{equation}
selecting the \(f\) eigenvectors with the lowest eigenvalues (ordered by real part).

The four-band model supplies the final HF labels --- $P(\bsl k)$ at filling $f=1$ and band energies \(\mathcal{E}\) --- from a translation-invariant HF calculation, by which we pretrain and validate our NN architectures. It provides a clean benchmark to quantify acceleration: we compare HF iteration counts when initialized with the NN-predicted final 1-RDM versus a random initial 1-RDM at a fixed tolerance.

\subsection{ML Results}\label{ML_toyModel}
Translation-invariant systems are simpler because the one-body correlator is diagonal in momentum
(\(\bsl k=\bsl k'\)), so inference reduces to a sequence of \(L^2\) matrices.
The lower dimensionality lets us use a conventional self-attention NN without neural
representation (SIREN). We again work on the discrete BZ with \(\bsl k\) in Eq.~\eqref{eq:k_expression}.
Index grid points by their integers \((l_1,l_2)\) with \(l_{1,2}\in\{0,\dots,L-1\}\), so that
\[
\bsl k(l_1,l_2)=\frac{l_1}{L}\,\bsl b_1+\frac{l_2}{L}\,\bsl b_2 .
\]
To encode relative position with periodic boundary conditions, use the wrapped (minimum-image) index differences
\begin{equation}
\Delta l_1 = \big((l_1-l_1' + \lfloor L/2\rfloor)\bmod L\big) - \lfloor L/2\rfloor,\quad
\Delta l_2 = \big((l_2-l_2' + \lfloor L/2\rfloor)\bmod L\big) - \lfloor L/2\rfloor,
\end{equation}
so we can define the 2D coordinate
\begin{equation}
    \bsl x_{(l_1,l_2),(l_1',l_2')} = \Big(\tfrac{\Delta l_1}{L},\,\tfrac{\Delta l_2}{L}\Big).
\end{equation}
For brevity we denote $\bsl x_{(l_1,l_2),(l_1',l_2')}$ as $\bsl x_{ij}$.

For the four-band model we keep the architecture of
\cref{subsec:transformer_arch} but, given the simplicity here, replace the custom
Fourier-based bias by a small MLP, following the formulation of \refcite{Li_2024_FIRE}.
That is,
\begin{equation}\label{eq:NN_f_theta}
B_{ij}
= f_{\theta}\big(\bsl{k}_i,\bsl{k}_j\big)
= \mathrm{GELU}\big(\bsl{x}_{ij} W_1 + \bsl{b}_1\big)\, W_2 + b_2 .
\end{equation} 
where \(\bsl{x}_{ij}\) represents the relative coordinate between two \(k\)-points in 2D momentum space and $W_1 \in \mathcal{M}_{2 \times d_\mathrm{MLP}}(\mathbb{R})$, $\bsl b_1 \in \mathbb{R}^{d_\mathrm{MLP}}$, $W_2 \in \mathcal{M}_{d_\mathrm{MLP} \times 1}(\mathbb{R})$, $\bsl b_2 \in \mathbb{R}$ are all filled with learnable parameters to be optimized during training, and $d_\mathrm{MLP}$ represents the size of the hidden layer in the MLP. \cref{eq:NN_f_theta} replaces \cref{eq:f_theta} for the bias function in the four-band model.
Thus, the output of $f_{\theta}$ is a scalar bias (that quantifies how closely related the corresponding $\bsl{k}$ points are), which is then added to the self-attention computation via \cref{eq:scaled_logits}, thereby injecting relative positional information into the NN.

\subsubsection{Hyperparameters}
To train the self-attention NN, we used a constant learning rate of $3 \times 10^{-4}$. A full list of the hyperparameters used can be seen in \cref{tab:toyModel_hyperparams}.

\begin{table}[hbt]
    \centering
    \setlength{\tabcolsep}{8pt} 
    \renewcommand{\arraystretch}{1.20} 
    \begin{tabular}{c|c}
         \toprule
         \textbf{Hyperparameter} & \textbf{Value} \\
         \midrule
         Epochs & $300$ \\
         $N$ & $3$ \\
         $N_{\text{head}}$ & $2$ \\
         $D$ & $32$\\
         $D_{\mathrm{ff}}$ & $256$ \\
         $d_{\mathrm{MLP}}$ & $32$ \\
         $p$ & $0.0$ \\
         \bottomrule
    \end{tabular}
    \caption{List of hyperparameters for training a self-attention NN on the four-band model. ``Epochs'' is the number of training iterations. $N$ is the number of attention layers in the NN and $D$ is the dimensionality (\ie\ the ``model width") as defined in \cref{subsubsec: Overall Structure consisting Attention Layers}. $N_{\text{head}}$ is the number of attention heads in a single attention block, as defined in \cref{subsubsec:attentionblock_multipleheads}. $D_{\mathrm{ff}}$ is the size of the hidden layer in the FFN, as defined in \cref{subsubsec:dropout+layernorm}. $p$ is the probability with which elements are set to $0$ in the Dropout operation, according to \cref{eq:Dropout}.}
    \label{tab:toyModel_hyperparams}
\end{table}

\subsubsection{Standardization}
For the translation-invariant case, given $O_{\bsl{k},\alpha\alpha'}^{\bsl{q}=0}$ we construct $X_{\mathrm{init}}$ and $X_{\mathrm{tgt}}$ as in \cref{eq:standardize} and \cref{eq:flattenO}. To standardize the data, we compute the mean and standard deviations across all features $f \in \{1, \dots 2\xi^2\}$. That is, for all $N_\text{sample}$ training samples with $\beta \in \{\text{init}, \text{tgt}\}$:
\begin{equation}\label{eq:toyModel_standardize}
    \mu_f^{(\beta)} \equiv \frac{1}{N_{\mathrm{sample}} L^2 \left(2\xi^2\right)} \sum_{\text{sample}}\sum_{i=1}^{L^2} \sum_{f=1}^{2\xi^2} X_{i,f}^{(\beta)}, 
    \qquad
    \left(\sigma_f^{(\beta)}\right)^2 \equiv \frac{1}{N_{\mathrm{sample}} L^2 \left(2\xi^2\right) - 1} \sum_{\text{sample}} \sum_{i=1}^{L^2}\sum_{f=1}^{2\xi^2} \left( X_{i,f}^{(\beta)} - \mu_f^{(\beta)} \right)^2.
\end{equation}
\cref{eq:toyModel_standardize} replaces \cref{eq:mean stand_dev} for computing $\mu_f^{(\beta)}$ and $\sigma_f^{(\beta)}$ in the four-band model. We then use these values to standardize $X_{\mathrm{init}}$ and $X_{\mathrm{tgt}}$ following \cref{eq:standardize}.

\subsubsection{Weight Initialization}
The initial values for all NN parameters were drawn from the default distributions in PyTorch's initialization scheme. Specifically, we use Kaiming (He) \cite{kaiming} initialization for all weight matrices and bias vectors in the input and output layers (\ie\ $W_{\text{in}}, b_{\text{in}}$ and $W_{\text{out}}, b_{\text{out}}$) as well as in the FFNs (see \cref{eq:FFN}) and the MLPs for $f_\theta$ (from \cref{eq:NN_f_theta}), and in the self-attention mechanism (\ie\ $ W_Q^{(h)},W_K^{(h)},W_V^{(h)},W_O$). Unlike the general case in \cref{eq:concat}, we do not use a learnable bias vector $b_O$ when mixing information across attention heads. For the LayerNorm operation (see \cref{eq:layernorm}), the weights $\gamma$ and the biases $\beta$ are initialized to all $1$'s and all $0$'s, respectively. Finally, we choose the learnable scalar $s$ in \cref{eq:scaled_logits} in each attention head to be initialized to $1$. All learnable parameters
are trained by backpropagation to minimize the loss.

\subsubsection{Results}
We train a self-attention model on 10{,}000 (input,label) 1-RDM pairs at $L=6$ and $L=8$, and evaluate on 1{,}000 held-out systems with $L\in\{10,12,\ldots,50\}$. 
For each test instance we run two HF solvers with identical tolerance and settings: 
(i) initialized from a random 1-RDM, and 
(ii) initialized from the NN-predicted final 1-RDM. 
We report the percent reduction in iteration count.
At $L=50$ the mean reduction is $91.63\%$. 
Results for all $L$ are in \cref{fig:toyModel_itersTest} and \cref{tab:toyModel_iters_summary}.

In order to test how well the NN generalizes to larger system sizes, we also computed the MSE loss between the predicted final 1-RDM from the NN and the true final 1-RDM for up to $L=50$, where the MSE is defined as in \cref{eq:MSE_def}. The results can be seen in \cref{fig:toyModel_generalizationTest}.

\begin{figure}[t]
    \centering
    \includegraphics[width=1.0\linewidth]{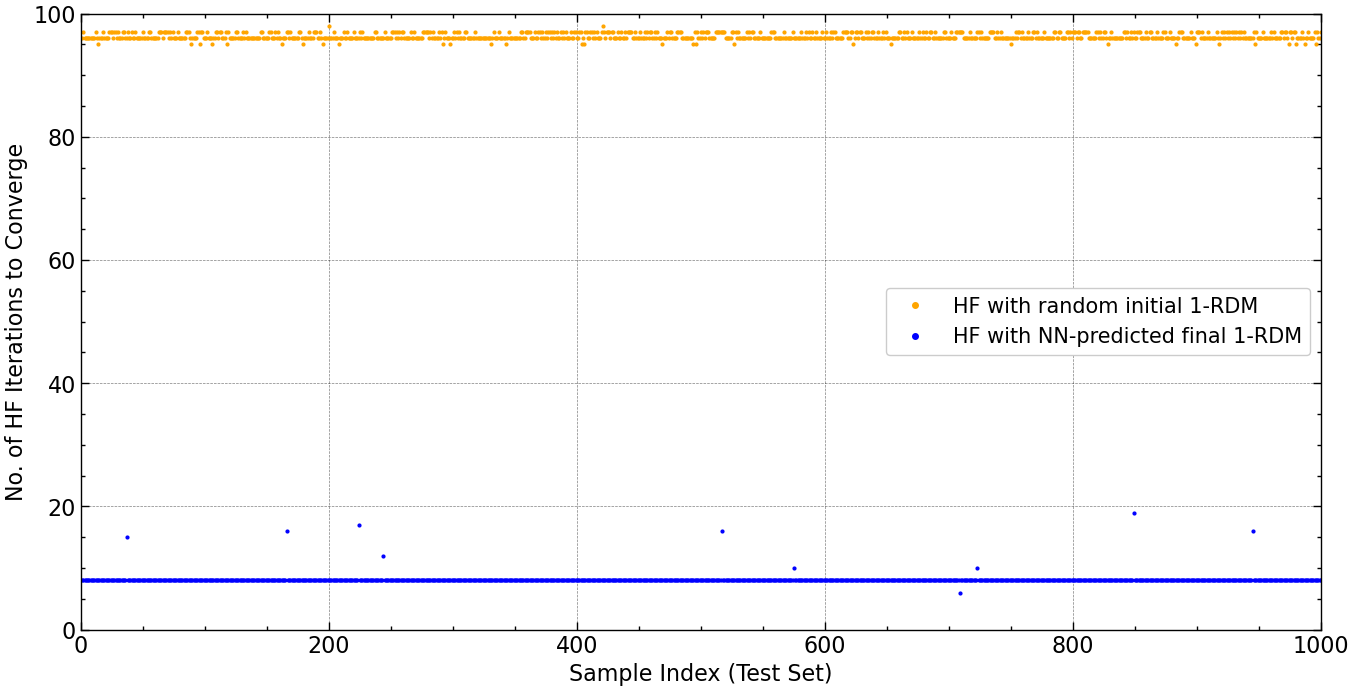}
    \caption{HF iteration test on translationally-invariant four-band model (see \crefrange{eq:toy-real}{eq:toy-update}) for $L=50$ using a self-attention NN trained only on $L=6$ and $L=8$. The x-axis enumerates the 1000 test samples; the y-axis shows the number of HF iterations required to reach the convergence criterion when using the predicted final 1-RDM from the NN as the initial 1-RDM (blue) and when using a random initial 1-RDM (orange).}
    \label{fig:toyModel_itersTest}
\end{figure}

\begin{table}[t]
    \centering
    \setlength{\tabcolsep}{8pt} 
    \renewcommand{\arraystretch}{1.20}  
    \begin{tabular}{@{}c|ccc@{}}
         \toprule
         $\bsl{L}$ & \textbf{Original} & \textbf{with NN} & \textbf{\% Reduction} \\
         \midrule
         $10$ & $97.62$ & $5.63$ & $94.24$ \\
         $12$ & $96.31$ & $7.00$ & $92.73$ \\
         $14$ & $96.58$ & $7.00$ & $92.75$ \\
         $16$ & $96.29$ & $7.00$ & $92.73$ \\
         $18$ & $96.31$ & $7.35$ & $92.36$ \\
         $20$ & $96.32$ & $7.96$ & $91.74$ \\
         $22$ & $96.28$ & $8.00$ & $91.69$ \\
         $24$ & $96.27$ & $8.00$ & $91.69$ \\
         $26$ & $96.23$ & $8.00$ & $91.69$ \\
         $28$ & $96.29$ & $8.00$ & $91.69$ \\
         $30$ & $96.28$ & $7.94$ & $91.75$ \\
         $32$ & $96.28$ & $8.00$ & $91.70$ \\
         $34$ & $96.24$ & $8.00$ & $91.69$ \\
         $36$ & $96.27$ & $8.00$ & $91.69$ \\
         $38$ & $96.27$ & $8.00$ & $91.69$ \\
         $40$ & $96.24$ & $8.02$ & $91.67$ \\
         $42$ & $96.25$ & $8.00$ & $91.69$ \\
         $44$ & $96.23$ & $8.08$ & $91.61$ \\
         $46$ & $96.24$ & $8.04$ & $91.64$ \\
         $48$ & $96.24$ & $8.04$ & $91.64$ \\
         $50$ & $96.30$ & $8.06$ & $91.63$ \\
         \bottomrule
    \end{tabular}
    \caption{Table showing results for the translationally-invariant HF calculation as described in \crefrange{eq:toy-real}{eq:toy-update}, using a self-attention NN trained on $L=6$ and $L=8$ and evaluated on $L=10, 12, \dots 50$. Here, ``Original" refers to the average number of HF iterations with a random initial 1-RDM, while ``with NN" refers to the average number of iterations when using the predicted final 1-RDM from the NN as the initial 1-RDM for HF. ``\% Reduction" refers to the percent difference between the two quantities.}
    \label{tab:toyModel_iters_summary}
\end{table}

\begin{figure}[t]
    \centering
    \includegraphics[width=1.0\linewidth]{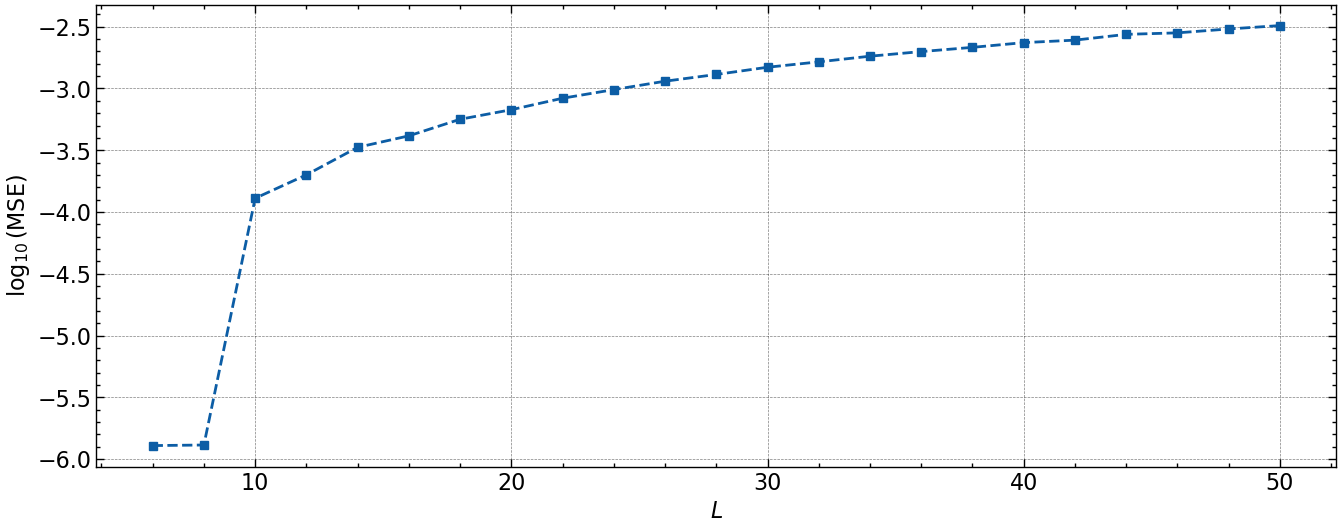}
    \caption{Generalization test for the translationally-invariant HF calculation (see \crefrange{eq:toy-real}{eq:toy-update}) using a self-attention NN trained only on $L=6$ and $L=8$, showing the MSE loss (defined in \cref{eq:MSE_def}) as a function of the system size $L$.}
    \label{fig:toyModel_generalizationTest}
\end{figure}

\section{Square-lattice Hubbard model}
\label{app:hubbard}
\subsection{Model Set-Up}
We study the spinful Hubbard model on a 2D square lattice \cite{1963RSPSA.276..238H, 1964RSPSA.277..237H} with size $N\equiv L \times L$. In real space, 
\begin{equation}
  H = -t \sum_{\langle \bsl r,\bsl r'\rangle,\sigma}\big(c^\dagger_{\bsl r\sigma} c_{\bsl r'\sigma} + \text{h.c.}\big)
  + U \sum_{\bsl r} n_{\bsl r\uparrow} n_{\bsl r\downarrow},
  \label{eq:hub-real}
\end{equation}
with $c_{\bsl r\sigma}$ the annihilation operator at site $\bsl r$ and spin $\sigma\in\{\uparrow,\downarrow\}$, and $n_{\bsl r\sigma}=c^\dagger_{\bsl r\sigma}c_{\bsl r\sigma}$. With the lattice Fourier transform
\begin{equation}
  c_{\bsl r\sigma}=\frac{1}{\sqrt{L^2}}\sum_{\bsl k} e^{i \bsl k\cdot \bsl r}\,c_{\bsl k\sigma},\quad
  c_{\bsl k\sigma}=\frac{1}{\sqrt{L^2}}\sum_{\bsl r} e^{-i \bsl k\cdot \bsl r}\,c_{\bsl r\sigma},\quad
  \bsl k=(k_x,k_y)\in[0,2\pi)^2 .
  \label{eq:fourier}
\end{equation}
The nearest-neighbor hopping $\boldsymbol\delta \in \{\pm \hat{\bsl{x}}, \pm \hat{\bsl{y}}\}$ gives
\begin{align}
H_0
&= -t \sum_{\langle \bsl r,\bsl r'\rangle,\sigma}
   \big(c^\dagger_{\bsl r\sigma} c_{\bsl r'\sigma} + \text{h.c.}\big)
   = -t \sum_{\bsl r,\sigma}\sum_{\boldsymbol\delta\in\{\pm\hat{\bsl x},\pm\hat{\bsl y}\}}
            c^\dagger_{\bsl r\sigma} c_{\bsl r+\boldsymbol\delta,\sigma} \\
&= -\frac{t}{N}\sum_{\bsl r,\sigma}\sum_{\boldsymbol\delta}
   \sum_{\bsl k,\bsl k'} e^{-i\bsl k\cdot \bsl r}\, c^\dagger_{\bsl k\sigma}\,
                                 e^{i\bsl k'\cdot (\bsl r+\boldsymbol\delta)} c_{\bsl k'\sigma} \\
&= -t \sum_{\bsl k,\sigma}
   \left(\sum_{\boldsymbol\delta} e^{\,i\bsl k\cdot \boldsymbol\delta}\right)
   c^\dagger_{\bsl k\sigma} c_{\bsl k\sigma}
   \qquad
   \left(\text{here} \quad \frac{1}{N}\sum_{\bsl r}e^{i(\bsl k'-\bsl k)\cdot\bsl r}
   = \delta_{\bsl k,\bsl k'}\right) \\
&= \sum_{\bsl k,\sigma} \varepsilon(\bsl k)\, c^\dagger_{\bsl k\sigma} c_{\bsl k\sigma},
\qquad
\varepsilon(\bsl k) = -t \sum_{\boldsymbol\delta} e^{\,i\bsl k\cdot \boldsymbol\delta} \\
&= -t\big(e^{ik_x}+e^{-ik_x}+e^{ik_y}+e^{-ik_y}\big)
 = -2t\big(\cos k_x+\cos k_y\big).
\end{align}
Choosing $2t=1$ yields
\begin{equation}
  \varepsilon(\bsl k)=-(\cos k_x+\cos k_y),
  \label{eq:hub-disp}
\end{equation}
so the single-particle block reads
\begin{equation}
  H_0(\bsl k)=\varepsilon(\bsl k)\mathbb{I}_2 ,
  \label{eq:hub-H0}
\end{equation}
with $\mathbb{I}_2$ the identity in spin space. The on-site repulsion is local in $\bsl{r}$:

\begin{align}
H_\text{int}
&= U\sum_{\bsl r} n_{\bsl r\uparrow} n_{\bsl r\downarrow} \nonumber= \frac{U}{L^2}\sum_{\bsl k,\bsl k',\bsl q}
c^{\dagger}_{\bsl k+\bsl q,\uparrow} c_{\bsl k,\uparrow}\,
c^{\dagger}_{\bsl k'-\bsl q,\downarrow} c_{\bsl k',\downarrow}.
\label{eq:hubbard-onsite-k}
\end{align}

The spin-resolved 1-RDM (occupied-band projector) in momentum space is
\begin{equation}
[O_\sigma]_{\bsl k\bsl k'} = \big\langle c^\dagger_{\bsl k'\sigma}\,c_{\bsl k\sigma}\big\rangle,
\qquad \sigma\in\{\uparrow,\downarrow\}.
\label{eq:hub-O-def}
\end{equation}
At temperature $T=0$ and for collinear states $O_\sigma$ is a projector onto the occupied subspace,
\begin{equation}
O_\sigma^\dagger=O_\sigma,\qquad O_\sigma^2=O_\sigma,\qquad 
\mathrm{Tr}\,O_\sigma = N_{\mathrm{occ},\sigma}.
\label{eq:hub-O-proj}
\end{equation}
We set the spin $U(1)$ symmetry not spontaneously broken, \ie, $[O]_{\uparrow\downarrow}=[O]_{\downarrow\uparrow}=0$. We further assume spin fillings to be
\begin{equation}
N_{\mathrm{occ},\uparrow}=N_{\mathrm{occ},\downarrow},
\label{eq:hub-equal-filling}
\end{equation}
but we do not impose translational symmetry.
For the on-site Hubbard interaction, it may generate a mixing between $\bsl k$ and $\bsl k+\bsl Q,$ if translational symmetry spontaneously breaks. At each iteration we diagonalize and refill the $N_{\mathrm{occ},\sigma}$ lowest eigenstates to update 1-RDM. This HF set-up provides true final 1-RDM labels 
(benchmarks to train and validate the ML model), and to quantify HF iteration savings at fixed discretization, filling, and \(U\).

\subsection{ML Results}
Despite spontaneous translation breaking, the order parameters of Hubbard model can be simplified to be similar enough to the translation-invariant case to use the same techniques according to \cref{subsubsec:physical set-up for attNN}. 
 
Label the $L\times L$ BZ grid by integer coordinates \cref{eq:transformedk}.

We investigated two approaches. Firstly, we trained a self-attention NN on the combined $L^2$-long sequence. Additionally, for this model we also trained SIRENs separately on the central diagonal and off-diagonal elements, and then combined the outputs to form the new 1-RDM. For both models, the training data was generated via HF calculations (for systems with small $L$) that had a convergence criterion of $2 \times 10^{-5}$. The trained model was then used to predict a density matrix for larger $L$ which was subsequently used as the initial guess for the density matrix in the HF algorithm, using the same convergence criterion.

The self-attention NN was trained only for $U=1$ (due to limited data), while we also trained additional SIREN models for $U=2$ and $U=3$. In each case, the resultant reduction in the number of HF iterations required for convergence was noted, where the reduction is defined as the percent difference between the average number of HF iterations before and after using the NN outputs as the initial guess for the density matrix.

The training process for both NN architectures is detailed below. \\

\subsubsection{Self-Attention NN}
For each $\bsl k$ we transform 1-RDM into a real vector of length $2\nint^{2}$
and standardize them as described in \cref{subsubsec:flatten_standardize_1RDM} and \cref{subsubsec:physical set-up for attNN}.

For the self-attention NN for the Hubbard model, we initialize $W_{\text{in}}$ and $W_{\text{out}}$ from a truncated normal distribution with $\sigma = 0.02$. Kaiming (He) normal initialization \cite{kaiming} is used for the learnable weight matrices in the feedforward layers, while Xavier (Glorot) initialization \cite{pmlr-v9-glorot10a} is used for each of the matrices $W_Q^{(h)}$, $W_K^{(h)}$, $W_V^{(h)}$, $W_O$ in the attention heads. The matrix of frequencies $W_r$ from \cref{eq:W_r_freqs} is initialized with values drawn from a standard normal distribution. Furthermore, the weights $\gamma$ for the LayerNorm operation from \cref{eq:layernorm}, the vector of coefficients $C$ (\cref{eq:inner_product_C}), the power $p$ (\cref{eq:f_theta}), and the learnable scalar $s$ (\cref{eq:scaled_logits}) are each initialized to be $1$. All bias vectors are set to $0$ initially. After initialization, these learnable parameters are updated by backpropagation to minimize the loss, as indicated in \cref{subsubsec: Overall Structure consisting Attention Layers}.

In order to train the NN, we used a custom learning rate schedule consisting of a linear warmup phase followed by cosine annealing (without restarts) \cite{loshchilov2017sgdr}, which is a well-established method for training self-attention models with adaptive optimizers. This learning rate cycle is repeated twice: initially, we train for $500$ epochs, and then for $5000$ epochs. Each time, the warmup phase lasts for $20$ epochs followed by cosine decay for the rest of the cycle. Thus, we train for a total of $5500$ training epochs. Nevertheless, we found that the loss decreased to a minimum around $3200$ epochs, which is the checkpoint used for evaluation. The self-attention NN was trained on $467$ samples (i.e. input/output pairs) for $L=8$ and $1001$ samples for $L=10$, and performance was observed to vary noticeably based on the random seed used (where ``random seed" refers to the initializer for the pseudorandom number generator(s) of the computer), corroborating the findings in \cite{zhou2024transformersachievelengthgeneralization}. A full list of hyperparameters can be seen in \cref{tab:hubbardModel_selfAttn_hyperparams}. \\

\begin{table}[t]
    \centering
    \setlength{\tabcolsep}{8pt} 
    \renewcommand{\arraystretch}{1.20}  
    \begin{tabular}{c|c}
         \toprule
         \textbf{Hyperparameter} & \textbf{Value} \\
         \midrule
         Epochs & $3201$ \\
         $N$ & $3$ \\
         $N_{\text{head}}$ & $4$ \\
         $D$ & $64$ \\
         $D_{\mathrm{ff}}$ & $128$ \\
         $m$ & $48$ \\
         $p$ & $0.1$ \\
         \bottomrule
    \end{tabular}
    \caption{List of hyperparameters for training a self-attention NN on the Hubbard model. ``Epochs'' is the number of training iterations. $N$ is the number of attention layers in the NN and $D$ is the dimensionality (\ie\ the ``model width") as defined in \cref{subsubsec: Overall Structure consisting Attention Layers}. $N_{\text{head}}$ is the number of attention heads in a single attention block and $m$ is the total number of frequencies, as defined in \cref{subsubsec:attentionblock_multipleheads}. $D_{\mathrm{ff}}$ is the size of the hidden layer in the FFN, as defined in \cref{subsubsec:dropout+layernorm}. $p$ is the probability with which elements are set to $0$ in the Dropout operation in \cref{eq:Dropout}.}
    \label{tab:hubbardModel_selfAttn_hyperparams}
\end{table}

\subsubsection{SIREN}
As aforementioned, each set of ``significant" points forms a tensor of shape $L \times L \times 2 \times 2$. To take full advantage of dimensionality reduction, we trained a separate SIREN for every index of the $2 \times 2$ matrix at each $\bsl{k}$ that had significantly nonzero elements. That is, each SIREN was trained on an $L \times L$ matrix (for the Hubbard model, each of the SIRENs was trained on $L=8$ ), and the outputs were then combined to reconstruct the original shape. Of course, just like with the Richardson model, we extend the size of the matrix to $(L+1) \times (L+1)$ by ``wrapping" around the first element of each dimension in order to satisfy normalization constraints, as described in \cref{subsubsec:Coordinate domain and periodic wrap and symmetry}. Due to the large search space of possible hyperparameter configurations, we again used TPE \cite{TPE_paper} through the Optuna and RayTune libraries to optimize hyperparameters individually for each SIREN model trained in this way.

\subsubsection{Results}
After training, we evaluate acceleration by re-running HF initialized with the NN-predicted final 1-RDM and reporting the percent reduction in iteration count relative to HF started from a random 1-RDM under the same tolerance and settings. Results are shown in \cref{tab:hubbard_summary}.

\begin{table}[t]
    \centering
    \setlength{\tabcolsep}{8pt} 
    \renewcommand{\arraystretch}{1.20}  
    \begin{tabular}{@{}rccccccccccc@{}}
         \toprule
         $L$ & $10$ & $12$ & $14$ & $16$ & $18$ & $20$ & $22$ & $24$ & $26$ & $28$ & $30$ \\
         \midrule
         \textbf{Self-Attention NN, }$\bsl{U=1}$ & \text{N/A} & $66.19$ & $73.51$ & $76.63$ & $80.63$ & $83.62$ & $84.93$ & $82.55$ & $81.55$ & $81.70$ & $79.89$ \\
         \textbf{SIREN, }$\bsl{U=1}$ & $60.79$ & $64.34$ & $73.25$ & $73.15$ & $73.27$ & $77.48$ & $88.70$ & $75.72$ & $77.21$ & $78.81$ & $79.25$ \\
         \textbf{SIREN, }$\bsl{U=2}$ & $82.94$ & $83.27$ & $83.16$ & $83.84$ & $83.41$ & $92.59$ & $86.27$ & $85.84$ & $86.33$ & $86.36$ & $92.78$ \\
         \textbf{SIREN, }$\bsl{U=3}$ & $81.25$ & $81.64$ & $82.30$ & $82.11$ & $82.61$ & $83.00$ & $82.79$ & $82.96$ & $82.97$ & $83.18$ & $83.24$ \\
         \bottomrule
    \end{tabular}
    \caption{\% Reduction across various system sizes ($10 \leq L \leq 30$) for the Hubbard model. For $U=1$, results are shown using both the self-attention NN and the SIREN, while for $U=2$ and $U=3$ we only obtained results for the SIREN due to limited training data.}
    \label{tab:hubbard_summary}
\end{table}

\fi


\end{document}